\documentclass[a4paper,12pt]{article}
\usepackage[usenames,dvipsnames]{color}

\usepackage{amsmath,graphicx,rotating,ctable,verbatim, amsfonts, dsfont,amssymb}
\usepackage{eurosym}
\usepackage[margin=1in]{geometry}
\usepackage{hyperref}
\usepackage{dcolumn}
\usepackage{bm,anyfontsize}
\usepackage{pgf}
\usepackage{upgreek}
\usepackage{natbib}
\usepackage{amsthm}
\usepackage{caption, subcaption,multirow,wrapfig}
\usepackage{setspace}

\usepackage[flushleft]{threeparttable}

\hypersetup{colorlinks=true,linkcolor=blue,urlcolor=cyan, pagebackref=true,pdfpagemode=UseNone}

\DeclareMathOperator*{\E}{\mathbb{E}}

\DeclareMathOperator{\sone}{\mathds{\footnotesize{1}}}

\newcolumntype{.}{D{.}{.}{-1}}
\newcolumntype{d}[1]{D{.}{.}{#1}}

\begin{document}

\title{{\fontsize{20}{25}\selectfont Regime and Treatment Effects in Duration Models: Decomposing Expectation and Transplant Effects on the Kidney Waitlist}\\
}
\author{ Stephen \textsc{Kastoryano}\thanks{University of Reading, IZA, e-mail:
\href{mailto:s.p.kastoryano@reading.ac.uk}{s.p.kastoryano@reading.ac.uk}.
 \newline Thanks to Jad Beyhum, Petyo Bonev, Christoph Rothe, Jaap Abbring, Sang Yoon (Tim) Lee, Bas van der Klaauw, Christoph Breunig, and Priya Urs for valuable comments and help.

  The data reported here have been supplied by the Hennepin Healthcare Research Institute (HHRI) as the contractor for the Scientific Registry of Transplant Recipients (SRTR). The interpretation and reporting of these data are the responsibility of the author(s) and in no way should be seen as an official policy of or interpretation by the SRTR or the U.S. Government. The research protocol was approved by the IRB of the University of Reading.}}

\date{\today}

\maketitle

\begin{abstract}

\thispagestyle{empty} \noindent

This paper proposes a causal decomposition framework for settings in which an initial regime randomization influences the timing of a treatment duration. The initial randomization and treatment affect in turn a duration outcome of interest. Our empirical application considers the survival of individuals on the kidney transplant waitlist. Upon entering the waitlist, individuals with an AB blood type, who are universal recipients, are effectively randomized to a regime with a higher propensity to rapidly receive a kidney transplant. Our dynamic potential outcomes framework allows us to identify the pre-transplant effect of the blood type, and the transplant effects depending on blood type. We further develop dynamic assumptions which build on the LATE framework and allow researchers to separate effects for different population substrata. Our main empirical result is that AB blood type candidates display a higher pre-transplant mortality. We provide evidence that this effect is due to behavioural changes rather than biological differences.
\end{abstract}

\vspace{1cm}



\newpage


\section{Introduction}
\setcounter{page}{1}

Experiments and quasi-experiments form the cornerstone of microeconometric evaluations. These methods use some exogenous variation to analyse the ex-post effect of a policy intervention, or treatment, on one or several outcomes of interest. Ex-post effects are usually measured within a specific information setting or are averaged across many information settings. For example, in randomized controlled trials, treatment is usually administered to previously unsuspecting subjects. In contrast, when analysing the effect of kidney transplants, different candidates may have widely varying expectations of when they may find a suitable donated kidney. Ex-post effects of a kidney transplant usually represent averages over all these individual beliefs.

An important policy question, when it comes to rolling out a new treatment or adapting an existing one, is whether these ex-post treatment effects remain invariant when changing the information setting. This question is particularly relevant in economics where we assume forward-looking agents can take the information on the assignment mechanism into account and change their expectations or behaviour prior to receiving the treatment (Rosenzweig and Wolpin, \hyperlink{RosWo2000}{2000}). These pre-treatment changes can alter the composition of individuals receiving treatment but also the ex-post effect of the treatment (Chassang, et.\ al., \hyperlink{ChaEA2012}{2012}). Static treatment evaluation methods are well developed but poorly accommodate these types of dynamic situations in which there is some time between an initial randomization to an (information) regime and the actual treatment intervention.

This paper adapts the analysis of heterogenous treatment effects to discuss the causal decomposition of a sequence of two randomizations when the treatment and outcome of interest are both duration variables. More specifically, we consider a situation in which individuals are randomized to a \emph{regime} upon entering a state at time $0$. The regime dictates a stochastic propensity for the timing of a future \emph{treatment} among agents. Thereafter, at different moments in time and depending upon their regime, surviving agents are randomized to actually receive treatment. This paper proposes a decomposition which allows researchers or policymakers to compare how the regime and the treatment influence an individual's probability of survival in the initial state.

\begin{figure}
  \begin{subfigure}{0.48\textwidth}
    \includegraphics[width=\linewidth]{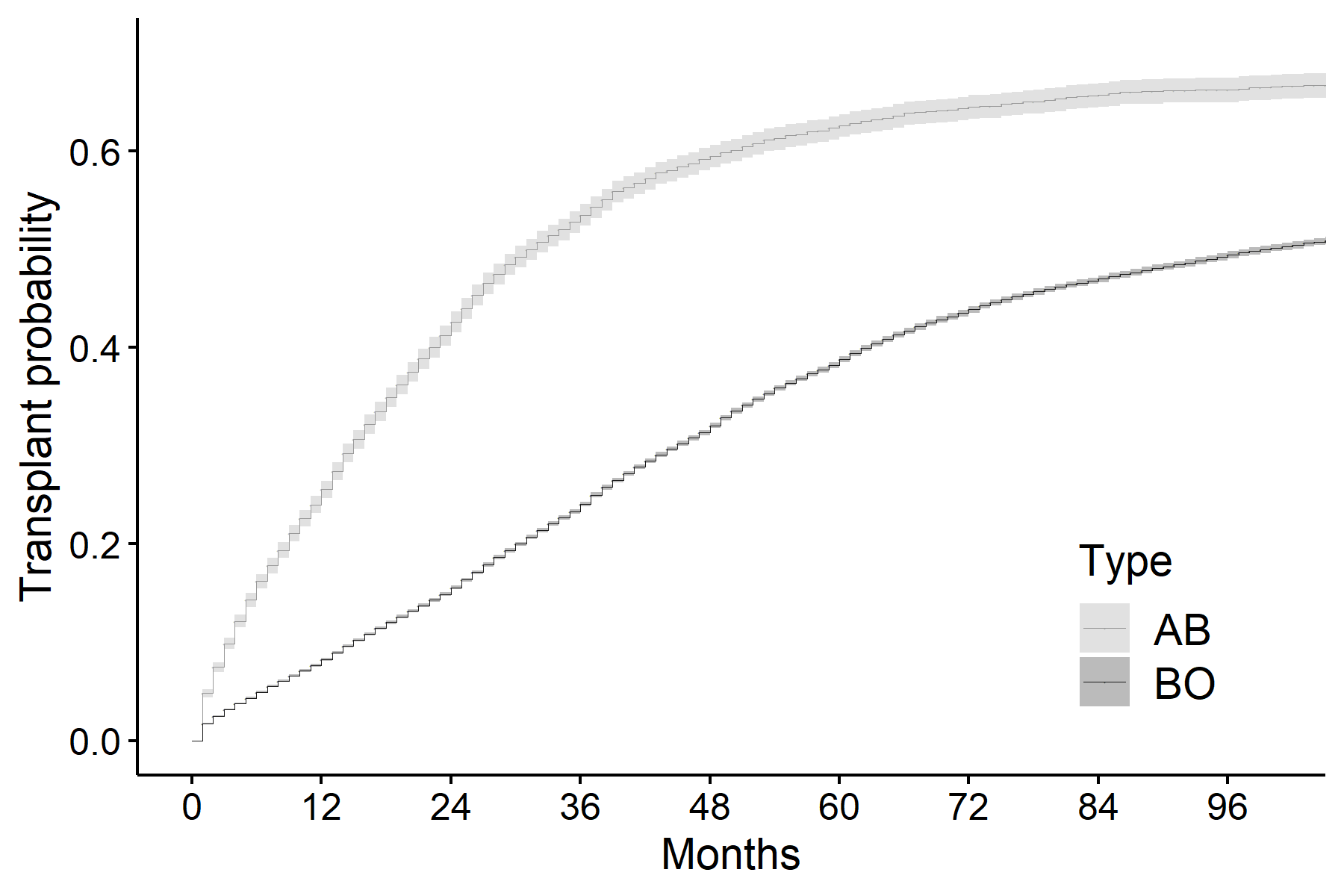}
    \caption{Transplant Probability} \label{fig:SurvT_main}
  \end{subfigure}%
  \hspace*{\fill}   
  \begin{subfigure}{0.48\textwidth}
    \includegraphics[width=\linewidth]{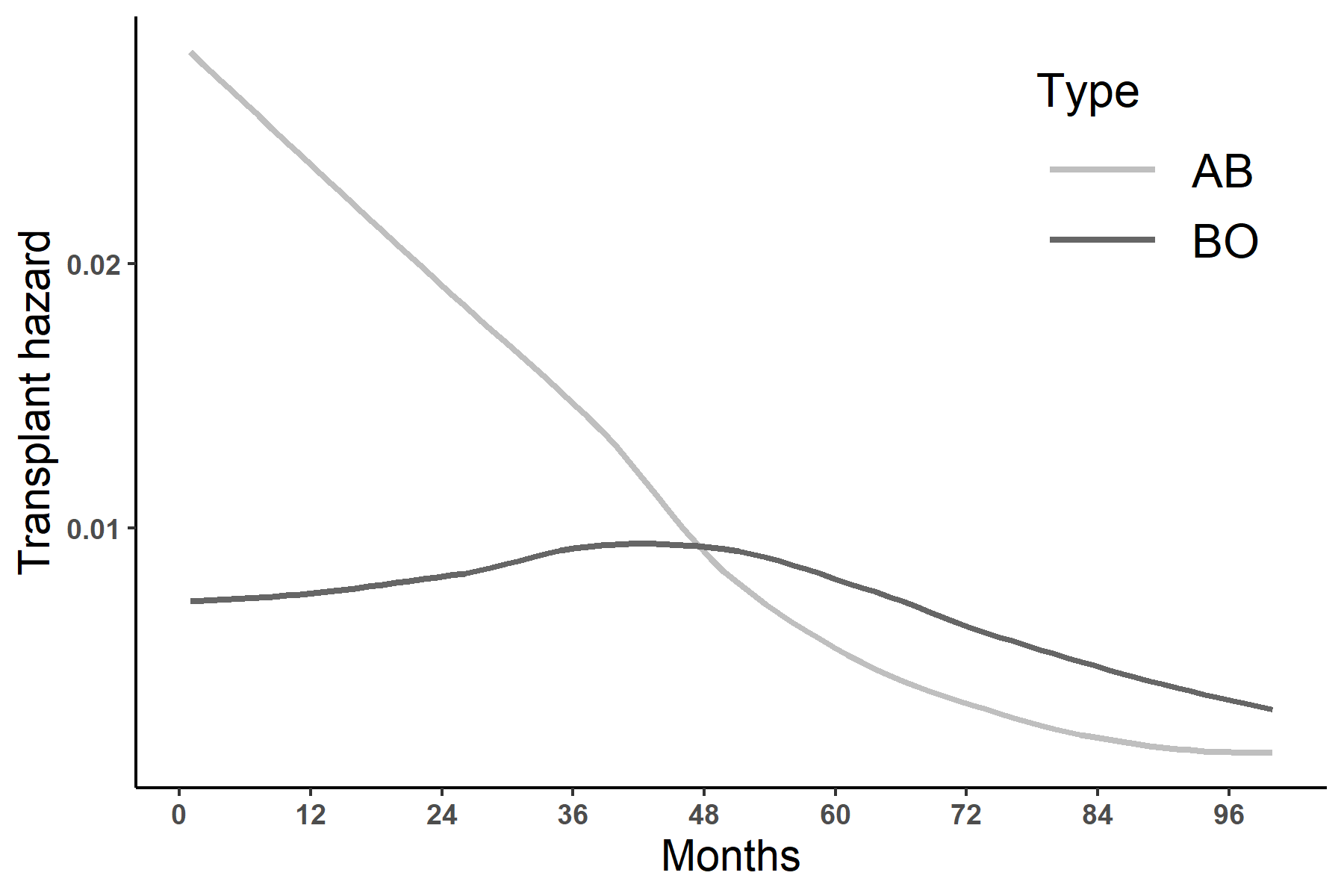}
    \caption{Transplant Hazard} \label{fig:Thaz_main}
  \end{subfigure}%
  \newline
  \begin{subfigure}{0.48\textwidth}
    \includegraphics[width=\linewidth]{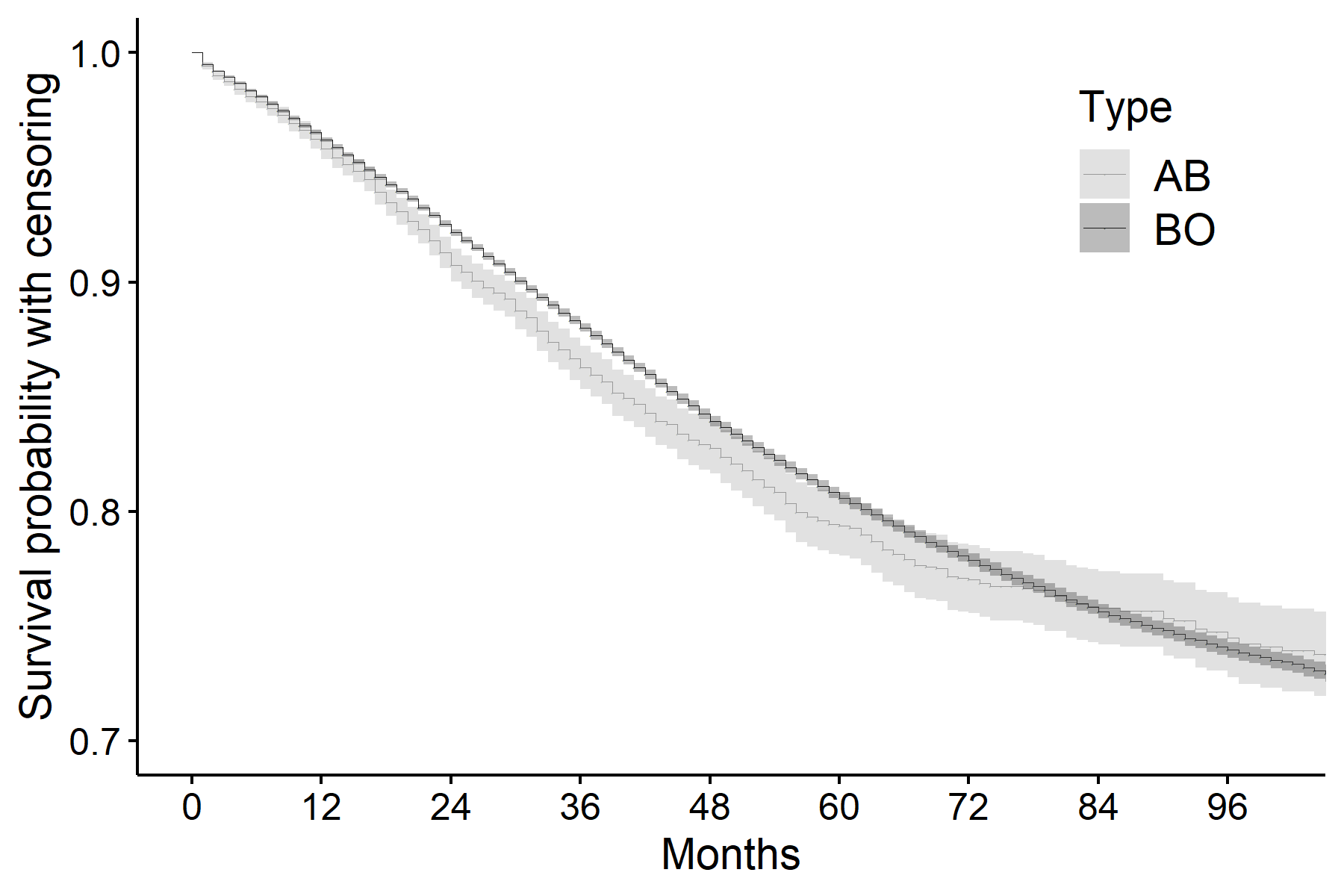}
    \caption{Pre-Transplant Survival} \label{fig:SurvC_main}
  \end{subfigure}%
  \hspace*{\fill}   
  \begin{subfigure}{0.48\textwidth}
    \includegraphics[width=\linewidth]{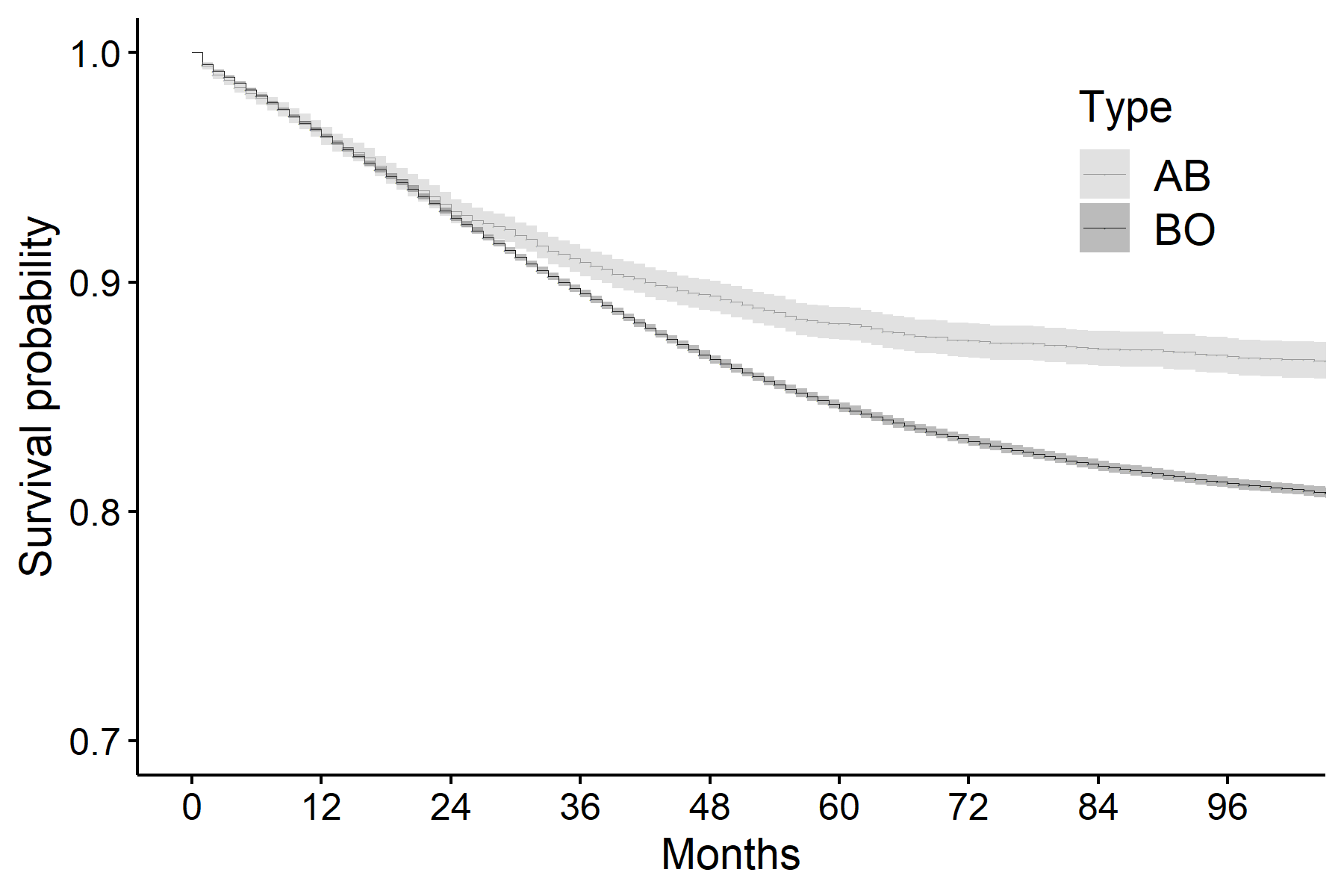}
    \caption{Survival} \label{fig:Surv_main}
  \end{subfigure}%

\caption{Survival of candidates on the kidney transplant waitlist \\
    \scriptsize{Based on selected sample of Scientific Registry of Transplant Recipients data. Selected sample is described in section \ref{sec:data}. $N_{B/O: no-Tr}=118,422$, $N_{B/O: Tr}=40,158$, $N_{AB: no-Tr}=4976$, $N_{AB: Tr}=4816$.}} \label{fig:SurvAll_main}
\end{figure}

We develop and apply our framework in the context of the survival of individuals on the kidney transplant waitlist. In our empirical application, we take the regime as the biological randomization of individuals to a given blood type. Blood type, as we will discuss, is seemingly random with respect to the survival outcome conditional on socio-economic characteristics. However, it stochastically influences the time until an individual receives an offer of a kidney. This can be seen in Figures \ref{fig:SurvT_main} and \ref{fig:Thaz_main}. Individuals on the kidney transplant waitlist with an AB blood type have a higher probability and a higher hazard of finding a suitable kidney than B and O blood types\footnote{AB blood types are universal recipients. O blood types are universal donors. B blood types tend to be under-represented in the pool of kidney donors. As a result, B and O blood types follow very similar treatment survival curves. In appendix \ref{sec: DataStats} we present the transplant Kaplan-Meier curves for each blood type separately.} (B/O hereafter) at every period over the first 4 years on the waitlist.

These differences are paired with curious pre-transplant trends in survival. Figure \ref{fig:SurvC_main} plots survival rates for AB-types and B/O-types when censoring the duration to death at the moment of receiving a kidney transplant. We see that over the first 4-5 years on the waitlist, the AB-type candidates have a significantly lower survival rate than B/O-type candidates prior to receiving a kidney transplant. One explanation for this previously undocumented difference could be behavioural. Individuals may be reacting to some notion of the relation between their blood type and their propensity to receive a kidney. Another explanation for this difference is simply mechanical. For example, the kidney allocation system prioritizes high potential life expectancy candidates when offering kidneys (Agarwal, et.\ al., \hyperlink{Agar2020}{2020}). This implies that observations censored by transplant in Figure \ref{fig:SurvC_main} would on average display higher survival rates. Since a larger share of these high potential survival observations are censored for the AB blood group, the difference in pre-treatment mortality trends between blood types may be largely due to selection. Without a formal causal framework, it is not possible to distinguish a behavioural or substantive explanation from a mechanical one. It is also not possible when looking at the survival rates of AB vs.\ B/O blood types depicted in Figure \ref{fig:Surv_main} to determine how much of the difference in survival is due to the positive effect of a transplant, and how much is due to causal pre-transplant differences in mortality.

Our dynamic potential outcomes framework allows researchers to formally decompose these different effects. More specifically, our framework draws from several strands of literature to provide non-parametric identification of: (i) the regime effect (blood type) on the probability of survival, (ii) the ex-post baseline effect of actually receiving treatment (a kidney transplant) on the probability of survival within a given regime, and (iii) the additional ex-post interaction effect of the regime and actually receiving treatment. This last effect may be non-zero if changes in behaviour, due to differences in blood type, result in different surviving populations of kidney recipients at any given time. It may also be non-zero if the changes in behaviour interact with the ex-post effect of the kidney transplant.

The paper begins by describing our setting and presents the relevant decomposition of causal effects. We then discuss a version of the well-known g-computation assumptions for discrete time identification developed by Robins (\hyperlink{Rob1986}{1986}; \hyperlink{Rob1997}{1997}), with initial contributions in economics for duration models by Abbring and van den Berg (\hyperlink{AbvdB2003a}{2003a}), Lechner (\hyperlink{Lec2009}{2009}) and Lechner and Miquel (\hyperlink{LeMiq2010}{2010}). We explain that these dynamic assumptions already allow us to identify policy relevant aggregate effects in the decomposition.

In some cases, researchers or policymakers may be additionally interested in effects for substrata of individuals. In particular, since many candidates on the kidney waitlist will die before receiving a transplant, researchers may be interested in evaluating the decomposition for individuals who actually receive a transplant. In addition, the initial set of assumptions does not allow researchers to disentangle whether varying treatment effects across regimes are due to differences in dynamic selection across regimes or due to substantive differences in treatment effects, known as state dependence.

To address these points in our decomposition, we develop dynamic assumptions in the spirit of the LATE literature (Imbens and Angrist, \hyperlink{ImAng1994}{1994}). The first assumption invokes a dynamic variation of a rank invariance assumption (Matzkin, \hyperlink{Mat2003}{2003}; Chernozhukov and Hansen, \hyperlink{ChHan2005}{2005}) on the timing of exit. This assumption acts in our framework as the reciprocal to the monotonicity assumption in the LATE framework by reducing four possible population substrata of survivors down to three. In a second step, researchers can call upon a variation of an exclusion restriction which imposes that ex-post effects, in contrast to potential outcomes, are invariant across regimes for each individual. Under the exclusion restriction assumption, we nullify any state dependence leaving the interaction effect to represent only differences in dynamic selection across regimes. These two assumptions also allow researchers to identify the three causal effects in our framework for the three substrata of the population identified by the rank invariance assumption. These additional assumptions are assumptions on unobserved variables. They cannot be externally manipulated in an experimental setting and should therefore be called upon with caution.

We propose an estimation method building on a semi-parametric proportional hazard model and present some simulation results to assess the performance of our estimator.

Our empirical application is intended as an intuitive illustration for the proposed framework. We provide guidelines for implementing the proposed estimation method and for interpreting the results. The main insight from our empirical application concerns expectation effects for candidates on the kidney waitlist. We find that individuals with an AB blood type have a lower survival rate pre-transplant than B/O types. Our set of results suggests that this effect is not driven by biological factors linked to blood type but is driven by changes in behaviour. We validate these results by exploiting a discontinuity in the kidney allocation system depending on the level of preexisting incompatible antibodies of some candidates. As in the case of blood-types, these differences in antibodies imply that, all else equal, some candidates are more likely to rapidly receive a transplant than others. We further exploit the fact that these discontinuities changed upon a kidney allocation reform in 2014. These results contribute to an ever growing field considering expectation effects in health economics (see Haaland, et.\ al., \hyperlink{Haetal2021}{2021} for a survey).\footnote{This body of research includes public messaging and perception effects in the context of the coronavirus pandemic (Fetzer, et.\ al., \hyperlink{Fetz2021}{2021}; Faia, et.\ al., \hyperlink{Faia2021}{2021}). It also includes information interventions which have been central to improving dietary and safety decisions in developing countries (Madajewicz, et.\ al., \hyperlink{Mada2007}{2007}; Fitzsimons, et.\ al., \hyperlink{Fetz2016}{2016}). The topic of expectation effects has received particular attention in regard to the relation between (sexual) behavioural and information about HIV (Dupas \hyperlink{Dupa2011}{2011}; Delavande, et.\ al., \hyperlink{Dela2016}{2016}).}

The proposed framework intends to find a middle ground between two strands of literature. First, we draw from static causal inference approaches, which are concerned with the use of a minimal set of assumptions to obtain credible causal effects. Secondly, we also wish to integrate the concerns of the macro-structural modelling approach, which emphasizes equilibrium effects of policy changes. The methods in this paper are relevant to researchers interested in decomposing causal effects in duration models with minimal functional form assumptions about agent utilities, expectations and underlying search processes. Topics of application include health interventions, development evaluations, active labor market programs, or topics in finance, among others.

The remainder of the paper proceeds as follows. We begin in the next section by briefly introducing the kidney transplant allocation system. Section \ref{sec:eval} introduces the potential outcomes framework, describes causal effects of interest, presents identifying assumptions, develops the non-parametric identification results, and describes our estimation approach. Section \ref{sec:empir} presents our empirical application and results.

\section{Kidney Transplant Allocation System}
\label{sec:back}

Transplantation is the most effective method to treat end-stage renal disease. Dialysis, the main alternative, is associated with higher mortality and higher overall costs (Wolfe, et.\ al., \hyperlink{Wolf2008}{2008}, Held, et.\ al., \hyperlink{Held2016}{2016}). The main problem with transplantation is the under-supply of healthy donated kidneys. In the United States, there are around 146,000 candidates on the kidney transplant waitlist, with around 40'000 added each year but only 25,000 of which receive a kidney transplant each year (OPTN/SRTR, \hyperlink{OPTN2019}{2019}).

Much of the early interest in kidney transplants from economists focused on kidney transplant exchange programs among living donors (see, e.g., Roth, et.\ al., \hyperlink{Roth2004}{2004}; Roth, et.\ al., \hyperlink{Roth2007}{2007}; Agarwal, et.\ al., \hyperlink{Agar2019}{2019}). These exchange programs, along with the larger market for living donor kidneys, represent only a fraction of recorded transplantations. The majority, over 70\%, of kidneys available for transplant are from deceased donors (OPTN/SRTR, \hyperlink{OPTN2019}{2019}).\footnote{Dickert-Conlin, et.\ al., (\hyperlink{Dick2019}{2019}) discuss how demand of kidneys to local supply while Teltser (\hyperlink{Telt2019}{2019}) provides further insight into how kidney exchanges increases the quantity and quality of donor kidneys.}

In the United States, the allocation of deceased kidneys to candidates on the waitlist is designed, coordinated, and administered by the Organ Procurement and Transplantation Network (OPTN). A person is allowed to register on the kidney transplant waitlist when their kidney function falls below 20\%. Upon entering the waitlist, OPTN collects in depth information about the candidate's health conditions,
immunological profile, and any other characteristic needed to compute priority.

When a potential kidney becomes available,\footnote{Deceased donor kidneys are usually gathered, after consent, from donors who experienced severe brain bleeding, have been declared brain dead, or for whom cardiac death is imminent.} OPTN gather extensive biological information about the kidney and the donor's medical history. Through its automated UNet system, OPTN then calculate a priority order for any candidate who is compatible with the potential kidney. The process balances equity and efficiency, prioritising candidates on the waitlist following strict criteria. Unlike some other organs, such as livers and hearts, priority is also not given to candidates who need a transplant most urgently. Prior to December 2014, the main prioritisation criteria were the time a candidate had been on the waitlist, as well as geographic proximity to the center holding the deceased donated kidney. Since 2014, in an effort to reduce the loss of healthy donated kidneys, healthier candidates were given higher priority for the highest potential longevity kidneys. Under the new allocation rules, less emphasis is placed on waiting time and geographical location (Israni, et.\ al., \hyperlink{Isra2014}{2014}). In addition, individuals with high calculated Panel Reactive
Antibodies (cPRA), a measure of immune sensitivity,\footnote{cPRA estimates the percentage of donors with whom a particular recipient would be incompatible due to preexisting antibodies.} were given a higher priority. This change in cPRA priority rules, which converted from a discontinuous cutoff to an exponential prioritization, will be exploited in a secondary part of our analysis. Key to our future assumptions, the UNet mechanism determining a kidney offer is a function of observed variables which are available to us in our data.

There also exists some discretion on the part of candidates concerning which kidney they are willing to accept. This has led to more recent work on optimal allocation mechanisms for the deceased donor waitlist given candidate discretion in rejecting offers (Zhang, \hyperlink{Zhan2010}{2010}; Agarwal, et.\ al., \hyperlink{Agar2018}{2018}; Agarwal, et.\ al., \hyperlink{Agar2020}{2020}; Agarwal, et.\ al., \hyperlink{Agar2021}{2021}). Most kidneys will be rejected automatically based on age, health, and kidney function criteria set by candidates upon entering the waitlist, and available in our data. In addition, some offered kidneys which pass the initial preference criteria will still be rejected by candidates, a decision typically made in consultation with a transplant surgeon (Gordon, et.\ al., \hyperlink{Gord2020}{2020}). Decisions to reject offers have no bearing on future offers and most rejected offers are for undesirable kidneys which, based on observed variables, pose high risks of future complications.\footnote{Several factors determine a kidney's quality in general. These include the donor's general health prior to death, their gender and age, as well as their kidney function and anatomy. Some donors die with infectious diseases which the candidate may contract if they proceed with a transplant. See Danovitch (\hyperlink{Dano2009}{2009}) for a comprehensive review of kidney biology.} As a result of rejected offers, upwards of 20\% of kidneys are never transplanted. We return to this choice problem when discussing our identifying assumptions.

Underlying the feasibility of a kidney transplant is the need for compatibility between the donor and the recipient's blood and human leukocyte antigen (HLA) tissue types. Without sufficiently high compatibility on both, a donated organ will generate an immune response and result in graft failure.\footnote{ABO incompatible transplants are possible but remain rare (de Weerd and Betjes, \hyperlink{deWe2018}{2018}).} As a result of differences in blood type, HLA-tissue type, and immune sensitisation (cPRA), candidates will face differences in their likelihood of finding a suitable donated kidney. To date, no study has considered differences in transplant perceptions or behaviour depending on candidates' blood type or immune sensitisation.

\section{Evaluation Framework}
\label{sec:eval}

In this section we introduce the potential outcomes notation, define causal effects and develop the causal decomposition framework. Throughout the discussion, we relate the framework and assumptions to our empirical setting of individuals on the kidney transplant waitlist.

\subsection{Potential Outcomes and Treatment Effects}

Our potential outcomes framework combines the dynamic presentation of Abbring and Heckman (\hyperlink{AbHec2007}{2007}) with the more familiar notation in static treatment and mediation analyses (Angrist, Imbens and Rubin, \hyperlink{AngEA1996}{1996}; Imai, Keele and Yamamoto, \hyperlink{ImaEA2010}{2010}). We follow each agent from the moment of entering an initial state which is set as $t = 0$. In our empirical setting this will be the moment an individual enters the kidney transplant waitlist, but in other settings it could be the moment of contracting an illness, becoming unemployed, or advertising a new product on the market. Since identification is non-parametric, we leave the conditioning on observed baseline ($t=0$) covariates implicit for notational convenience. However, we do refer to covariates when discussing the problems of intermediate or time-varying covariates.

\subsubsection{Regime and Potential Treatment Duration}

At $t=0$, agents are randomized to one of two regimes denoted by the random variable $Z$. $Z$ is observed by the econometrician and may be partially or fully known by the agents. We focus on the most simple setting in which agents can either be assigned to a baseline regime $Z=0$ or to an alternative regime $Z=1$. The regime is a special type of randomization which influences the timing of a future treatment by introducing administrative or other constraints. In our kidney transplant setting, $Z=0$ for AB blood type individuals who face a high probability of rapidly finding a suitable kidney. $Z=1$ for B/O blood type individuals who have a lower probability of rapidly finding a suitable kidney.

To define durations to treatment we use a potential outcomes notation. Let the random variable $S^z$ be the potential duration to treatment, or equivalently the treatment time, had an agent been subject to regime $Z=z$. The treatment time can take on any value $s > 0$. Given our description of the data generating process, our setup parallels a characteristic from the instrumental variable setting in that $\Pr(S^0 \leq s)=\Pr(S^1 \leq s)$ does not hold for all $s$. We consider the simplified setting in which there is only a single binary treatment which once allocated remains permanently thereafter. In our kidney transplant setting, $s$ is the duration of time on the waitlist until an individual receives a kidney transplant. One important peculiarity about the timing of treatment is that it is stochastic from the point of view of the agent, even if it might be deterministic with respect to a data generating process.  So a person may know that they have a low probability of finding a suitable kidney, but will not know the exact time $s$ at which they will receive a kidney transplant.

\subsubsection{Potential Exit Duration}

We further define the potential outcome $T^{z,s}$ as the duration to exit had the agent been subject to regime $Z=z$, and had he been treated at $S^z=s$. In our example, $T^{1,6 \text{ months}}$ would be the potential time to death for a person on the kidney transplant waiting list with B/O blood type ($Z=1$), and who received a kidney transplant after $6$ months on the waitlist ($S^1=6$ months). Several comparisons can be made based on this notation. In this paper, our object of interest is the probability of survival past a period $\tau$ given treatment at $s$. Any comparison therefore requires formulating counterfactual potential outcomes of the form $\Pr(T^{z,s} > \tau)$.\footnote{In this paper our outcome of interest is the probability of survival instead of the expected exit duration $\E[T^{z,s} ]$. Focusing on the expected duration to exit may be hampered by the fact that in many duration settings, as is the case in our application, a large fraction of exit outcomes are censored so the right tail of the exit distribution will be poorly approximated. Also, the expected potential exit duration outcomes can all be expressed as functions of the probability to exit, $\Pr(T^{z,s} \leq \tau)= 1-\Pr(T^{z,s} > \tau)$. Alternatively, some studies focus on the relative effects of the potential hazard as causal effects. The drawback of focusing directly on the hazard to infer about causal effects is that its magnitude is difficult to interpret for cost-benefit analyses without transforming it into a survivor function.}

The usual problem with formulating causal comparisons with this potential outcome is that counterfactuals are never observed. In static settings this is because an individual can only be observed in one of two states of the world: treated or non-treated. In our dynamic setting there are a possibly infinite number of unobserved counterfactuals, one for each treatment time $s$ in each of the regimes $z$. Furthermore, there is dynamic selection in the sense that unobserved variables will influence the timing of exit, and some agents will exit before ever receiving treatment. In our kidney transplant setting, this means the population which receives a kidney transplant at $s$ is not necessarily the same in terms of unobservables as the population which receives a kidney transplant at any other time $s^*$, nor as the population of individuals who entered the kidney transplant waitlist at $t=0$. These factors combined imply that for any comparison $\Pr(T^{z,s}  > \tau)-\Pr(T^{z,s^*}  > \tau)$ of treated populations at two points $s,s^* < \tau$ in a given regime $z$, we cannot know whether a difference in effect is due to an actual change in the causal effect of treatment, known as state dependence, or due to a change in the composition of treated due to dynamic selection (Heckman and Singer, \hyperlink{HeSin1984}{1984}). Such comparisons are therefore of limited policy relevance.

Instead, we use the shorthand notation $T^{z,\infty}$ to define the potential duration to exit had the agent been exposed to regime $Z=z$ and had not received treatment in finite time ($s \to \infty$), or, more concretely, by the time $\tau$ at which the outcome is evaluated. The `no-anticipation' assumption of Abbring and van den Berg (\hyperlink{AbvdB2003a}{2003a}), $\Pr(T_i^{z,s} > \tau)=\Pr(T_i^{z,s'} > \tau)$ for all $s, s' < \tau$, is often elicited to define this counterfactual. In our setting, $T^{z,\infty}$ can be understood as the reciprocal to the non-treated potential outcome in a static setting.

\subsubsection{Regime and Treatment Effect Decomposition}

Using this notation, we can already define our decomposition at a treatment time s:
\begin{equation} \label{eq:decomp} \small
\begin{split}
\Pr(T^{z,s}  > \tau)&=\beta_0 + \beta_z z + \beta_s \sone(S=s) + \beta_{zs}z \sone(S=s)\\
\text{with  } &\beta_0= \Pr(T^{0,\infty}  > \tau)\\
&\beta_z= [\Pr(T^{1,\infty}  > \tau)-\Pr(T^{0,\infty}  > \tau)]\\
&\beta_s=[\Pr(T^{0,s}  > \tau)-\Pr(T^{0,\infty}  > \tau)]\\
&\beta_{zs}=[(\Pr(T^{1,s}  > \tau)-\Pr(T^{1,\infty}  > \tau)) - (\Pr(T^{0,s}  > \tau)-\Pr(T^{0,\infty}  > \tau))]
\end{split}
\end{equation}

This decomposition parallels the well known decomposition when evaluating heterogenous treatment effects. The above formulation may not however be the most policy relevant decomposition. A more relevant decomposition should average effects over a treatment time interval, such as treatment before period $s$. When producing the decomposition relative to treatment in regime $Z=1$, the average effects over the treatment interval $(0,s]$ are given by $\sum_{t=0}^{s} \cdot \Pr(S^1  = s) \Pr(T^{z,s}  > \tau)$. Our average policy parameters of interest would then be $\beta_0$, $\beta_z$, $\beta_{(0,s]} = \sum_{t=0}^{s} \Pr(S^1  = s) \cdot \beta_s$ and $\beta_{z(0,s]} = \sum_{t=0}^{s} \Pr(S^1  = s) \cdot \beta_{zs}$.

As in a heterogeneity decomposition, $\beta_0$ represents the average probability of survival past $\tau$ had all candidates been of AB blood type and had none of them received a kidney transplant. $\beta_z$ represents the difference in the probability of survival past $\tau$ for candidates with a B/O blood type relative to those with an AB blood type, withholding any effects of actually receiving a kidney transplant. $\beta_{(0,s]}$ represents the difference in the probability of survival past $\tau$ for candidates with blood type AB receiving a kidney transplant before $s$ relative to candidates who did not receive a kidney transplant by $\tau$. $\beta_{z(0,s]}$ represents the additional difference in the probability of survival past $\tau$ for B/O type candidates who receive a kidney transplant before $s$ relative to $\beta_{(0,s]}$.

\subsection{Non-Parametric Identification of Causal Effects}

In this section, we discuss the nonparametric identification of dynamic treatment effects originally formulated in Robins (\hyperlink{Rob1986}{1986}; \hyperlink{Rob1997}{1997}). The evaluation framework will be presented in discrete time with exits and treatment times $t\in \mathbb{N}$ and $s\in\mathbb{N}$. While not fundamental to identification, discrete time allows for a more intuitive presentation which parallels assumptions from the static causal inference setting. To reduce notational burden we assume the observed period interval is the smallest discrete unit of time.

For each agent $i$, the econometrician would like to observe the joint distribution $(Z_i, S_i, T_i, X_i)$ where $Z_i$ is the regime assignment, $S_i$ the time to treatment, $T_i$ the time to exit, and $X_i$ is a set of baseline covariates at $t=0$. However, this joint distribution is usually not fully observed for agents who exit before treatment. For these agents we only know that $T_i<S_i$. We suppress hereafter the subscript $i$.

\subsubsection{Dynamic Unconfoundedness Assumption}
\label{se:Unconfound}

The first identification requirement is that the regime is randomized and that treatments are randomized on survivors. For this we invoke the following dynamic unconfoundedness assumptions,
\begin{center} \begin{minipage}{16.3cm}
\begin{flushleft}
\emph{\textbf{Assumption A.I: Dynamic Unconfoundedness:} For  $z=0,1$, $s\in \mathbb{N}$, $t\in \mathbb{N}$, $s\geq t$, denote by $\{T^{z,s}\}$, $\{S^z\}$ the sets of all permutations of potential variables, then,}
\[
\begin{split}
 (\{T^{z,s}\},\{S^z\})  \hspace{2mm}  &\!\perp\!\!\!\perp   \hspace{2mm}   Z \\
\{T^{z,s}\}  \hspace{2mm}  &\!\perp\!\!\!\perp   \hspace{2mm}   \sone(S = s) \hspace{1mm}  | \hspace{1mm}   S \geq t, T \geq t, Z=z
\end{split}
\]
\end{flushleft}
\end{minipage}
\end{center}
In the context of our study, the first assumption states that the blood type is randomized at $t=0$ with respect to the potential time to death and potential time until receiving a kidney transplant. The second assumption is a dynamic version of an unconfoundedness assumption which says that receiving a kidney transplant is randomized for the survivors on the waitlist who did not yet receive a kidney transplant. Recall that $X$ is implicit in the conditioning set so we can allow selection on these covariates into treatment for the untreated survivors or onto the regime. In addition, of importance in this paper, it should be noted that assumption A.I does not exclude that agents can manipulate their exit time based on some knowledge of the treatment distribution under one or the other of the regimes $z$.

Assumption A.I was first proposed by Robins (\hyperlink{Rob1997}{1997}) and introduced into the economics literature by Lechner (\hyperlink{Lec2009}{2009}). It implicitly contains structural assumptions concerning the timing of causes and effects within a period. In our formulation of the conditioning set, $S\geq t$, we allow treatment at period $t$ to influence exit in the same period. So, receiving a kidney transplant can influence survival immediately. If there is censoring in the sample, one can add a similar dynamic unconfoundedness condition under which right censored observations are dynamically missing (completely) at random.

In our empirical setting, assumption A.I states that a person's blood type is independent of their potential death and transplant dates. ABO blood type have been identified as risk factors in specific disease processes such as vascular disease and malignancy (see e.\ g.\ Wu, et.\ al., \hyperlink{Wu2008}{2008}; Sun, et.\ al., \hyperlink{Sun2015}{2015}) as well as for some infectious diseases (Cooling, \hyperlink{Cool2015}{2015}). However, on the whole, from the current author's assessment, the literature on ABO blood type as a risk factor does not consistently indicate any one blood type as leading to a higher overall mortality. That being said, ABO blood types are known to be correlated to demographic factors such as race, which are associated with a higher mortality. To account for these correlations we include a set of baseline covariates, including race and education, which may be correlated with blood type. We assume the first independence assumption of A.I will hold conditional on this set of baseline covariates.

The second part of assumption A.I should be questioned in dynamic settings and its credibility hangs on the researcher's knowledge of the assignment mechanism to treatment as well as knowledge of which variables candidates may be systematically reacting to. If our treatment of interest were the offer of a kidney, as opposed to the transplant of a kidney, we could make a relatively strong case for this assumption to hold. OPTN, through the UNet system, allocate kidney offers to candidates on the waitlist purely based on candidate and donor characteristics, and registered restrictions, which we observed in our data. In addition, because this algorithm is complex, and not readily accessible in its details, we can assume that there is no predetermined time at which candidates will be offered a suitable kidney from the point of view of candidates or transplant surgeons. Beyond these two factors, it is always important to know which exogenous variation is exploited to determine the treatment allocation among comparable individuals.

In our setting, the exogenous variation could be said to stem from regional differences in kidney availability or in donor-recipient compatibility criteria, such as tissue type. One could for instance argue that there exist surviving agents who are similar on all measures but the one living marginally closer to the donor is prioritised to receive the kidney. If such arguments are not sufficient, one can argue that HLA tissue type compatibility requirements will result in one candidate receiving the transplant while an otherwise comparable set of candidates remain on the waitlist.\footnote{HLA tissue types are not known to be correlated to blood types (Erikoglu, \hyperlink{Erik2011}{2011}).} Exogeneity of this variation would also require that candidates do not change their behaviour based on knowledge of their tissue type. In contrast to blood type, this could be argued on the grounds that HLA compatibility is multidimensional and more complicated than that of blood type. It may therefore not be a salient heuristic for predicting the propensity to rapidly receive a transplant. All of these scenarios provide support for the credibility of our assumptions when considering the offer of a kidney. However, in our data we only observe the actual kidney transplant, as opposed to the offer of a kidney.

When considering the time to a kidney transplant as the treatment of interest, our main tenet is that all candidate or surgeon decisions to accept or reject a kidney are a function of the OPTN allocation variables which we control for. Under this assumption, we can draw on the same exogeneity arguments as above to justify the second part of A.I. While this line of argument is imperfect, Agarwal et.\ al., (\hyperlink{Agar2021}{2021}) (Table III) offer some descriptive statistics supporting this assumption. They show that kidney acceptance rates are strikingly similar across candidates of different health measures and of different ages. Since behavioural phenomena such as risk aversion are correlated to age and health, we would expect these unobserved phenomena to induce noticeable differences in acceptance rates were they central to the decision to reject an offer. Large differences in acceptance rates seem instead to result mostly from observed differences in the compatibility of the donated kidney and the health of the donor.

When it comes to identifying dynamic treatment effects in non-experimental dynamic settings, there are some proposed alternatives to the above assumptions, but all of these come with their own complications. For example, if one includes features of the donor or the donated kidney in the set of baseline control variables $X$, then one is controlling for future endogenous intermediate variables, with all the problems this entails. We also cannot include time varying candidate characteristics. If one adjusts dynamically for the set of time specific control variables $X_t$, then the interpretation of the causal effects of interest will depend on the included intermediate variables. Taking an instrumental variable approach to account for possible selection on unobservables would also result in two problems. First, there would be the usual problem that the resulting ex-post transplant effects on survival would depend on the particular choice of instrument (Heckman and Vytlacil, \hyperlink{Heck2005}{2005}). This leaves their interpretation for general policy recommendations questionable. Second, even if one finds an instrument with a policy relevant interpretation, making any comparisons of transplant effects across regimes would require that marginal treatment effects are invariant across regimes, a difficult argument to defend.

In an experimental setup, assumption A.I can be made credible by externally randomizing treatment sequentially on survivors. It can also be achieved by randomizing at $t=0$ all future treatment times and withholding this information from agents.\footnote{Such a setup is for example presented in Kastoryano and van der Klaauw (\hyperlink{KavdK2022}{2022}) in an active labour market setting in which and external firm is contracted by the unemployment benefits agency to randomize unemployed individuals to enter a program based on a select number of observed baseline variables.} For example, a researcher at baseline can randomize individuals to two groups, one which is told they have a low chance to receive treatment (or are told nothing), another which is told at baseline that they are likely to receive a future treatment. Individuals from each group are then subsequently treated according to their prescribed group.

\subsubsection{Overlapping Support Assumption}

In addition, we impose that regimes, treatments, and the decomposition can be evaluated (Robins, \hyperlink{Rob1997}{1997}; Lok, Gill, Van der Vaart and Robins, \hyperlink{LokEA2004}{2004}),
\begin{center} \begin{minipage}{16.3cm}
\begin{flushleft}
\emph{\textbf{Assumption A.II: Overlapping Support:}} For $z=0,1$, $t  \in (0,s]$, $s\in \mathbb{N}$, $t\in \mathbb{N}$,
\[
\begin{split}
(i)~~~ &0 < \Pr(Z=z) < 1 \\
(ii)~~~ &0<\Pr(S = t| S \geq t, T \geq t,  Z=z) < 1 \\
\end{split}
\]
\end{flushleft}
\end{minipage}
\end{center}
This overlapping support assumption guarantees first, $A.II(i)$, that we observe agents under both regimes. $A.II(ii)$ provides that there is no time before period $s$ at which the treatment is allocated to all, or none, of the untreated survivors.

Assumption $A.II(i)$ amounts to the usual overlap assumption in static settings. The validity of assumption $A.II(ii)$ depends on several aspects of the data generating process and the data itself. In practice, this assumption is rendered more credible when the assignment mechanism to treatment includes stochastic components. This is the case in our empirical setting since the UNet allocation system of the OPTN does not predict the exact time at which a candidate on the waitlist will find a suitable donated kidney. We can draw from similar exogeneity arguments as in the previous section to justify the random variation necessary for these assumptions to hold.\footnote{Beyond the credibility of the treatment assignment mechanism, assumption $A.II(ii)$ also imposes strong requirements on the data since the data is generated from a branching process. If the researchers are analysing data with relatively small time intervals, then they impose a large set of intervals at which the overlap conditions must hold. If, on the other hand, they analyse time periods with relatively large intervals, then they risk counfounding within intervals due to dynamic selection. In our empirical application, we set the unit of a time period to two months in order to balance both considerations.}

\subsubsection{SUTVA Assumption}

We add to these two assumptions Rubin's Stable Unit Treatment Value Assumption (SUTVA, Rubin, \hyperlink{Rub1980}{1980}) which we rejoin with Robins' consistency assumption (Robins, \hyperlink{Rob1997}{1997}; Murphy, \hyperlink{Mur2003}{2003}) in the following,
\begin{center} \begin{minipage}{16.3cm}
\begin{flushleft}
\emph{\textbf{Assumption A.III: SUTVA:} For $z=0,1$, $s  \in \mathbb{N}$, }
\[
\begin{split}
 & S^z=S   \qquad if~Z=z\\
 & T^{z,s}  =T   \qquad if~S=s,Z=z
\end{split}
\]
\end{flushleft}
\end{minipage}
\end{center}
In our empirical setting, this assumption states that the potential duration until an individual receives a kidney transplant and the potential time to death equal their corresponding observed duration to transplant and duration to death in our sample. The above assumption does not allow potential outcomes for an agent to depend on the observed or counterfactual outcomes of any other agent, or that the setting from which the data were collected does not represent the hypothetical causal setting of interest. Since our data includes the universe of candidates on the waitlist in the US, and our goal is to decompose causal effects for that population, A.III will hold.

\subsubsection{Identification of Regime and Treatment Effects}

Under $A.I-A.III$, we can identify the causal decomposition of interest in equation \ref{eq:decomp}.\\
\begin{center} \begin{minipage}{16.3cm}
\begin{flushleft}
\emph{\textbf{Proposition 1.} Under Assumptions $A.I-A.III$ we can identify the causal decomposition effects,}
\begin{equation} \label{eq: prop1} \small
\begin{split}
&\beta_0 = \prod_{t=0}^{\tau} \Pr(T > t|S > t,T \geq t,Z=0)\\
&\beta_z = \prod_{t=0}^{\tau} \Pr(T > t |S > t,T \geq t,Z=1) - \prod_{t=0}^{\tau} \Pr(T > t |S > t,T \geq t,Z=0) \\
&\beta_s =\prod_{t=0}^{\tau} \Pr(T > t|S=s,T \geq t,Z=0) - \prod_{t=0}^{\tau} \Pr(T > t |S > t,T \geq t,Z=0)\\
&\beta_{zs} =\prod_{t=0}^{\tau} \Pr(T > t|S=s,T \geq t,Z=1) - \prod_{t=0}^{\tau} \Pr(T > t |S > t,T \geq t,Z=1) - \beta_s\\
&\text{where we use a shorthand notation,}\\
&\prod_{t=0}^{\tau} \Pr(T > t|S=s,T \geq t,Z=z)\equiv\\
& \qquad \qquad \prod_{t=s}^{\tau} \Pr(T > t|S=s,T \geq t,Z=z)\cdot \prod_{t=0}^{s-1} \Pr(T > t|S>t,T \geq t,Z=z)
\end{split}
\end{equation}

\end{flushleft}
\end{minipage}
\end{center}

In order to identify $\beta_{(0,s]}$ and $\beta_{z(0,s]}$ we must also identify the probability of
treatment at period $s$ under regime $z$,
\begin{center} \begin{minipage}{16.3cm}
\begin{flushleft}
\emph{\textbf{Corollary 1.} Under Assumptions $A.I-A.III$ we can identify,}
\begin{equation*} \label{eq: corol}
\Pr(S^z = s  )=\Pr(S = s |S \geq s,T \geq s,Z=z)  \cdot \prod_{t=0}^{s-1} \Pr(S > t |S \geq t,T \geq t,Z=z)
\end{equation*}
\end{flushleft}
\end{minipage}
\end{center}

We present the well-known proofs to Proposition 1 and Corollary 1 in appendix \ref{app: identNP}. The above equations are adapted versions of Robins' (\hyperlink{Rob1997}{1997}) and Gill and Robins' (\hyperlink{GiRob2001}{2001}) ``g-computation formula'' which also admits non-binary and non-permanent treatments as well as non-duration outcomes. As in the general formula, our version allows for causal dependence of treatment histories on outcomes but also of outcome histories on the treatment.

\subsection{Identifying Effects on Substrata}

The decomposition identified in Proposition 1 has the advantage of relying on relatively limited assumptions for our empirical setting, which is why it provides the basis for our main results. More generally, when the treatment allocation process is experimentally manipulated, these assumptions provide reliable grounds for internally valid causal effects. However, one disadvantage of this decomposition is in the interpretation of the ex-post interaction effect $\beta_{zs}$. The main problem is that the group receiving treatment at $s$ in regime $Z=0$ may not be comparable to the group receiving treatment at $s$ in regime $Z=1$. Due to dynamic selection, these two groups of survivors at $s$ may be composed of individuals differing in their unobserved characteristics (Ham and Lalonde, \hyperlink{HaLal1996}{1996}; Eberwhein, Ham, Lalonde, \hyperlink{EbeEA1997}{1997}). This is problematic. For example, in our empirical setting we would like to know when comparing the transplant effect of AB and B/O blood types whether differences in survival are due to substantive effects (state dependence) or simply due to the fact that the groups of survivors who receive a transplant differ in their composition (dynamic selection).

To separate dynamic selection from substantive differences in the ex-post effect of a kidney transplant between regimes, we need to build on our previous assumptions. In this paper, we propose dynamic assumptions which parallel the LATE framework. To introduce these assumptions, we need to explicitly discuss unobserved variables. We represent the initial conditions at $t=0$ of an agent by $u \in \mathcal{U}$, which is the realization of a one-dimensional non-negative unobserved random variable $U$. $U$ is a single index scalar representing all variables unobserved by the econometrician which are relevant to the duration to exit outcome of an agent.

\subsubsection{Dynamic Rank Invariance Assumption}

A first assumption required to disentangle state dependence from dynamic selection is a dynamic rank invariance assumption. This assumption draws from the quantile treatment effect literature (Chernozhukov and Hansen, \hyperlink{ChHan2005}{2005}). We impose that agents exit in finite time\footnote{Without this assumption, we can still define the rank invariance assumption at any time $s$ for the set of agents who would have exited before time $s$ under $Z=0$ and $Z=1$.}, $T^{z,s}<\infty$, and that the following rank invariance assumption on the exit sequence holds,
\begin{center} \begin{minipage}{16cm}
\begin{flushleft}
\emph{\textbf{Assumption A.IV: Dynamic Rank Invariance:} For any two agents $u_j$, $u_k \in \mathcal{U}$},
\[
\begin{split}
&T_j^{0,\infty} < T_k^{0,\infty}~~~ \text{iff}~~~  T_j^{1,\infty} < T_k^{1,\infty}\\
\text{and} \qquad &T_j^{0,\infty} = T_k^{0,\infty}~~~ \text{iff}~~~  T_j^{1,\infty} = T_k^{1,\infty}
\end{split}
\]
\end{flushleft}
\end{minipage}
\end{center}
This is a new application of a rank invariance assumption in the dynamic treatment effects literature. It states that the exit order with respect to $u$ is preserved across regimes for not yet treated agents. This assumption also qualifies the definition of $U$ by requiring that, if assumption $A.IV$ can hold conditional on observed baseline covariates, then the ordering of realizations $u$ at $t=0$ is defined such that assumption $A.IV$ must hold. Returning to our kidney waitlist application, the rank invariance assumption imposes that, withholding the effect of the transplant, the sequence of death would be the same had individuals expected to receive a kidney transplant rapidly (AB type) as had those same individuals expected to wait a long time for a suitable kidney (B/O type). This assumption serves in our dynamic setting as the dynamic counterpart to the monotonicity assumption in the LATE framework.

Under the rank invariance assumptions, and assuming for presentation that $\Pr(T^{1,\infty} \geq s)> \Pr(T^{0,\infty}  \geq s)$, we can distinguish three distinct groups in the population. A first group, characterised by $\{T^{1,\infty} < s, T^{0,\infty}< s\}$, never survives to receive treatment at $s$. A second group, $\{T^{1,\infty}\geq s, T^{0,\infty} \geq s\}$, survives up until $s$ in both regimes. A third group $\{T^{1,\infty}\geq s, T^{0,\infty} <s\}$, survives up until $s$ in regime $Z=1$ but does not survive until $s$ in regime $Z=0$. We call these three groups the \emph{never-survivors} (ns), \emph{always-survivors} (as) and \emph{complier-survivors} (cs), respectively.

Assumption A.IV alone already allows us to separately identify certain causal effects for substrata in the decomposition of Proposition 1. Assuming for presentation\footnote{We provide the derivations for the general case in appendix \ref{app: identNP}.} that $\Pr(T^{0,\infty}  > s) \geq \Pr(T^{1,\infty}  > \tau)$, we can identify the effects on substrata,

\begin{equation} \label{eq:LARTE1} \small
\begin{split}
&\beta_0 =\Pr(T^{0,\infty}  > \tau)=\Pr(T^{0,\infty}  > \tau|as)+\Pr(T^{0,\infty}  > \tau|cs)+\Pr(T^{0,\infty} > \tau|ns)\\
& \qquad =\prod_{t=s}^{\tau} \Pr(T > t|S > t,T \geq t,Z=0)\cdot \Pr(as) \\
&\beta_z =[\Pr(T^{1,\infty}  > \tau)-\Pr(T^{0,\infty}  > \tau)]= [\Pr(T^{1,\infty} > \tau |as)-\Pr(T^{0,\infty}  > \tau |as)]\cdot \Pr(as)\\
& \quad  =(\prod_{t=s'}^{\tau} \Pr(T > t |S > t,T \geq t,Z=1) - \prod_{t=s}^{\tau} \Pr(T > t |S > t,T \geq t,Z=0))\cdot \Pr(as)\\
&\beta_s =[\Pr(T^{0,s}  > \tau)-\Pr(T^{0,\infty}  > \tau)]=\Big(\Pr(T^{0,s}   > \tau | as)-\Pr(T^{0,\infty}   > \tau |as)\Big)\cdot \Pr(as)\\
&\quad =\Big(\prod_{t=s}^{\tau} \Pr(T > t|S=s,T \geq t,Z=0) - \prod_{t=s}^{\tau} \Pr(T > t |S > t,T \geq t,Z=0)\Big)\cdot \Pr(as)\\
& \Pr(as)= \Pr(T^{1,\infty}\geq s, T^{0,\infty} \geq s)=\prod_{t=0}^{s-1} \Pr(T > t |S > t,T \geq t,Z=0)\\
&\Pr(cs) =\Pr(T^{1,\infty}\geq s, T^{0,\infty}< s)=\prod_{t=0}^{s-1} \Pr(T > t |S > t,T \geq t,Z=1)-\Pr(as)\\
&\Pr(ns) = 1- \Pr(as) - \Pr(cs)
\end{split}
\end{equation}

It is worth noting that even if assumption A.IV is violated, the above effects will only be slightly biased as long as the probability of being a \emph{defier-survivor}, characterised by $\{T^{1,\infty}\geq s, T^{0,\infty} <s\}$, is relatively small, the effects for defier-survivors are similar to those of other substrata, and effects of a kidney transplant are relatively smooth over time.\footnote{This can be seen in our simulation exercise in appendix \ref{app: simResults}.}

A.IV is an assumption on unobserved variables so it cannot be tested directly and should be regarded with scrutiny. The credibility of the rank invariance assumption in applications can be strengthened by including control variables which can proxy for behavioural preferences such as risk-aversion or time-inconsistent preferences, among others. It is also possible to evaluate the credibility of this assumption by comparing the untreated survival distributions across regimes.\footnote{See Frandsen and Lefgren (\hyperlink{FrLef2018}{2018}), and Dong and Shen (\hyperlink{DoShe2018}{2018}) for tests of rank-invariance assumptions in non-duration outcome models.} In particular, if these distributions display sharp spikes in one regime but not the other, this should raise concerns about the credibility of the assumption.

\subsubsection{Relative Exclusion Restriction Assumption}

To identify the substrata effects of the interaction effect $\beta_{zs}$, we must additionally impose a variation on the classical exclusion restriction. This exclusion restriction on effects can be stated as follows in our setting,

\begin{center} \begin{minipage}{16.3cm}
\begin{flushleft}
\emph{\underline{\textbf{Assumption A.V: Relative Exclusion Restriction}}\\ \vspace{2mm} For $z=0,1$, $s  \in \mathbb{N}$, it holds for each agent that,}
\begin{equation*}
  T_i^{0,s}-T_i^{0,\infty}=T_i^{1,s}-T_i^{1,\infty}
\end{equation*}
\end{flushleft}
\end{minipage}
\end{center}

Calling upon the relative exclusion restriction assumption, it follows that the interaction effect $\beta_{zs}$ of Proposition 1 corresponds to the following effect,
\[ \small
\begin{split} \label{eq:LARTE2}
\beta_{zs}&= [(\Pr(T^{1,s}  > \tau)-\Pr(T^{1,\infty}  > \tau)) - (\Pr(T^{0,s}  > \tau)-\Pr(T^{0,\infty}  > \tau))]\\
&= [\Pr(T^{1,s}   > \tau| cs)-\Pr(T^{1,\infty}  > \tau | cs)]\cdot \Pr(cs)
\end{split}
\]

A.V implies that the difference in ex-post effect is purely due to a change in the composition of treated at $s$ across regimes. In our empirical setting, $\beta_{zs}$ represents the effect of the kidney transplant in regime $Z=1$ for individuals who survived until $s$ under regime $Z=1$, but would have died prior to $s$ under regime $Z=0$.

It is important to note that A.V is less restrictive than a classical exclusion restriction of the form $T_i^{0,s}=T_i^{1,s}$ for all $s>0$. This classical exclusion restriction would likely be violated in most dynamic settings. In our application, the relative exclusion restriction A.V allows a candidate's expectations and behaviour to change depending on their blood type. It does not, however, allow a different interaction effect between expectations and actually receiving a kidney transplant. The relative exclusion restriction will hold in our empirical setting if we assume a kidney transplant generates a proportional shift in health relative to baseline mortality regardless of the individual's blood type.

\subsection{Estimation Approach}
\label{subsec:estim}

In this section we propose a simple estimation method which builds on a semi-parametric proportional hazard model. We choose a continuous time estimator rather than a discrete time one mainly for computational reasons. Also, for settings in which identifying assumptions are likely to hold and time intervals are small, continuous and discrete time estimators produce similar results in treatment effect duration models (Kastoryano and van der Klaauw, \hyperlink{KavdK2022}{2022}).\footnote{Kastoryano and van der Klaauw (\hyperlink{KavdK2022}{2022}) also provide non-parametric discrete time duration estimators adaptable to our setting and discuss the tradeoffs between discrete and continuous time estimation approaches.}  We write the joint exit and treatment hazards in the proportional hazard specification,
\[ \small
\begin{split}
\theta_t^{T}(x,z,s)&= \lambda^T_t(z,\mathds{1}(s \leq t))\exp( x^\prime \beta^T)
\\
\theta_t^{S}(x,z)&= \lambda^S_t(z) \exp(x^\prime \beta^S)
\end{split}
\]
For the estimation, we first parameterize the duration dependence functions $\lambda^T_t(z,\mathds{1}(s \leq t))$ and $\lambda^S_t(z)$ as separate piecewise constant baseline hazards depending on $Z$ and whether treatment occurred $\mathds{1}(s \leq t)$. We then estimate the parameters of the joint density of the above by Maximum Likelihood.\footnote{R programming code is available at \url{www.skastoryano.com}.}

Each treated causal effect taking the form $\prod_{t=\tau_1}^{\tau_2} \Pr(T > t |S = s,T \geq t,Z=z)$ is then computed as,
\[
\sum_{\{i\}}\hat{w}_i(s)\exp(-\int_{\tau_1}^{\tau_2} \hat{\theta}_t^{T}(x_i,z,s) dt) = \sum_{\{i\}}\hat{w}_i(s)\exp(-\int_{\tau_1}^{\tau_2} \hat{\lambda}^T_t(z_i,\mathds{1}(s_i \leq t))\exp( x_i^\prime \hat{\beta}^T) dt)
\]
While non-treated causal effects taking the form $\prod_{t=\tau_1}^{\tau_2} \Pr(T > t |S > t,T \geq t,Z=z)$ are computed as,
\[
\sum_{\{i\}}\hat{w}_i(s)\exp(-\int_{\tau_1}^{\tau_2} \hat{\theta}_t^{T}(x_i,z,S>\tau_2) dt) = \sum_{\{i\}}\hat{w}_i(s)\exp(-\int_{\tau_1}^{\tau_2} \hat{\lambda}^T_t(z_i,0)\exp( x_i^\prime \hat{\beta}^T) dt)
\]
with weights given by,
\[
\hat{w}_i(s)=\frac{\hat{\theta}_s^{S}(x_i,1)\exp(-\int_{0}^{s} \hat{\theta}_\uptau^{S}(x_i,1)d\uptau)}{\sum_{\{i\}} \hat{\theta}_s^{S}(x_i,1)\exp(-\int_{0}^{s} \hat{\theta}_\uptau^{S}(x_i,1)d\uptau)}\\
\]
To obtain the expected value of the effects over an interval of treatment times $[1,s]$, one should replace the denominators in the weights with a sum over $t=1,\ldots,s$, $\sum_{t=1}^{\overline{s}}\sum_{\{i\}} \hat{\theta}_t^{S}(x_i,1)\exp(-\int_{0}^{t} \hat{\theta}_\uptau^{S}(x_i,1)d\uptau)$. Finally, we use the delta method to compute standard errors around the causal effects of interest.

In appendix \ref{app: simResults} we provide some simulation results assessing the robustness of our estimation method on simulated data generated from a dynamic discrete choice model. Our estimator performs well despite its low level of complexity when the underlying data generating process is not excessively non-linear.

\section{Decomposing Effects in Kidney Transplant Setting}
\label{sec:empir}

In this section we illustrate the proposed framework with our empirical application on expectation and transplant effects for individuals on the kidney transplant waitlist in the United States. We present the different types of average effects from our framework, discuss the practical implementation of our empirical method, and pay particular attention to pre-transplant effects.

\subsection{Data}
\label{sec:data}

This study used data from the Scientific Registry of Transplant Recipients (SRTR). The SRTR data system includes data on all donor, wait-listed candidates, and transplant recipients in the US, submitted by the members of the Organ Procurement and Transplantation Network (OPTN). The Health Resources and Services Administration (HRSA), U.S. Department of Health and Human Services provides oversight to the activities of the OPTN and SRTR contractors.

The dataset we use contains detailed information on candidate and donor characteristics, as well as information on the date a candidate enters the waitlist, when they receive a kidney transplant, and the date of death. In our initial analysis we include all individuals who entered the waitlist between December 1st 2002 and December 1st 2014, the month in which the new kidney allocation system was implemented. We exclude any candidate who was scheduled to receive multiple transplants and candidates who were under 18 years of age, as both of these groups follow special allocation rules. In order to uphold the credibility of our dynamic unconfoundedness and rank invariance assumptions, we also exclude candidates who receive a transplant from a living donor since those individuals effectively jump the line based on unobservables.\footnote{Note that this data selection may be endogenous to the blood type if candidates with a different propensity to finding a suitable kidney have different propensities to seeking out living donors. In appendix \ref{sec: Robust} we show that our results remain robust when including candidates who received a kidney transplant from a living donor.} For similar reasons, we only keep individuals who are listed to be on dialysis upon entering the waitlist. Finally, we only include first spells on the waitlist thereby excluding repeated observations for individuals who experience graft failure or renewed kidney failure. Appendix \ref{sec: DataClean} provides a full description of our data selection.

We take the unit of time to be two months. Our outcome duration $T$ of interest is the time from the moment a candidate with kidney failure enters the waitlist until the time of death. In the SRTR data, information on the date of death is obtained from social security records (Massie, et.\ al., \hyperlink{Mass2014}{2014}).

The duration to treatment $S$ is the time until a candidate receives a kidney transplant. The regime randomization $Z$ in our study is the biological randomization to a blood type AB or B/O. We consider B and O blood type individuals as a single group based on the similarity of their hazard to treatment. This can be observed in appendix \ref{sec: DataStats} Figure \ref{fig:SurvAll_main}. AB blood type individuals who are universal recipients have the highest propensity to rapidly receive a kidney transplant. B and O blood type candidates have the lowest propensity to receive a transplant, and blood type A individuals have a propensity lying between the two. We exclude blood type A individuals from the main analysis sample in order to emphasize the distinction in treatment propensities between AB and B/O blood type individuals. We also exclude a limited group of individuals with rare subgroup blood types.

In the estimation we control for an extensive set of baseline variables described with summary statistics in appendix \ref{sec: DataStats}. We include almost every variable that is used to determine the Kidney Donor Profile Index (KDPI) and Estimated Post Transplant Survival (EPTS) calculations. These are the main values used to match donors to candidates. They include age, diabetes status, whether the person had a previous organ transplant, number of previous organ transplants, BMI and race. We only exclude certain variables which present a large fraction of missing values.\footnote{These are: hypertension (26\% missing), Creatinine measure of candidate (33\% missing).} In the blood type analysis, we also do not include the small fraction of individuals with a first cPRA score above 0, excluding anyone immune sensitised. These individuals are included in a secondary analysis when further explaining pre-transplant effects. At no point do we control for intermediate variables which require stringent assumptions to be considered exogenous. In our initial analysis we censor the death outcome on December 1st 2014 to avoid interference due to changes in the allocation system. We further assume that censoring of duration variables is missing at random conditional on the included covariates.

\subsection{Empirical results}

\subsubsection{Main Decomposition Effects}

Table \ref{ta:Effmain} presents our main decomposition results of blood type and kidney transplant effects on survival. The first two columns evaluate the decomposition for candidates who received a kidney transplant within the first two years on the waitlist, on the probability of survival beyond $\tau=4$ years. The first column does not include any covariates. In the estimation, we specify the segments of the duration dependence terms to be one year each. This choice is guided by the desire to calibrate our model on the observed patterns of Figure \ref{fig:SurvT_main} and \ref{fig:SurvC_main} in our estimation without covariates. The estimates of $\beta_z$ and $\alpha_z=\Pr(S^{AB} = 24\hspace{1mm}months  )-\Pr(S^{B/O} = 24\hspace{1mm}months  )$, the effect of the regime on the probability of transplant, approximate well the observed patterns in Figure \ref{fig:SurvT_main} and \ref{fig:SurvC_main}. 

The second column includes the full set of baseline control variables described in appendix \ref{sec: DataStats}. Estimates of effects are slightly larger in magnitude when including covariates. In column 2, the estimated probability of survival beyond $4$ years for candidates with AB blood type who did not receive a transplant within that time is $\beta_0=80.6$ percent. Candidates with a B/O blood type face a $\beta_z=4.1$ percentage points higher pre-transplant probability of survival. These pre-transplant effects of blood type have not been documented before.

Several explanations are possible. A first possibility is that individuals see their blood type as a heuristic which proxies their chances of rapidly finding a suitable kidney. Candidates with an AB blood type, knowing they are universal recipients, are less diligent about their overall health, leading to a higher pre-transplant mortality than B/O types. A second possibility is that candidates are directly responding to a signal concerning their treatment hazard which may, for instance, be conveyed to them by their physician, or by the monthly number, and quality, of offered kidneys. Alternatively, it may be that there are yet undocumented biological interactions between a person's blood type and their survival when suffering from renal failure. A last possibility is that we are simply not capturing all confounders jointly determining survival and treatment durations. If that is the case, then $\beta_z$ is a spurious non-causal effect. We will offer suggestive evidence through our subsequent results in section \ref{se:pre-treat} that the measured effect is neither biological nor spurious.

\begin{table} [!h] \scriptsize
\begin{center}
\caption{Causal Effect Decomposition}
\label{ta:Effmain}
\begin{tabular}{l cccccccccc}
 \hline \hline \\
   & \multicolumn{1}{c}{(1)}     &    \multicolumn{1}{c}{(2)} &
  \multicolumn{1}{c}{(3)}   &&   \multicolumn{1}{c}{(4)}  & \multicolumn{1}{c}{(5)} & \multicolumn{1}{c}{(6)}  &  \multicolumn{1}{c}{(7)}  \\
  \hline
  \\[0.5ex]
$\beta_0$	&	0.820	&	0.806	&	0.730	&&	0.887	&	0.944	&	0.902	&	0.940		\\[-0.3ex]
	& (0.014) & (0.088) & (0.133) && (0.050) & (0.021) & (0.045) & (0.025)	\\[-0.3ex]
	& [0.000] & [0.000] & [0.000] && [0.000] & [0.000] & [0.000] & [0.000]	\\[0.3ex]
$\beta_z$	&	0.018	&	0.041	&	0.038	&&	0.038	&	0.011	&	0.022	&	0.012		\\[-0.3ex]
	& (0.011) & (0.007) & (0.016) && (0.013) & (0.003) & (0.009) & (0.005)	\\[-0.3ex]
	& [0.102] & [0.000] & [0.014] && [0.003] & [0.000] & [0.020] & [0.009]	\\[0.3ex]
$\beta_{(0,s]}$	&	0.127	&	0.150	&	0.226	&&	0.075	&	0.038	&	0.066	&	0.041		\\[-0.3ex]
	& (0.013) & (0.066) & (0.109) && (0.033) & (0.014) & (0.029) & (0.017)	\\[-0.3ex]
	& [0.000] & [0.023] & [0.038] && [0.022] & [0.008] & [0.025] & [0.015]	\\[0.3ex]
$\beta_{z(0,s]}$	&	-0.001	&	-0.032	&	-0.028	&&	-0.022	&	-0.004	&	-0.013	&	-0.006		\\[-0.3ex]
	& (0.010) & (0.006) & (0.018) && (0.007) & (0.002) & (0.005) & (0.002)	\\[-0.3ex]
	& [0.892] & [0.000] & [0.117] && [0.002] & [0.040] & [0.013] & [0.007]	\\[0.3ex]
	&		&		&		&&		&		&		&			\\
$\alpha_{z}$	&	0.264	&	0.263	&	0.246	&&	0.036	&	0.139	&	0.002	&	0.007		\\[-0.3ex]
	& (0.015) & (0.120) & (0.126) && (0.019) & (0.087) & (0.007) & (0.008)	\\[-0.3ex]
	& [0.000] & [0.029] & [0.051] && [0.058] & [0.109] & [0.779] & [0.389]	\\[0.3ex]
	&		&		&		&&		&		&		&			\\
$N_{Z=0: no-Tr}$	&	4976	&	4514	&	4306	&&	2179	&	1566	&	91,795	&	69,799		\\
$N_{Z=0: Tr}$	&	4816	&	4179	&	4387	&&	608	&	858	&	21,991	&	16,848		\\
$N_{Z=1: no-Tr}$	&	118,422	&	107,024	&	98,326	&&	7037	&	7354	&	7037	&	7354		\\
$N_{Z=1: Tr}$	&	40,158	&	34,424	&	43,122	&&	1652	&	1733	&	1652	&	1733		\\[0.3ex]

\hline \hline
\end{tabular}
\end{center}
Standard errors in parenthesis. P-values in brackets. Column (1) presents pre Dec. 2014 decomposition for transplant within first two years on waitlist on survival past 4 years with $Z=0$:AB-blood types, $Z=1$:B/O-blood types, without covariates. Column (2) presents same effects with covariates included. Column (3) presents decomposition effects pre-2014 for transplant within first 2 years on waitlist on survival past 7 years with covariates.  Column (4) presents pre Dec. 2014 decomposition for transplant within first two years on waitlist on survival past 3 years with $Z=0$:high-cPRA, $Z=1$:low-cPRA. Column (5) presents same cPRA decomposition post Dec. 2014. Column (6) presents pre Dec. 2014 decomposition for transplant within first two years on waitlist on survival past 3 years with $Z=0$:0-cPRA, $Z=1$:low-cPRA. Column (7) presents same cPRA decomposition post Dec. 2014.
\end{table}

The third row of column 2 in Table \ref{ta:Effmain} shows the average effect on survival of receiving a kidney transplant within the first $2$ years on the waitlist for candidates with an AB blood type. We find that a kidney transplant increases the probability of survival by $\beta_{(0,s]}=15.0$ percentage points. For the B/O blood type group, the effect of a transplant is slightly lower by $\beta_{z(0,s]}=-3.2$ percentage points. 

Considering these results only in terms of percentage points may, however, not provide the full picture. These effects ignore differences in the base pre-transplant survival rate. The causal effects in terms of percentage of the base rate show that $\beta_{(0,s]}/(1-\beta_0)=77.3$ percent of AB blood type individuals who receive a transplant are prevented from dying within 4 years. For B/O blood types, $(\beta_{(0,s]}+\beta_{z(0,s]})/(1-\beta_0-\beta_z)=77.1$ percent of individuals who receive a transplant are prevented from dying within 4 years. We see that despite the seemingly large transplant effect in terms of percentage points, these differences are small when considered in percentage relative to their respective base survival rates. 

In appendix \ref{sec: Robust}, we show that the effects remain robust when analysing the sample of individuals entering the waitlist after 2014, when comparing AB types to O blood types only, and when including candidates who obtained a transplant from living donors.

Under the additional dynamic rank invariance assumption A.IV and the relative exclusion restriction A.V, we can discuss the effects in column 2 of Table \ref{ta:Effmain} for specific substrata. $\beta_0$ in our setting is purely due to a $\E[\beta_0 | a-s]=0.840$ $(0.074)$ $[0.000]$ survival probability for always-survivors. As presented in equation \ref{eq:LARTE1}, this survival probability is equivalent to the weighted average of the period $s$ survival probabilities, divided by the probability of being an always-survivor at $s$. Similarly, we can obtain effects for always survivors $\E[\beta_z | a-s]=0.043$ $(0.009)$ $[0.000]$ and $\E[\beta_{(0,s]} | a-s]=0.137$ $(0.071)$ $[0.053]$. In addition, we can obtain effects for complier-survivors, $\E[\beta_{(0,s]} | c-s]=31.519$ $(249.134)$ $[0.899]$ and $\E[\beta_{z(0,s]} | c-s]=-37.064$ $(223.630)$ $[0.868]$. These are effects on O/B type candidates who survived until $s$, but who would have died prior to $s$ had they been of AB blood type. These effects are poorly estimated due to the low share of complier-survivors in the first 24 months which results in a division with a near 0 denominator.\footnote{It should also be noted that the substrata effects are averages over a changing subpopulation over periods $(0,s]$.} All remaining substrata effects of $\beta_z$, $\beta_{(0,s]}$ and $\beta_{z(0,s]}$ on never-survivors and complier-survivors are also identified and equal to $0$. Given the specificities of our setting, the results of column 2 should be seen as more policy relevant than those on substrata discussed in this paragraph.

Column 3 explores the effects of the decomposition when increasing the outcome period $\tau$ to 7 years. Despite the trends in pre-transplant survival rate observed in Figure \ref{fig:SurvC_main}, we find when including covariates in the estimation that $\beta_z$ remains steady over time.\footnote{We show in appendix \ref{sec: Robust}, that our model estimates the pre-transplant effect to be near 0 as in Figure \ref{fig:SurvC_main} when excluding covariates.} Due to the longer post-transplant time horizon, we also find larger transplant effects of $\beta_{(0,s]}=22.6$ percentage points.

\subsubsection{Explaining Pre-Transplant Effects using CPRA}
\label{se:pre-treat}

To obtain further insight into pre-transplant effects, we exploit a changing discontinuity in the allocation rules based on the candidates' calculated Panel Reactive Antibodies (cPRA), a measure of immune sensitivity for transplants. cPRA, ranging from a low of 0 to a high of 1, estimates the percentage of donors with whom a particular recipient would be incompatible due to anti-HLA antibodies. Anti-HLA antibodies are usually generated as a result of a blood transfusion, pregnancy, or a previous transplant (Weinstock and Schnaidt, \hyperlink{Wein2019}{2019}).\footnote{Anti-HLA antibodies result from exposure to proteins which appear similar to tissue type. In some occasions, the origin of anti-HLA antibodies is unknown. A person who has developed anti-HLA antibodies is unlikely to be able to remove them, although small variation is possible.} These antibodies are not known to affect survival but are an important impediment to a successful transplant (Naji et.\ al., \hyperlink{Naji2017}{2017}).

Due to their low matchability, high cPRA candidates have since 2009 received priority in the kidney allocation system (Cecka, \hyperlink{Ceck2010}{2010}). Prior to the 2014 kidney allocation system reform, candidates with a cPRA above 80\% received 4 additional points in the allocation system.
\begin{wrapfigure}{r}{0.5\linewidth}
\centering
\includegraphics[width=0.5\textwidth]{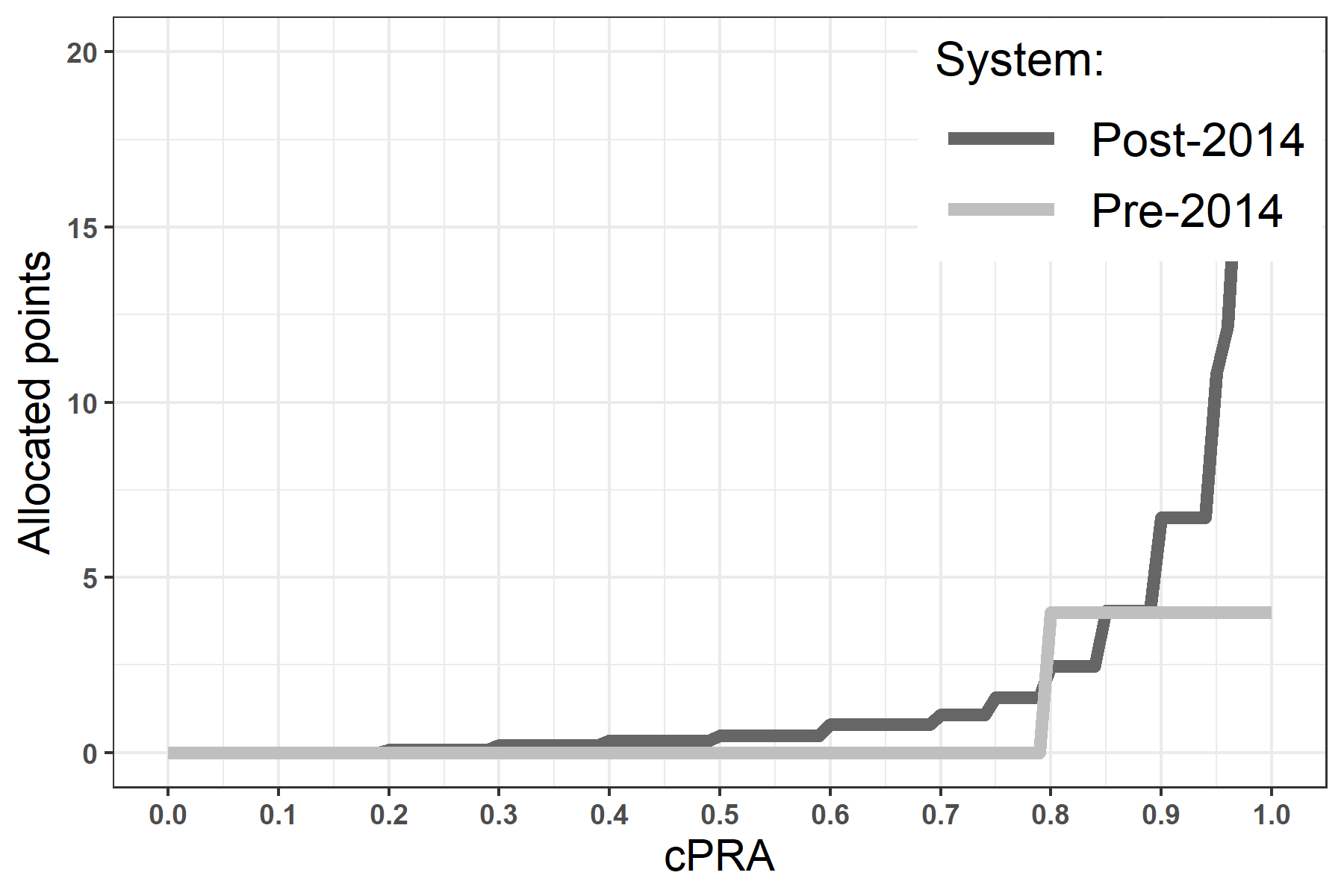}
\caption{cPRA points pre- and post- 2014 reform}
\label{fig:cPRA_points}
\end{wrapfigure} This discontinuous rule still left the most highly sensitised candidates with low prospects of finding a suitable kidney, and was deemed unfair for sensitised individuals below the 80\% threshold. In response, the kidney allocation system reform of 2014 implemented an exponential point system as displayed in Figure \ref{fig:cPRA_points} (Stegall et.\ al., \hyperlink{Steg2017}{2017}).

\begin{figure} [!h]
  \begin{subfigure}{0.48\textwidth}
    \includegraphics[width=\linewidth]{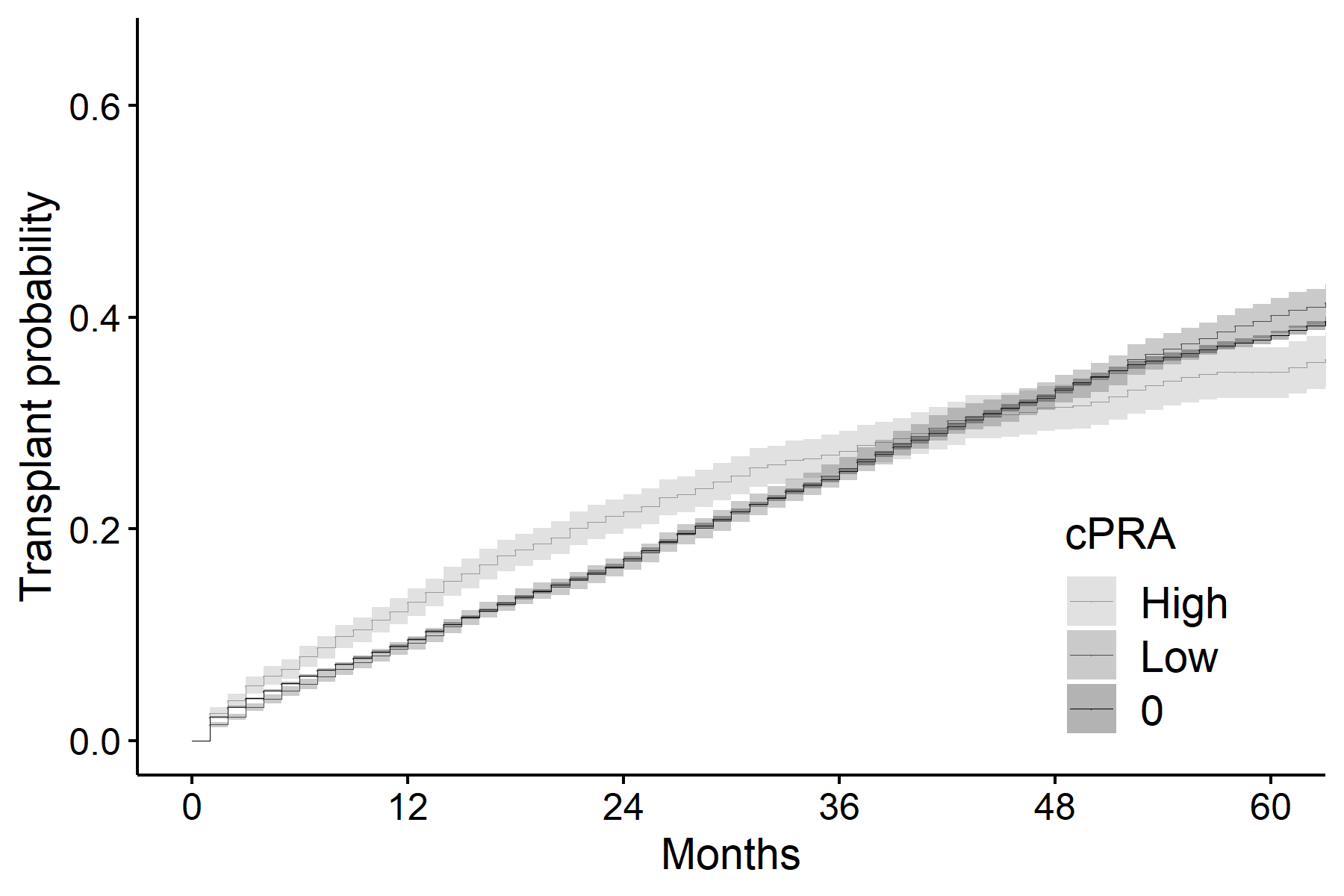}
    \caption{Transplant Probability Pre-Reform} \label{fig:SurvT_mainPRAall}
  \end{subfigure}%
  \hspace*{\fill}   
    \begin{subfigure}{0.48\textwidth}
    \includegraphics[width=\linewidth]{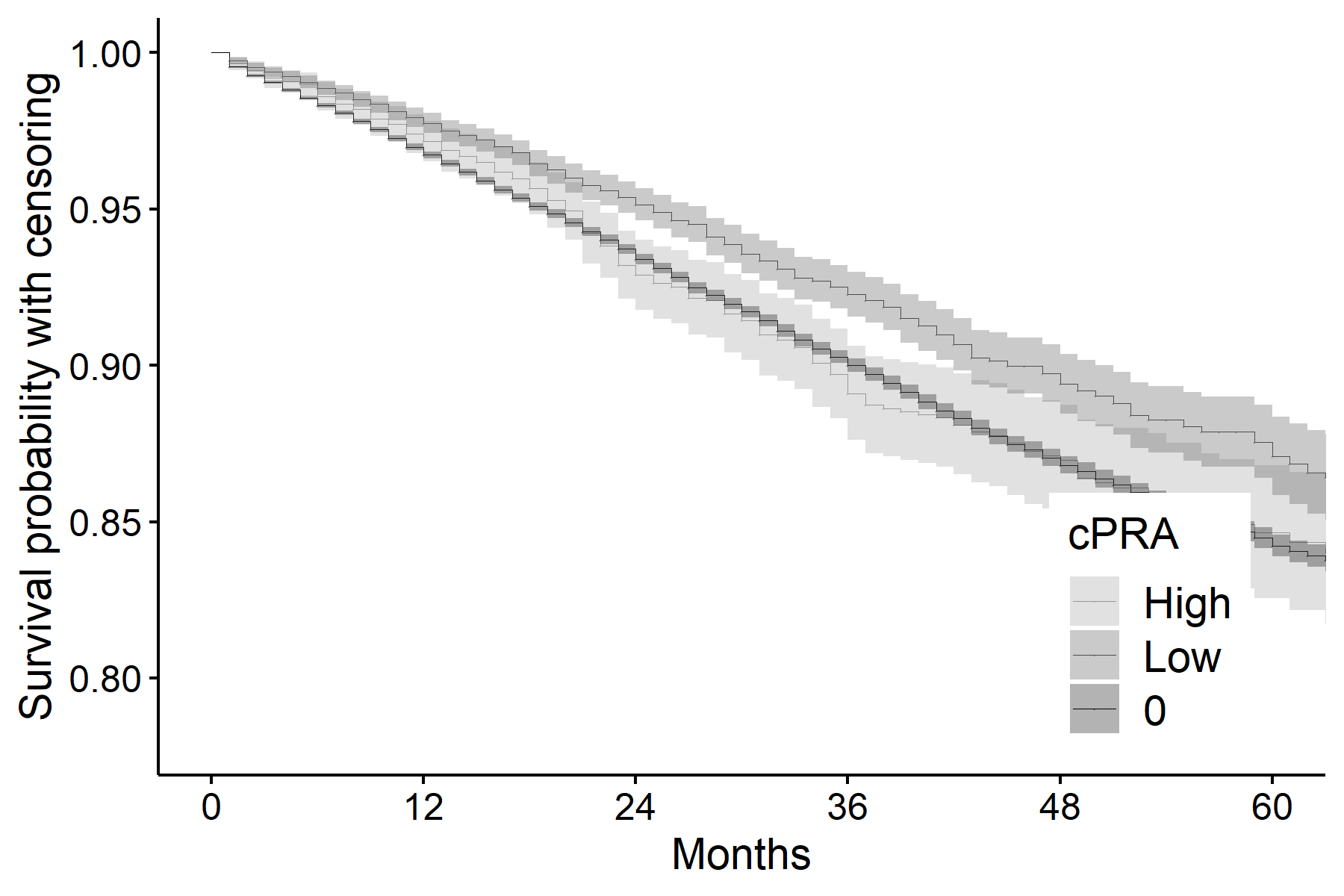}
    \caption{Pre-Transplant Survival Pre-Reform} \label{fig:SurvC_mainPRAall}
  \end{subfigure}%
  \newline
  \begin{subfigure}{0.48\textwidth}
    \includegraphics[width=\linewidth]{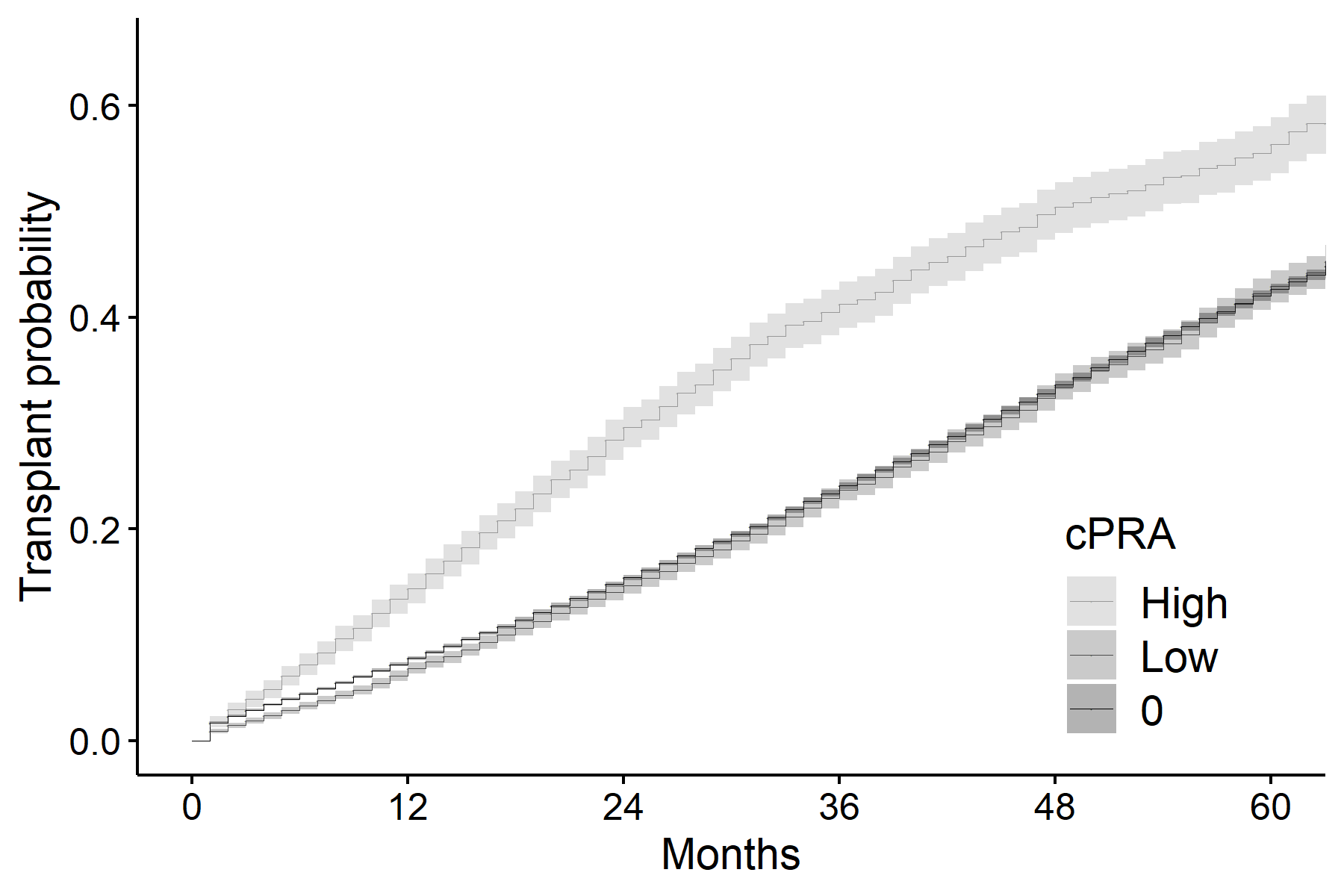}
    \caption{Transplant Probability Post-Reform} \label{fig:SurvT_mainPRAall_post2015}
  \end{subfigure}%
  \hspace*{\fill}   
  \begin{subfigure}{0.48\textwidth}
    \includegraphics[width=\linewidth]{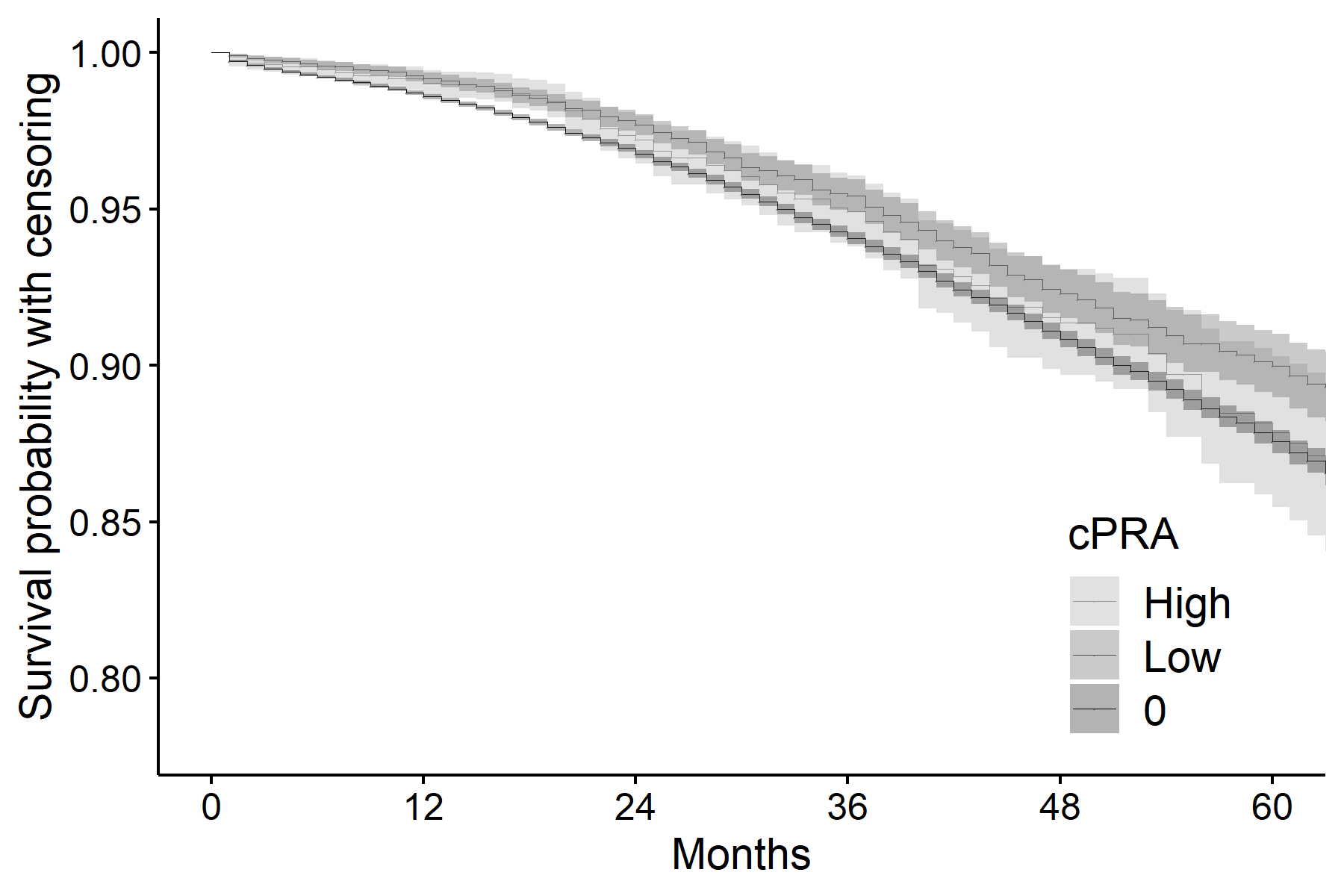}
    \caption{Pre-Transplant Survival Post-Reform} \label{fig:SurvC_mainPRAall_post2015}
  \end{subfigure}%

\caption{Survival of candidates on the kidney transplant waitlist, by cPRA group \\
    \scriptsize{Based on selected sample of SRTR data. 0-cPRA: cPRA$=0$; Low-cPRA: $0<$cPRA$<0.8$; High-cPRA: $0.8 \leq$cPRA. Excludes candidates who previously received a non-renal transplant}} \label{fig:SurvPRA}
\end{figure}

We can observe the consequences of the discontinuity and the reform on transplant rates and pre-transplant survival in Figure \ref{fig:SurvPRA}. In this figure, we define the cPRA measure, which can vary over time, as the first registered cPRA for each candidate. We then separate the cPRA into three groups. A `high-cPRA' group with cPRA above 80\%, a `low-cPRA' group with cPRA between 0 and 80\%, and a `0-cPRA' group with cPRA$=0$. The cPRA analysis excludes candidates who previously received a non-renal transplant, but we show in appendix \ref{sec: DataStats} that survival curves remain similar when including these observations. Further descriptions of the data and its selection are offered in appendix \ref{sec: DataClean}.

Prior to December 2014, Figure \ref{fig:SurvT_mainPRAall} shows that, due to the prioritisation rules, candidates with a cPRA above 80\% had a slightly higher transplant probability within the first 2.5-3 years on the waitlist than other candidates. Thereafter, the transplant probability of the high-cPRA group reduces due their poor matchability. Following the 2014 reform, Figure \ref{fig:SurvT_mainPRAall_post2015} shows that the probability of transplant becomes distinctly higher for the high-cPRA group. 

Pre-transplant survival patterns also change. Prior to the reform, Figure \ref{fig:SurvC_mainPRAall} shows that the high-cPRA group and 0-cPRA groups display similar pre-transplant survival patterns. The low-cPRA group, conflictingly, shows a higher survival rate than the other groups. After the reform, Figure \ref{fig:SurvC_mainPRAall_post2015} shows that survival rates increase more for the high-cPRA group and the 0-cPRA group than for the low-cPRA group. As a result, the differences in pre-transplant survival reduce.

Columns 4-7 of Table \ref{ta:Effmain} reproduce these same results formally within our framework. In columns 4-5, $Z=0$ for the high-cPRA group, cPRA$\geq 0.8$, and $Z=1$ for the low-cPRA group, $0<$cPRA$<0.8$. We evaluate the survival outcome at $\tau=3$ years, at which point pre-transplant differences are salient, and again calibrate the duration dependence terms in the no-covariate model (presented in appendix \ref{sec: Robust}) to match the survival patterns of Figure \ref{fig:SurvPRA}. Looking at candidates who entered the waitlist prior to December 2014, we see that low-cPRA candidates are $\beta_z=3.8$ percentage points more likely to survive beyond $3$ years than high-cPRA candidates pre-transplant. High- and low-cPRA candidates entering the waitlist after the December 2014 reform show a reduced difference in pre-transplant survival rates of $\beta_z=1.1$ percentage points.

Columns 6-7 of Table \ref{ta:Effmain} reproduce these results but comparing the low-cPRA group to the 0-cPRA group. In the analysis, $Z=0$ for the 0-cPRA group, cPRA$=0$, and $Z=1$ for the low-cPRA group, $0<$cPRA$<0.8$. We see that the results from columns 6-7 closely follow those from columns 4-5. The effect for $\beta_z$ pre-reform is slightly smaller in magnitude at $\beta_z=2.2$ percentage points. The post-reform estimate remains the same when comparing the low-cPRA group to the high-cPRA group or to the 0-cPRA group (1.1 vs.\ 1.2 percentage points). 

These results, along with the observed changes in trends in Figures \ref{fig:SurvC_mainPRAall} and \ref{fig:SurvC_mainPRAall_post2015}, provide several insights. First, they show that the 0-cPRA group and the high-cPRA group follow comparable pre-transplant survival trends both before and after the 2014 reform. The changes in trend also indicate that low-cPRA individuals, who before the reform show higher pre-transplant survival, display less divergent post-reform trends compared to the other two groups.

Combining the results offers some insight into pre-transplant effects. If general unmeasured confounders were jointly determining the transplant hazard and pre-transplant survival, then we should observe similar patterns in the blood type analysis as in the cPRA analysis. This is not the case. In the blood type analysis we find that higher transplant rates are associated with lower pre-transplant survival rates. In the cPRA analysis, we find that transplant rates have diverging effects on pre-transplant survival rates post-reform depending on the pre-reform baseline transplant rates. Low-cPRA candidates, who may be primed to expect a lower probability of receiving a transplant than the 0-cPRA group, face a higher pre-transplant survival rate than the 0-cPRA group. When low-cPRA candidates are signaled after the 2014 reform that they will receive preferential allocation, this difference in pre-transplant survival reduces by half. These differences and changes are present despite both groups having, in effect, the same transplant rates as 0-cPRA candidates as observed in Figures \ref{fig:SurvT_mainPRAall}-\ref{fig:SurvT_mainPRAall_post2015}.  The differences are also consistent with the patterns observed between blood types where lower perceived transplant rates lead to higher pre-transplant survival rates. However, high-cPRA candidates, who benefit from a further increase in their probability of transplant following the reform, do not display any differences in their pre-transplant survival rates relative to the 0-cPRA group.

Several behavioural explanations are consistent with the observed patterns. It may be that candidates are adapting their health behaviour based on salient heuristics. In the blood type analysis, the heuristic is simply the blood type. In the cPRA analysis, candidates with low-cPRA adapt their health behavior following the reform which offers some, albeit small, priority in the allocation system. However, high-cPRA candidates, who already have priority in the allocation system, do not respond to the reform which switches from a discrete jump in allocation points to a more complicated exponential increase in allocation points. Another behavioural explanation consistent with observed patterns is that candidates are not changing their health behaviour uniformly due to a heuristic, but responding to changes in the number, and quality, of kidney offers sent to them. They may also be responding directly to communications with their transplant surgeon. The dataset used in this study, along with other datasets which include offers (see Agarwal, et.\ al., \hyperlink{Agar2020}{2020}), can allow more in depth insights into the factors driving changes in health behaviour. We leave these further inquiries to future research.

\section{Conclusion}

This paper adapts the discussion of treatment heterogeneity models to a setting with an initial regime randomization which affects a later duration to treatment. We propose a dynamic potential outcome framework, building on dynamic variations of well-known assumptions, which allows researchers to decompose the effects of the regime and treatment on an duration outcome. Our empirical results discuss previously unexplored expectation effects for candidates on the kidney transplant waitlist. We find that candidates who, due to their blood type, have a higher propensity to receive a transplant, also have a higher pre-transplant mortality within the first four years on the transplant waitlist. Our results suggest these effects are behavioural rather than biological. Exploiting a changing discontinuity in the allocation system based on preexisting antibodies, we also show that changes in expectation effects depend on the complexity of signaled information.

\section*{References}
\newlength{\leftlocal}
\setlength{\leftlocal}{\leftmargini}
\addtolength{\leftmargini}{-.5\leftmargini}
\begin{description}
\newlength{\labellocal}
\setlength{\labellocal}{\labelwidth} \setlength{\labelwidth}{5pt}
\newlength{\itemlocal}
\setlength{\itemlocal}{\itemsep} \setlength{\itemsep}{0pt}

\item \hypertarget{AbvdB2003a}{Abbring, J.H., \& van den Berg, G.J. (2003a). The nonparametric identification of treatment effects in duration models. \emph{Econometrica}, 71(5): 1491-1517.}

\item \hypertarget{AbHec2007}{Abbring, J.H., \& Heckman, J.J. (2007). Econometric evaluation of social programs, part III: Distributional treatment effects, dynamic treatment effects and dynamic discrete choice, and general equilibrium policy evaluation, Chapter 72 in J.J. Heckman and E. Leamer, editors. \emph{Handbook of Econometrics, Volume 6B}, 5145-5303, Elsevier Science, Amsterdam.}

\item \hypertarget{Agar2018}{Agarwal, N., Ashlagi, I., Somaini, P., \& Waldinger, D. (2018). Dynamic Incentives in Wait List Mechanisms. \emph{AEA Papers and Proceedings}, (Vol. 108, pp. 341-47).}

\item \hypertarget{Agar2019}{Agarwal, N., Ashlagi, I., Azevedo, E., Featherstone, C. R., \& Karaduman, {\"O}. (2019). Market failure in kidney exchange. \emph{American Economic Review}, 109(11), 4026-70.}

\item \hypertarget{Agar2020}{Agarwal, N., Hodgson, C., \& Somaini, P. (2020). Choices and outcomes in assignment mechanisms: The allocation of deceased donor kidneys (No. w28064). \emph{National Bureau of Economic Research}.}

\item \hypertarget{Agar2021}{Agarwal, N., Ashlagi, I., Rees, M. A., Somaini, P., \& Waldinger, D. (2021). Equilibrium allocations under alternative waitlist designs: Evidence from deceased donor kidneys. \emph{Econometrica}, 89(1), 37-76.}

\item \hypertarget{AngEA1996}{Angrist, J.D., Imbens G. W., \& Rubin, D. B. (1996), Identification of Causal Effects Using Instrumental Variables. \emph{Journal of the American Statistical Association}, 91(434): 444-455.}

\item \hypertarget{Ceck2010}{Cecka, J. M. (2010). Calculated PRA (CPRA): the new measure of sensitization for transplant candidates. \emph{American journal of transplantation}, 10(1), 26-29.}

\item \hypertarget{ChaEA2012}{Chassang, S. , Padró I Miquel, G., \& Snowberg, E. (2012). Selective Trials: A Principal-Agent Approach to Randomized Controlled Experiments. \emph{American Economic Review}, 102(4): 1279-1309.}

\item  \hypertarget{ChHan2005}{Chernozhukov, V., \& Hansen, C. (2005). An IV model of quantile treatment effects. \emph{Econometrica}, 73(1): 245-261.}

\item \hypertarget{Cool2015}{Cooling, L. (2015). Blood groups in infection and host susceptibility. \emph{Clinical microbiology reviews}, 28(3), 801-870.}

\item \hypertarget{Dano2019}{Danovitch, G. M. (Ed.). (2009). \emph{Handbook of kidney transplantation}. Lippincott Williams \& Wilkins.}

\item \hypertarget{Dela2016}{Delavande, A., \& Kohler, H. P. (2016). HIV/AIDS-related expectations and risky sexual behaviour in Malawi. \emph{The Review of Economic Studies}, 83(1), 118-164.}

\item \hypertarget{Dick2019}{Dickert-Conlin, S., Elder, T., \& Teltser, K. (2019). Allocating scarce organs: How a change in supply affects transplant waiting lists and transplant recipients. \emph{American Economic Journal: Applied Economics}, 11(4), 210-39.}

\item \hypertarget{Dupa2011}{Dupas, P. (2011). Do teenagers respond to HIV risk information? Evidence from a field experiment in Kenya. \emph{American Economic Journal: Applied Economics}, 3(1), 1-34.}

\item \hypertarget{EbeEA1997}{Eberheim, C., Ham, J.C., \&  LaLonde, R.J. (1997). The impact of being offered and receiving classroom training on the employment histories of disadvantaged women. \emph{Review of Economics Studies}, 64(4): 655-682.}

\item \hypertarget{Erik2011}{Erikoglu, M., B\"{u}y\"{u}kdogan, M., \& T\"{u}lin, C.O.R.A. (2011). The relationship between HLA antigens and blood groups. \emph{European Journal of General Medicine}, 8(1), 65-68.}

\item \hypertarget{Faia2021}{Faia, E., Fuster, A., Pezone, V., \& Zafar, B. (2021). Biases in information selection and processing: Survey evidence from the pandemic (No. w28484). \emph{National Bureau of Economic Research}.}

\item \hypertarget{Fetz2021}{Fetzer, T., Hensel, L., Hermle, J., \& Roth, C. (2021). Coronavirus perceptions and economic anxiety. \emph{Review of Economics and Statistics}, 103(5), 968-978.}

\item \hypertarget{Fitz2016}{Fitzsimons, E., Malde, B., Mesnard, A., \& Vera-Hernández, M. (2016). Nutrition, information and household behavior: Experimental evidence from Malawi. \emph{Journal of Development Economics}, 122, 113-126.}

\item \hypertarget{GiRob2001}{Gill, R.D. \& Robins, J.M. (2001). Causal inference for complex longitudinal data: the continuous case. \emph{Annals of Statistics}, 29(6): 1785-1811.}

\item \hypertarget{Gord2020}{Gordon, E. J., Knopf, E., Phillips, C., Mussell, A., Lee, J., Veatch, R. M., ... \& Reese, P. P. (2020). Transplant candidates’ perceptions of informed consent for accepting deceased donor organs subjected to intervention research and for participating in posttransplant research. \emph{American Journal of Transplantation}, 20(2), 474-492.}

\item \hypertarget{Haetal2021}{Haaland, I., Roth, C., \& Wohlfart, J. (2021). Designing information provision experiments. \emph{Journal of Economic Literature}, forthcoming.}

\item \hypertarget{HaLal1996}{Ham, J.C.,, \& LaLonde, R.L. (1996). The effect of sample selection and initial conditions in duration models: evidence from experimental data on training. \emph{Econometrica}, 64(1): 175-205.}

\item \hypertarget{Hart2021}{Hart, A., Lentine, K. L., Smith, J. M., Miller, J. M., Skeans, M. A., Prentice, M., ... \& Snyder, J. J. (2021). OPTN/SRTR 2019 annual data report: kidney. \emph{American Journal of Transplantation}, 21, 21-137.}

\item \hypertarget{HeSin1984}{Heckman, J.J., \& Singer, B. (1984). A method for minimizing the impact of distributional assumptions in econometric models for duration data. \emph{Econometrica}, 52(4): 271-320.}

\item \hypertarget{Heck2005}{Heckman, J.J., \& Vytlacil, E. (2005). Structural equations, treatment effects, and econometric policy evaluation 1. \emph{Econometrica}, 73(3), 669-738.}

\item \hypertarget{Held2016}{Held, P. J., McCormick, F., Ojo, A., \& Roberts, J. P. (2016). A cost‐benefit analysis of government compensation of kidney donors. \emph{American Journal of Transplantation}, 16(3), 877-885.}

\item  \hypertarget{ImaEA2010}{Imai, K., Keele, L., \& Yamamoto, T. (2010). Identification, Inference and Sensitivity Analysis for Causal Mediation Effects. \emph{Statistical Science}, 25(1): 51-71.}

\item \hypertarget{ImAng1994}{Imbens, G., \& Angrist, J. (1994). Identification and estimation of local average treatment effects. \emph{Econometrica}, 62(2): 467-475.}

\item \hypertarget{Isra2014}{Israni, A. K., Salkowski, N., Gustafson, S., Snyder, J. J., Friedewald, J. J., Formica, R. N., ... \& Kasiske, B. L. (2014). New national allocation policy for deceased donor kidneys in the United States and possible effect on patient outcomes. \emph{Journal of the American Society of Nephrology}, 25(8), 1842-1848.}

\item \hypertarget{KavdK2022}{Kastoryano, S., \& van der Klaauw, B. (2022). Dynamic evaluation of job search assistance. \emph{Journal of Applied Econometrics}, 37(2), 227-241.}

\item \hypertarget{Lec2009}{Lechner, M. (2009). Sequential causal models for the evaluation of labor market programs.
\emph{Journal of Business and Economic Statistics}, 27(1): 71-83.}

\item \hypertarget{LeMiq2010}{Lechner, M., \& Miquel, R. (2010). Identification of the effects of dynamic treatments by sequential conditional independence assumptions. \emph{Empirical Economics}, 27:71-83.}

\item \hypertarget{LokEA2004}{Lok, J., R. Gill, van der Vaart, A., \& Robins, J. (2004). Estimating the causal effect of a time-varying treatment on time-to-event using structural nested failure time models. \emph{Statictica Neerlandica}, 58(3): 271-295.}

\item \hypertarget{Mada2007}{Madajewicz, M., Pfaff, A., Van Geen, A., Graziano, J., Hussein, I., Momotaj, H., Sylvi, R., \& Ahsan, H. (2007). Can information alone change behavior? Response to arsenic contamination of groundwater in Bangladesh. \emph{Journal of development Economics}, 84(2), 731-754.}

\item \hypertarget{Mass2014}{Massie, A. B., Kuricka, L. M., \& Segev, D. L. (2014). Big data in organ transplantation: registries and administrative claims. \emph{American Journal of Transplantation}, 14(8), 1723-1730.}

\item \hypertarget{Mat2003}{Matzkin, R.L. (2003). Nonparametric estimation of nonadditive random functions. \emph{Econometrica}, 71: 1339–1375.}

\item \hypertarget{Mur2003}{Murphy, S.A. (2003).Optimal dynamic treatment regimes. \emph{Journal of the Royal Statistical Society: Series B (Statistical Methodology)}, 65(2): 331-355.}

\item \hypertarget{Naji2017}{Naji M., Stanton A.D., Ekwenna O., Mitro G., Rees M., Ortiz, J. (2017). Alemtuzumab Equalizes Short Term Outcomes in High Risk PRA Patients: Long Term Outcomes Suffer. \emph{Journal of Clinical Experimental Transplantation}, 2: 117.}

\item \hypertarget{Rob1986}{Robins, J.M. (1986). A new approach to causal inference in mortality studies with sustained exposure periods -- application to control of healthy worker survivor effect. \emph{Mathematical Modelling}, 7(9-12): 1393-1512.}

\item \hypertarget{Rob1997}{Robins, J.M. (1997). Causal inference from complex longitudinal data'', in M. Berkane (eds.). \emph{Latent Variable Modelling and Applications to Causality. Lecture Notes in Statistics (120)}, Springer-Verlag, 69-117, New York.}

\item  \hypertarget{RosWo2000}{Rosenzweig, M.R., \& Wolpin, K.I. (2000). Natural ``natural experiments'' in economics. \emph{Journal of Economic Literature}, 38(4): 827-874.}

\item \hypertarget{Roth2004}{Roth, A. E., S{\"o}nmez, T., \& {\"U}nver, M. U. (2004). Kidney exchange. \emph{The Quarterly journal of economics}, 119(2), 457-488.}

\item \hypertarget{Roth2007}{Roth, A. E., S{\"o}nmez, T., \& {\"U}nver, M. U. (2007). Efficient kidney exchange: Coincidence of wants in markets with compatibility-based preferences. \emph{American Economic Review}, 97(3), 828-851.}

\item \hypertarget{Rub1980}{Rubin D.B. (1980). Comment on: ``Randomization analysis of experimental data in the fisher randomization
test by D. Basu''. \emph{Journal of the American Statistical Association}, 75:591–593.}

\item \hypertarget{Steg2017}{Stegall, M. D., Stock, P. G., Andreoni, K., Friedewald, J. J., \& Leichtman, A. B. (2017). Why do we have the kidney allocation system we have today? A history of the 2014 kidney allocation system. \emph{Human immunology}, 78(1), 4-8.}

\item \hypertarget{Sun2015}{Sun, W., Wen, C. P., Lin, J., Wen, C., Pu, X., Huang, M., ... \& Chow, W. H. (2015). ABO blood types and cancer risk—a cohort study of 339,432 subjects in Taiwan. \emph{Cancer epidemiology}, 39(2), 150-156.}

\item \hypertarget{Telt2019}{Teltser, K. F. (2019). Do kidney exchanges improve patient outcomes?. \emph{American Economic Journal: Economic Policy}, 11(3), 427-53.}

\item \hypertarget{deWe2018}{de Weerd, A. E., \& Betjes, M. G. (2018). ABO-incompatible kidney transplant outcomes: a meta-analysis. \emph{Clinical Journal of the American Society of Nephrology}, 13(8), 1234-1243.}

\item \hypertarget{Wein2019}{Weinstock, C., \& Schnaidt, M. (2019). Human leucocyte antigen sensitisation and its impact on transfusion practice. \emph{Transfusion Medicine and Hemotherapy}, 46(5), 356-369.}

\item \hypertarget{Wolf2008}{Wolfe, R. A., McCullough, K. P., Schaubel, D. E., Kalbfleisch, J. D., Murray, S., Stegall, M. D., \& Leichtman, A. B. (2008). Calculating life years from transplant (LYFT): methods for kidney and kidney‐pancreas candidates. \emph{American Journal of Transplantation}, 8(4p2), 997-1011.}

\item \hypertarget{Wu2008}{Wu, O., Bayoumi, N., Vickers, M. A., \& Clark, P. A. B. O. (2008). ABO (H) blood groups and vascular disease: a systematic review and meta‐analysis. \emph{Journal of thrombosis and haemostasis}, 6(1), 62-69.}

\item \hypertarget{Zhan2010}{Zhang, J. (2010). The sound of silence: Observational learning in the US kidney market. \emph{Marketing Science}, 29(2), 315-335.}

\end{description}
\setlength{\leftmargini}{\leftlocal}
\setlength{\labelwidth}{\labellocal}
\setlength{\itemsep}{\itemlocal}

\newpage

\appendix

\begin{flushleft}
\LARGE{\textbf{APPENDIX}}
\end{flushleft}

\section{Identification of Causal Effects}
\label{app: identNP}

In our derivation of identification results we maintain the overlap assumptions in A.II throughout and apply SUTVA/consistency A.III whenever converting potential variables to observed variables. All identification results remain the same when there is censoring if we assume censoring always occurs before treatment and exit at time $t$, and censored observations are dynamically missing at random. We derive effects in discrete time $t \in\mathbb{N}$ and $s \in\mathbb{N}$.

\begin{flushleft}
\emph{\textbf{Proof of Proposition 1 and Corollary 1}}:
\end{flushleft}
Using the dynamic unconfoundedness assumption A.I, we can straightforwardly derive all causal effects from the general form:
\begin{equation*} \small
\begin{split}
\Pr(T^{z,\infty}  > \tau) &= \Pr(T^{z,\infty}  > \tau |S > \tau,  Z=z) \\
&= \prod_{t=0}^{\tau} \Pr(T > t|S > t,T \geq t,Z=z)\\
\Pr(T^{1,s}  > \tau) &= \Pr(T^{z,s}  > \tau |S =s,  Z=z) \\
&=\prod_{t=s}^{\tau} \Pr(T > t|S=s,T \geq t,Z=z)\cdot \prod_{t=0}^{s-1} \Pr(T > t|S>t,T \geq t,Z=z) \\
\Pr(S^z = s  )&= \Pr(S^z = s  | Z=z) \\
&=\Pr(S = s |S \geq s,T \geq s,Z=z)  \cdot \prod_{t=0}^{s-1} \Pr(S > t |S \geq t,T \geq t,Z=z)
\end{split}
\end{equation*}

For the identification of effects for substrata, we first add the additional dynamic rank invariance assumption A.IV. Note that under A.IV, and assuming for presentation that $\Pr(T^{1,\infty} \geq s)> \Pr(T^{0,\infty}  \geq s)$, there is a unique time $s'>s$ in regime $Z=1$ at which the distribution of unobserved variables for non-treated survivors is the same as that for non-treated survivors at time $s$ in regime $Z=0$, $F(u| T^{0,\infty} \geq s)= F(u| T^{1,\infty} \geq s')$. This time point $s'$ is defined as that at which  $\Pr(T^{1,\infty} \geq s')= \Pr(T^{0,\infty}  \geq s)$ holds. We can then identify,

\begin{equation*} \small
\begin{split}
\beta_0 &=\Pr(T^{0,\infty}  > \tau)=\Pr(T^{0,\infty}  > \tau|as)+\Pr(T^{0,\infty}  > \tau|cs)+\Pr(T^{0,\infty} > \tau|ns)\\
&=\prod_{t=s}^{\tau} \Pr(T > t|S > t,T \geq t,Z=0)\cdot \Pr(as) \\
\beta_z  &=[\Pr(T^{1,\infty}  > \tau)-\Pr(T^{0,\infty}  > \tau)]\\
& \qquad  = [\Pr(T^{1,\infty}  > \tau |as)-\Pr(T^{0,\infty}  > \tau |as)]\cdot \Pr(as)\\
& \qquad \qquad \qquad + [\Pr(T^{1,\infty}   > \tau |cs)-\Pr(T^{0,\infty}  > \tau |cs)]\cdot \Pr(cs)\\
& \quad  =\Big(\sone(s' > \tau) + \sone(s' \leq \tau)\cdot \prod_{t=s'}^{\tau} \Pr(T > t |S > t,T \geq t,Z=1)\\
&\qquad \qquad \qquad \qquad \qquad \qquad \qquad \qquad - (\prod_{t=s}^{\tau} \Pr(T > t |S > t,T \geq t,Z=0)\Big)\cdot \Pr(as) \\
& \qquad \qquad  + \big(\sone(s' >\tau)\cdot \big( \prod_{t=0}^{\tau} \Pr(T > t |S > t,T \geq t,Z=1)- \Pr(as)\big)/\Pr(cs) \big)\cdot \Pr(cs)\\
\beta_s &=[\Pr(T^{0,s}  > \tau)-\Pr(T^{0,\infty}  > \tau)]\\
& \qquad  =\Big(\Pr(T^{0,s}   > \tau | as)-\Pr(T^{0,\infty}   > \tau |as)\Big)\cdot \Pr(as)\\
&\quad =\Big(\prod_{t=s}^{\tau} \Pr(T > t|S=s,T \geq t,Z=0) - \prod_{t=s}^{\tau} \Pr(T > t |S > t,T \geq t,Z=0)\Big)\cdot \Pr(as)\\
& \Pr(as)= \Pr(T^{1,\infty}\geq s, T^{0,\infty} \geq s)=\prod_{t=0}^{s-1} \Pr(T > t |S > t,T \geq t,Z=0)\\
&\Pr(cs) =\Pr(T^{1,\infty}\geq s, T^{0,\infty}< s)=\prod_{t=0}^{s-1} \Pr(T > t |S > t,T \geq t,Z=1)-\Pr(as)\\
&\Pr(ns) = 1- \Pr(as) - \Pr(cs)
\end{split}
\end{equation*}

Adding assumption A.V we also obtain:
\[ \small
\begin{split}
\beta_{zs}&= [(\Pr(T^{1,s}  > \tau)-\Pr(T^{1,\infty}  > \tau)) - (\Pr(T^{0,s}  > \tau)-\Pr(T^{0,\infty}  > \tau))]\\
&= [\Pr(T^{1,s}   > \tau| cs)-\Pr(T^{1,\infty}  > \tau | cs)]\cdot \Pr(cs)
\end{split}
\]
All unlisted effects for other substrata are also identified and equal to 0.

\newpage

\section{Additional Descriptive Figures}
\label{sec: DataStats}

\begin{figure} [!h]
  \begin{subfigure}{0.48\textwidth}
    \includegraphics[width=\linewidth]{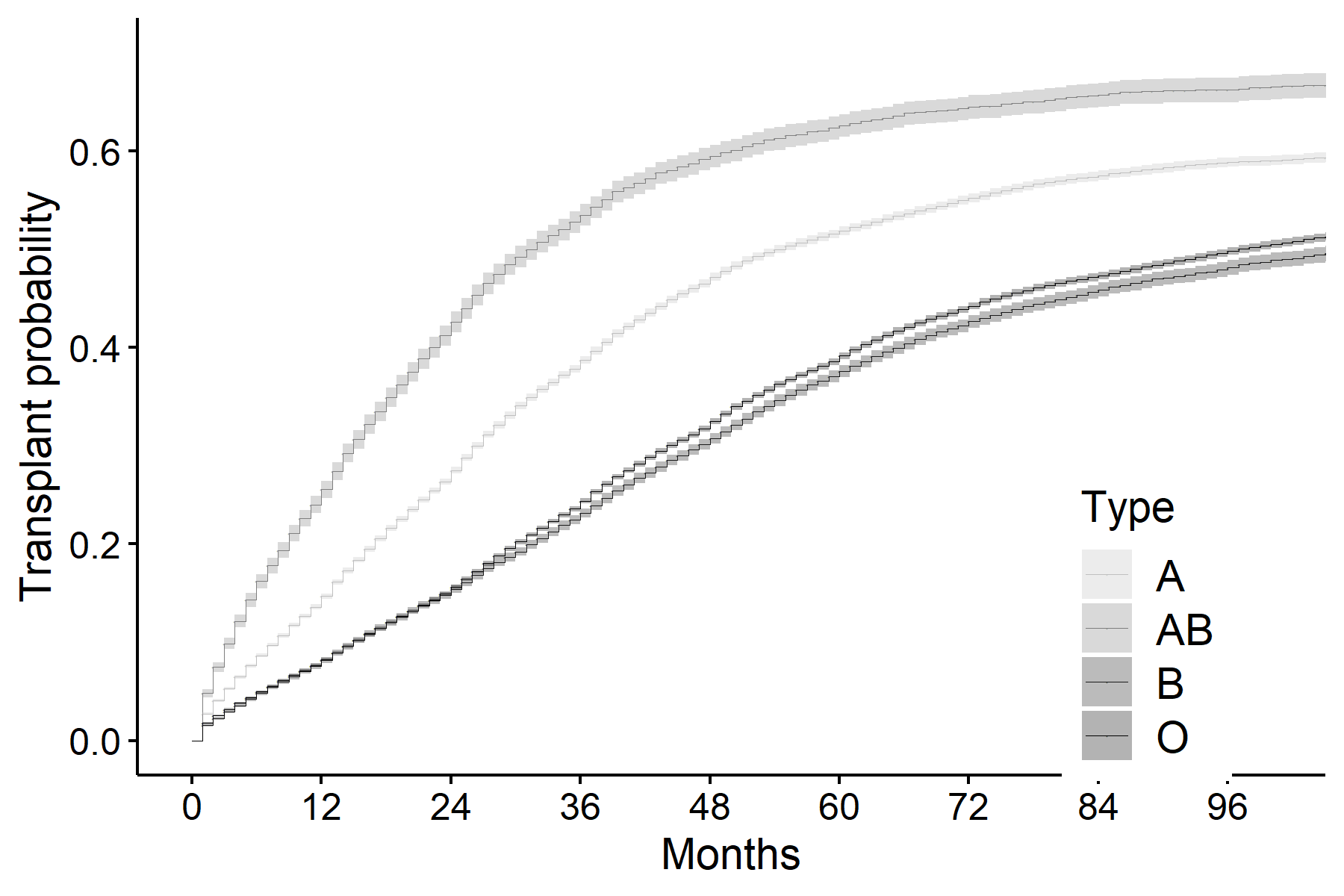}
    \caption{No-Transplant Probability} \label{fig:SurvT_all}
  \end{subfigure}%
  \hspace*{\fill}   
  \begin{subfigure}{0.48\textwidth}
    \includegraphics[width=\linewidth]{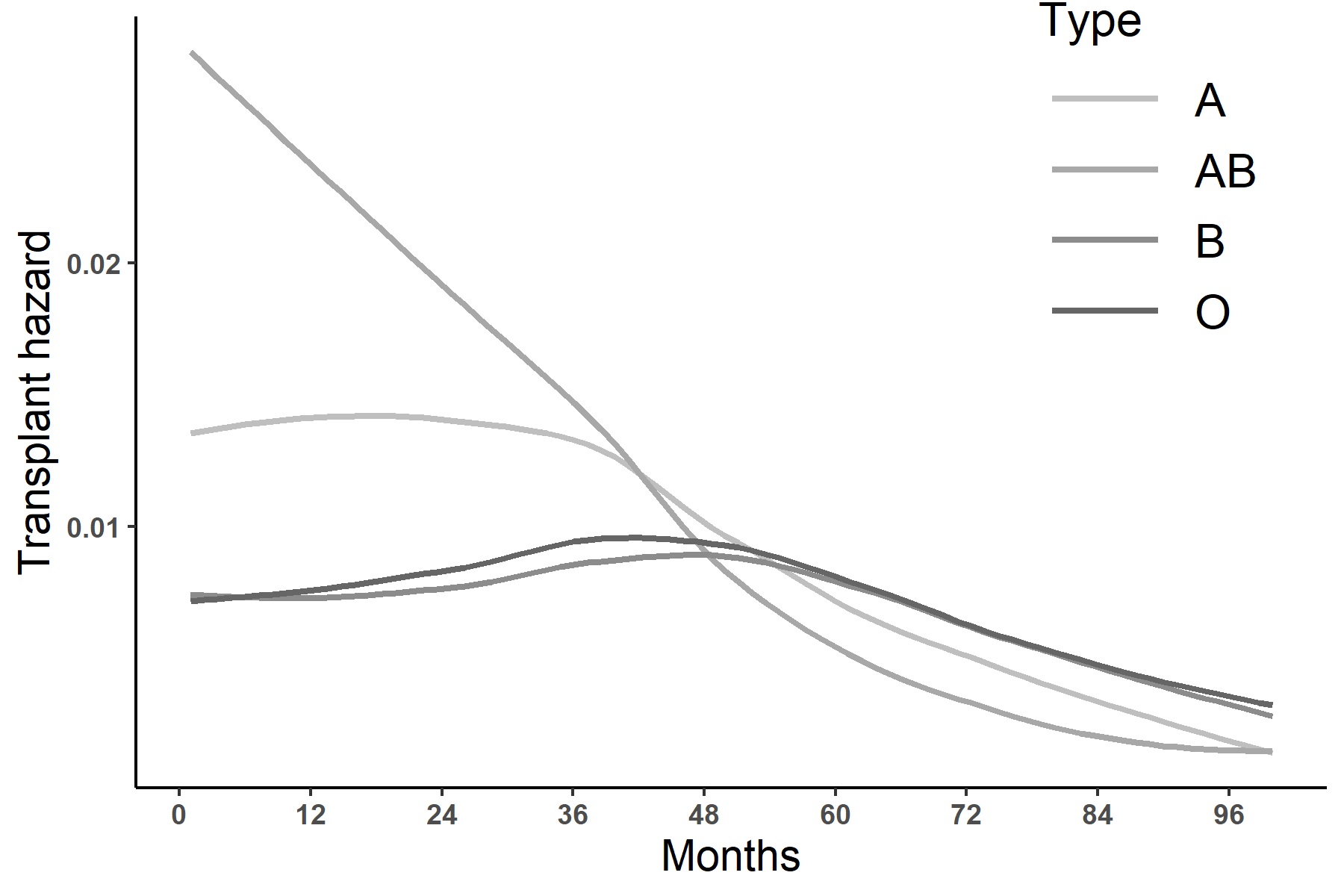}
    \caption{Transplant Hazard} \label{fig:Thaz_all}
  \end{subfigure}%

\caption{No-Transplant Probability and Transplant Hazard for major ABO blood types \\
    \scriptsize{Based on selected sample of SRTR data. $N_{A}=58,666$, $N_{AB}=7203$, $N_{B}=29,181$, $N_{O}=97,170$.}} \label{fig:SurvAll_main}
\end{figure}

\begin{figure} [!h]
  \begin{subfigure}{0.45\textwidth}
    \includegraphics[width=\linewidth]{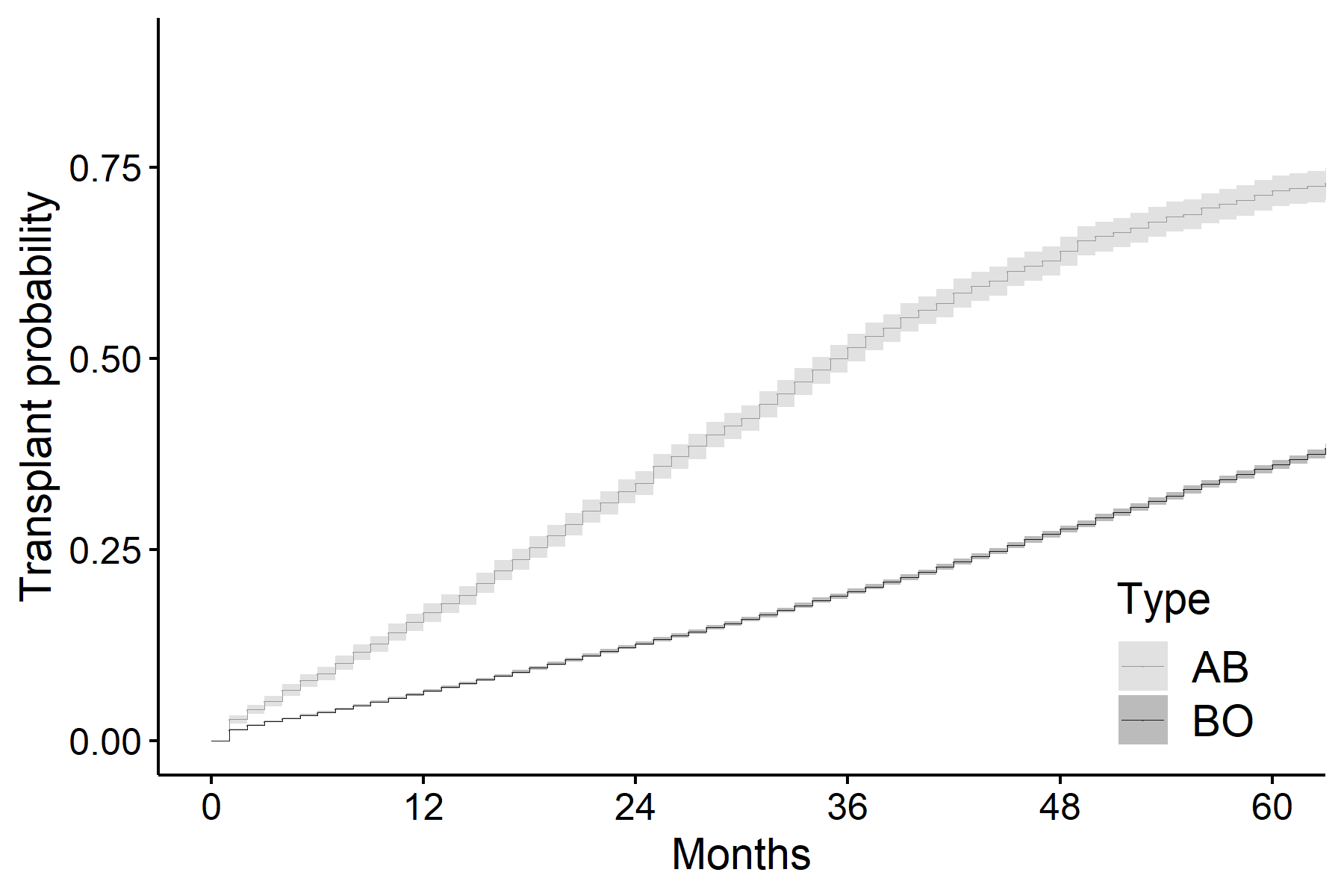}
    \caption{No-Transplant Probability} \label{fig:SurvT_post2015}
  \end{subfigure}%
  \hspace*{\fill}   
  \begin{subfigure}{0.45\textwidth}
    \includegraphics[width=\linewidth]{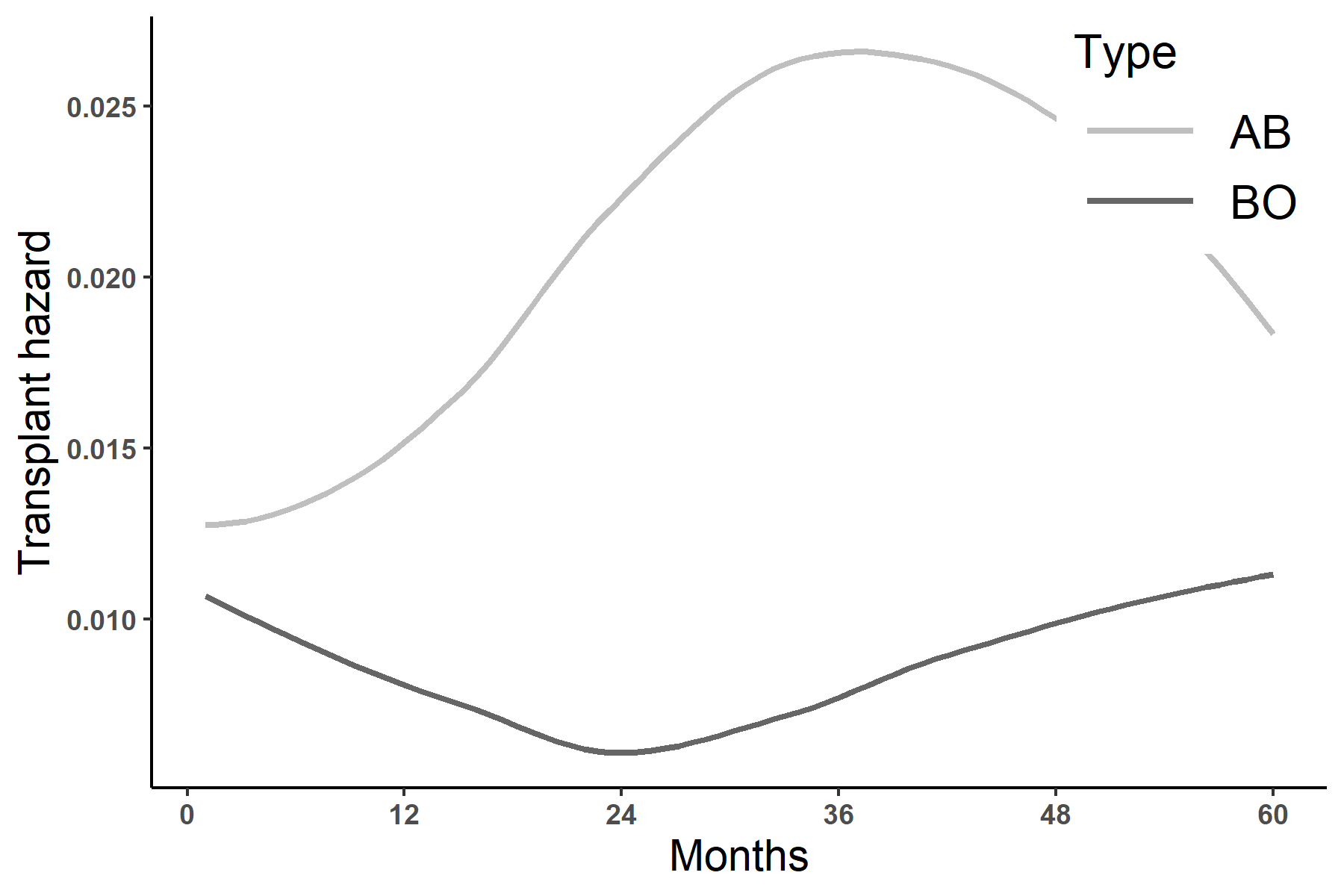}
    \caption{Transplant Hazard} \label{fig:Thaz_post2015}
  \end{subfigure}%
  \newline
  \begin{subfigure}{0.45\textwidth}
    \includegraphics[width=\linewidth]{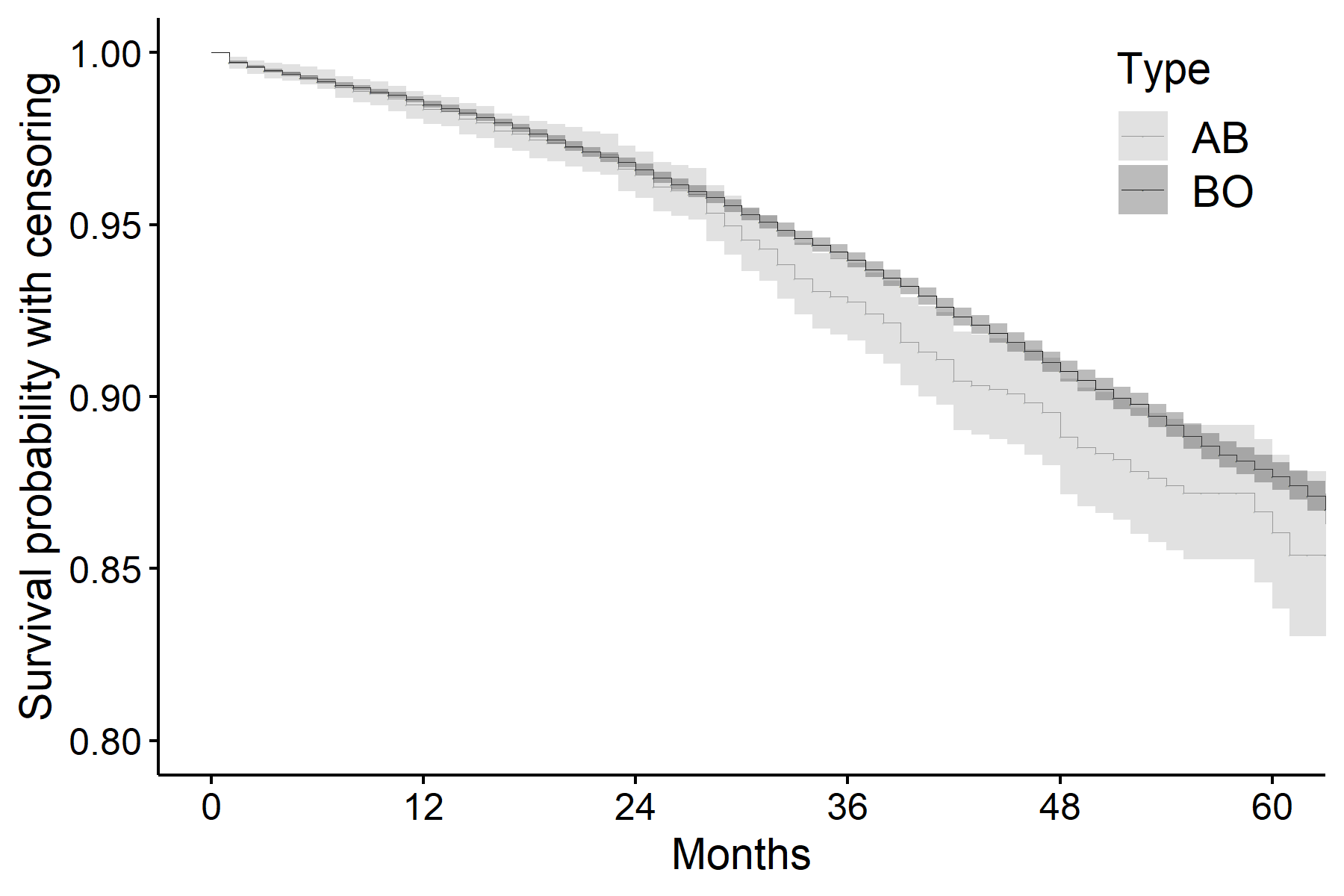}
    \caption{Pre-Transplant Survival} \label{fig:SurvC_post2015}
  \end{subfigure}%
  \hspace*{\fill}   
  \begin{subfigure}{0.45\textwidth}
    \includegraphics[width=\linewidth]{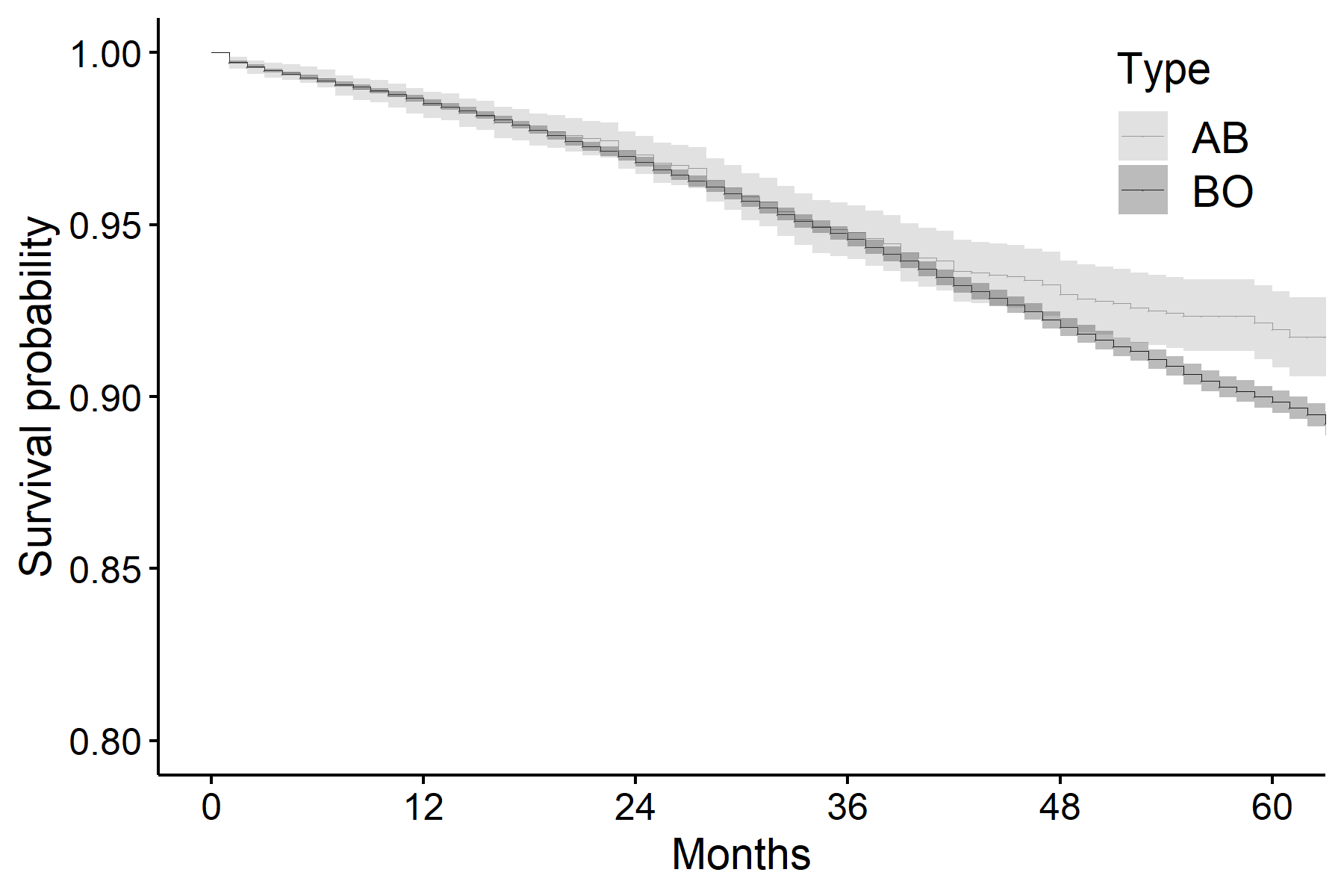}
    \caption{Survival} \label{fig:Surv_post2015}
  \end{subfigure}%

\caption{Survival of candidates post-December 2014 reform\\
    \scriptsize{Based on selected sample of SRTR data described in section \ref{sec:data}. $N_{B/O: no-Tr}=52,215$, $N_{B/O: Tr}=12,652$, $N_{AB: no-Tr}=2156$, $N_{AB: Tr}=1937$.}} \label{fig:SurvAll_post2015}
\end{figure}

\begin{figure}
  \begin{subfigure}{0.45\textwidth}
    \includegraphics[width=\linewidth]{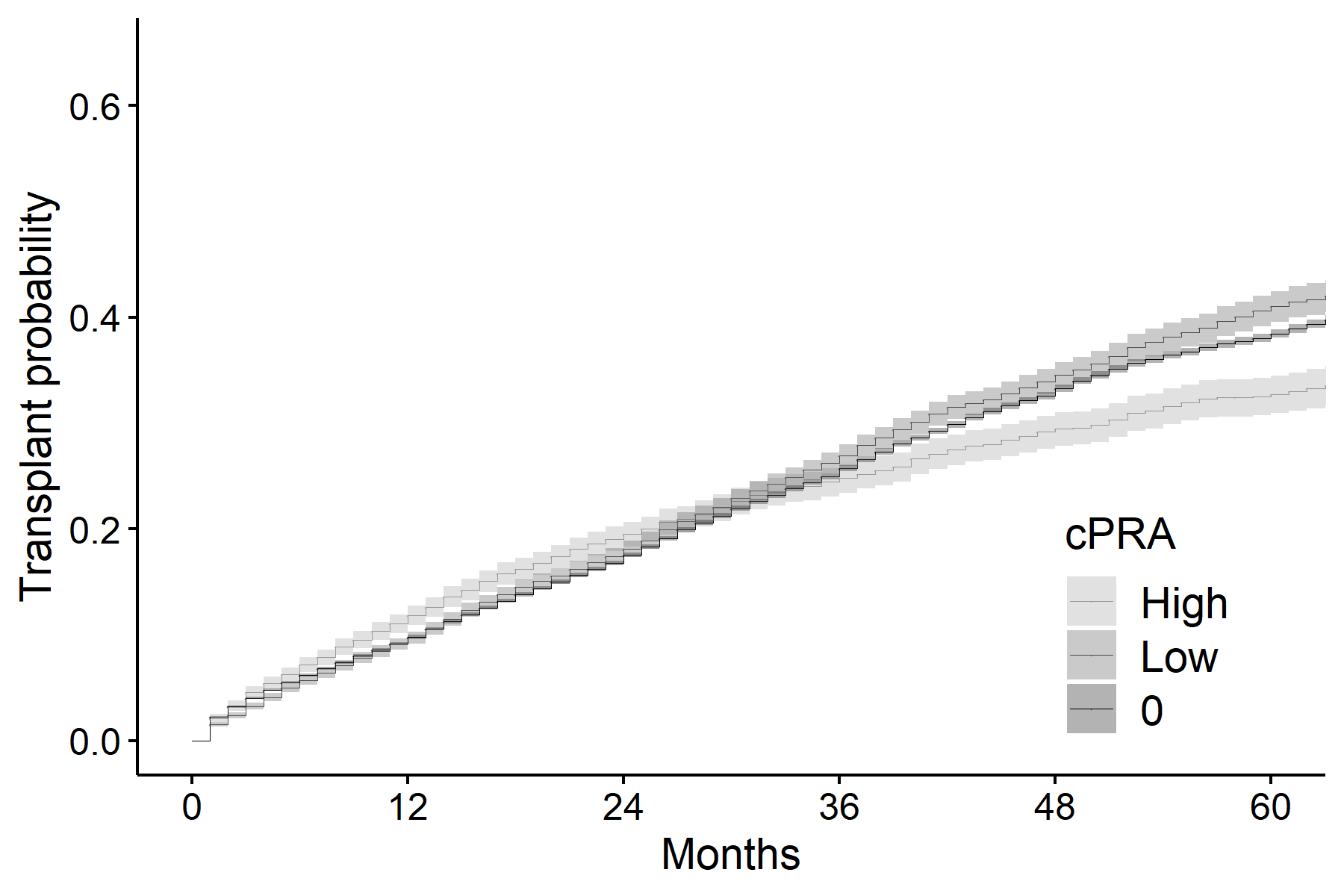}
    \caption{Transplant Probability Pre-Reform} \label{fig:SurvT_mainPRAall_wTX}
  \end{subfigure}%
  \hspace*{\fill}   
    \begin{subfigure}{0.45\textwidth}
    \includegraphics[width=\linewidth]{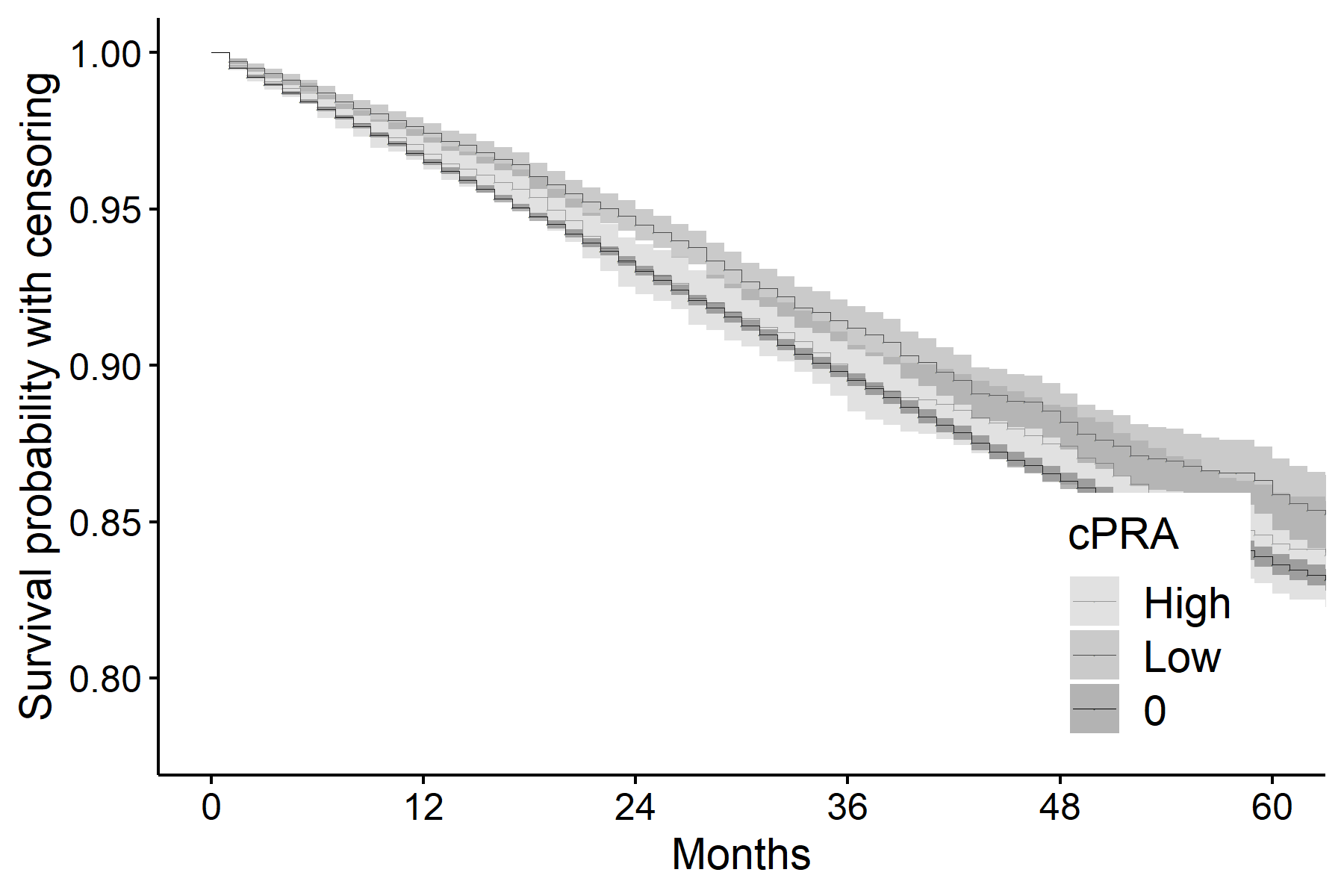}
    \caption{Pre-Transplant Survival Pre-Reform} \label{fig:SurvC_mainPRAall_wTX}
  \end{subfigure}%
  \newline
  \begin{subfigure}{0.45\textwidth}
    \includegraphics[width=\linewidth]{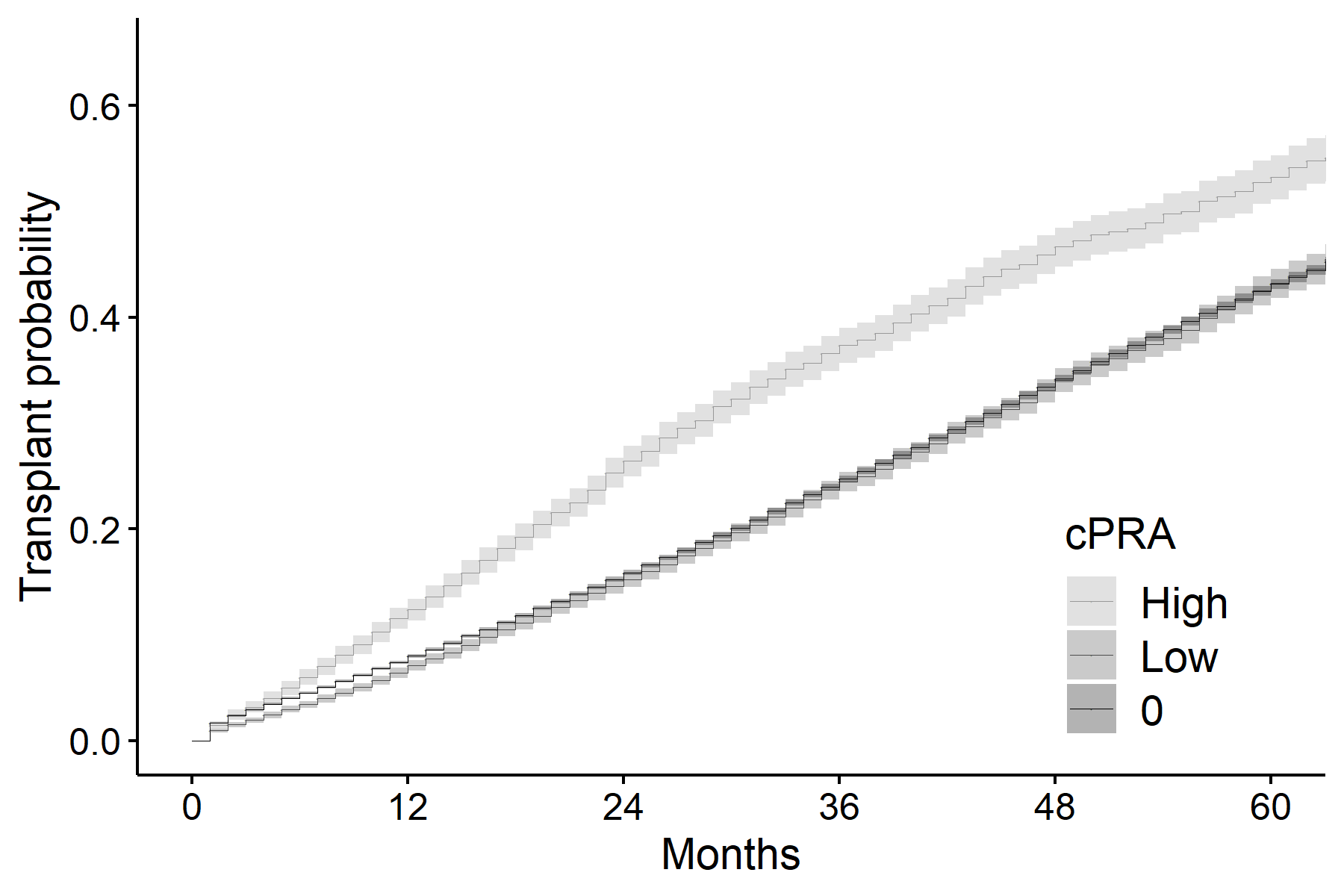}
    \caption{Transplant Probability Post-Reform} \label{fig:SurvT_mainPRAall_post2015_wTX}
  \end{subfigure}%
  \hspace*{\fill}   
  \begin{subfigure}{0.45\textwidth}
    \includegraphics[width=\linewidth]{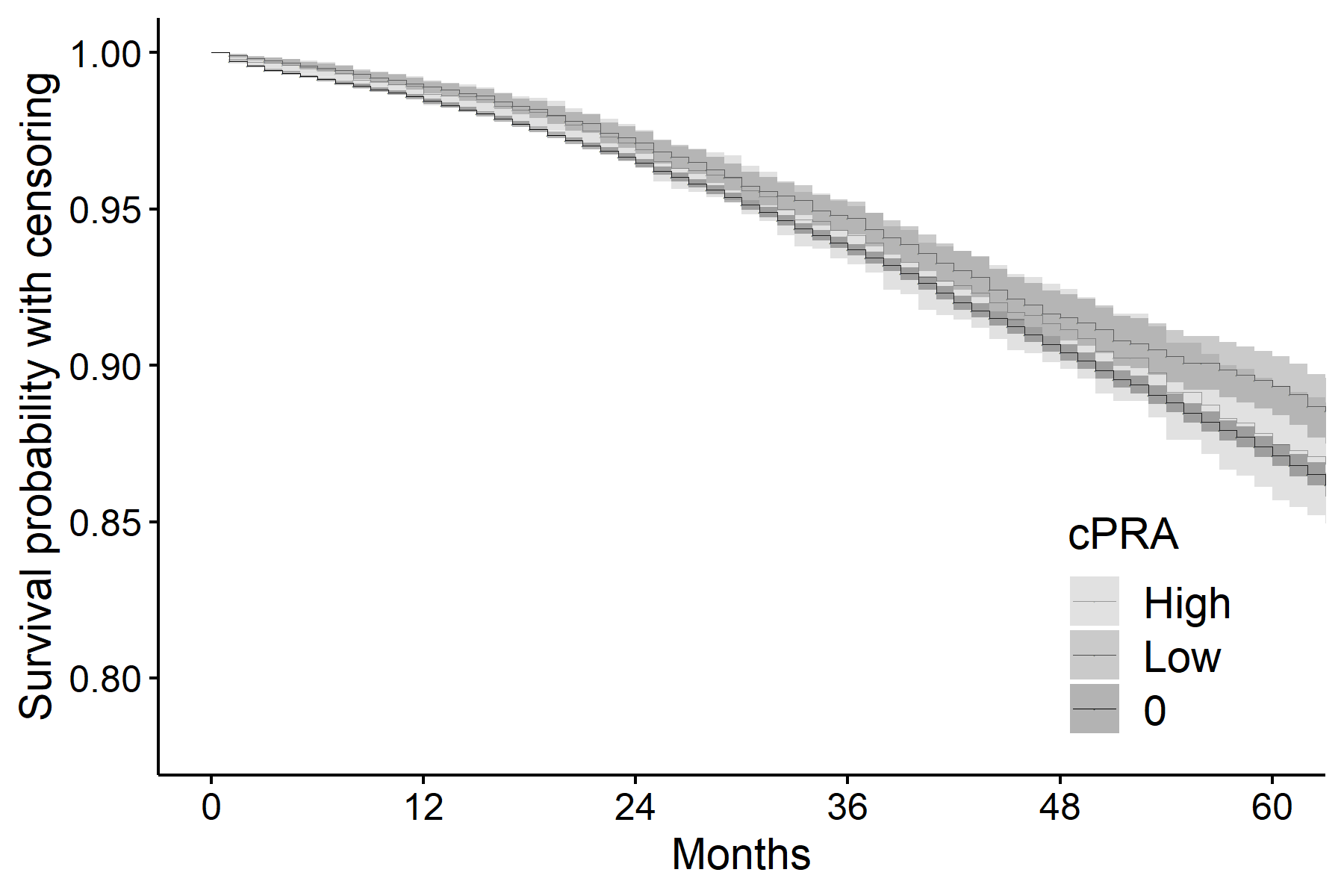}
    \caption{Pre-Transplant Survival Post-Reform} \label{fig:SurvC_mainPRAall_post2015_wTX}
  \end{subfigure}%

\caption{Survival of candidates on the kidney transplant waitlist, by cPRA group including previously transplanted candidates \\
    \scriptsize{Based on selected sample of SRTR data. 0-cPRA: cPRA$=0$; Low-cPRA: $0<$cPRA$<0.8$; High-cPRA: $0.8 \leq$cPRA.}} \label{fig:SurvPRA_wTX}
\end{figure}

\newpage

\section{Descriptive Statistics and Data selection}
\label{sec: DataClean}

\begin{table} [!h] \scriptsize
\begin{center}
\caption{Descriptive Statistics}
\label{ta:DescStats}
\begin{tabular}{l ccd{1}cccccc}
 \hline \hline \\
   & \multicolumn{2}{c}{Blood Type Analysis}     & &   \multicolumn{6}{c}{cPRA Analysis}   \\
  \cline{2-3}  \cline{5-10} \\[-0,5ex]
    & \emph{AB pre} & \emph{B/O pre} & & \emph{High pre} &  \emph{Low pre} & \emph{0 pre} &  \emph{High post} &  \emph{Low post} & \emph{0 post}  \\

  \hline
  \\[0.5ex]

$Age$	&	52.772	&	52.117	&&	52.455	&	52.958	&	53.415	&	53.012	&	53.427	&	53.938		\\[-0.3ex]
	& (12.754) & (12.730) && (11.952) & (12.657) & (12.619) & (12.056) & (12.783) & (12.868)	\\[0.3ex]
$Acpt_{Hep B}$	&	0.544	&	0.568	&&	0.542	&	0.540	&	0.594	&	0.588	&	0.558	&	0.668		\\[-0.3ex]
	& (0.498) & (0.495) && (0.498) & (0.498) & (0.491) & (0.492) & (0.497) & (0.471)	\\[0.3ex]
$Acpt_{HCV+}$	&	0.050	&	0.069	&&	0.072	&	0.075	&	0.077	&	0.262	&	0.327	&	0.305		\\[-0.3ex]
	& (0.219)	 & (0.253) && (0.259) & (0.264) & (0.266) & (0.440) & (0.469) & (0.461)	\\[0.3ex]
$Coll_{HS or less}$	&	0.518	&	0.563	&&	0.535	&	0.524	&	0.541	&	0.447	&	0.441	&	0.452		\\[-0.3ex]
	& (0.500) & (0.496) && (0.499) & (0.499) & (0.498) & (0.497) & (0.497) & (0.498)	\\[0.3ex]
$Coll_{any}$	&	0.254	&	0.240	&&	0.265	&	0.265	&	0.248	&	0.293	&	0.271	&	0.259		\\[-0.3ex]
	& (0.436) & (0.427) && (0.442) & (0.442) & (0.432) & (0.455) & (0.445) & (0.438)	\\[0.3ex]
$Coll_{deg}$	&	0.227	&	0.197	&&	0.199	&	0.211	&	0.212	&	0.260	&	0.288	&	0.289		\\[-0.3ex]
	& (0.419) & (0.397) && (0.400) & (0.408) & (0.409) & (0.439) & (0.453) & (0.453)	\\[0.3ex]
$BMI$	&	28.490	&	28.508	&&	28.641	&	29.099	&	29.145	&	28.993	&	29.555	&	29.380		\\[-0.3ex]
	& (5.867) & (5.767) && (5.655) & (5.586) & (5.669) & (5.594) & (5.413) & (5.553)	\\[0.3ex]
$PrevMalign$	&	0.062	&	0.057	&&	0.064	&	0.064	&	0.067	&	0.075	&	0.076	&	0.084		\\[-0.3ex]
	& (0.241) & (0.233) && (0.245) & (0.244) & (0.250) & (0.263) & (0.265) & (0.278)	\\[0.3ex]
$Asian$	&	0.117	&	0.068	&&	0.065	&	0.068	&	0.068	&	0.069	&	0.070	&	0.086		\\[-0.3ex]
	& (0.321) & (0.251) && (0.246) & (0.252) & (0.252) & (0.254) & (0.255) & (0.281)	\\[0.3ex]
$Black$	&	0.360	&	0.360	&&	0.412	&	0.418	&	0.333	&	0.427	&	0.386	&	0.274		\\[-0.3ex]
	& (0.480) & (0.480) && (0.492) & (0.493) & (0.471) & (0.495) & (0.487) & (0.446)	\\[0.3ex]
$White$	&	0.508	&	0.550	&&	0.499	&	0.487	&	0.576	&	0.481	&	0.521	&	0.615		\\[-0.3ex]
	& (0.500) & (0.498) && (0.500) & (0.500) & (0.494) & (0.500) & (0.500) & (0.487)	\\[0.3ex]
$Other$	&	0.016	&	0.022	&&	0.024	&	0.028	&	0.022	&	0.022	&	0.023	&	0.025		\\[-0.3ex]
	& (0.124) & (0.147) && (0.153) & (0.164) & (0.148) & (0.148) & (0.151) & (0.155)	\\[0.3ex]
$Female$	&	0.378	&	0.380	&&	0.901	&	0.508	&	0.362	&	0.898	&	0.474	&	0.359		\\[-0.3ex]
	& (0.485) & (0.485) && (0.299) & (0.500) & (0.481) & (0.303) & (0.499) & (0.480)	\\[0.3ex]
$Diabetes$	&	0.393	&	0.385	&&	0.393	&	0.431	&	0.439	&	0.388	&	0.432	&	0.444		\\[-0.3ex]
	& (0.488) & (0.487) && (0.488) & (0.495) & (0.496) & (0.487) & (0.495) & (0.497)	\\[0.3ex]
$Transp$	&	0.114	&	0.099	&&	-	&	-	&	-	&	-	&	-	&	-		\\[-0.3ex]
	& (0.318) & (0.299) && (-) & (-) & (-) & (-) & (-) & (-)	\\[0.3ex]
$Blood-A$	&	0.000	&	-	&&	0.311	&	0.310	&	0.314	&	0.292	&	0.325	&	0.319		\\[-0.3ex]
	& (0.000) & (-) && (0.463) & (0.462) & (0.464) & (0.455) & (0.468) & (0.466)	\\[0.3ex]
$Blood-B$	&	0.000	&	-	&&	0.148	&	0.165	&	0.149	&	0.167	&	0.158	&	0.150		\\[-0.3ex]
	& (0.000) & (-) && (0.355) & (0.371) & (0.356) & (0.373) & (0.365) & (0.357)	\\[0.3ex]
$Blood-AB$	&	0.000	&	-	&&	0.038	&	0.038	&	0.040	&	0.036	&	0.038	&	0.039		\\[-0.3ex]
	& (0.000) & (-) && (0.190) & (0.191) & (0.195) & (0.187) & (0.190) & (0.195)	\\[0.3ex]
$Blood-O$	&	0.000	&	-	&&	0.503	&	0.488	&	0.497	&	0.506	&	0.480	&	0.492		\\[-0.3ex]
	& (0.000) & (-) && (0.500) & (0.500) & (0.500) & (0.500) & (0.500) & (0.500)	\\

\hline \hline
\end{tabular}
\end{center}
Means for each group presented with standard deviations in parenthesis. $Age$: Candidate age at listing,
$Acpt_{Hep B}$: Will accept an Hepatitis B Core Antibody Positive Donor?,
$Acpt_{HCV+}$: Will accept an HCV Positive donor? , $Coll_{HS or less}$: Attended High school or less, $Coll_{any}$
Attended college/technical school, $Coll_{deg}$: Obtained at least Associate/Bachelor degree, $BMI$,  $PrevMalign$: Any previous Malignancy
$Asian$: Race-Asian, $Black$: Race-Black, $White$: Race-White, $Other$: Race-Other Non-White, $Female$, $Diabetes$: Had Diabetes upon entering the waitlist, $Transp$: Received previous transplant (other than kidney transplant),
$Blood-A$: A-Blood type, $Blood-B$: B-Blood type, $Blood-AB$: AB-Blood type, $Blood-O$: O-Blood type
\end{table}

\begin{flushleft}
\textbf{Blood type analysis data selection:}
\end{flushleft}
Our initial data contains 1,010,051 observations. We remove individuals with no activation date (954,406), and select only people set to receive a kidney transplant (894,372). We then select only individuals entering the waitlist between December 1st 2002 and December 1st 2014 (429,645). Among these we only keep the first observed kidney transplant (351,547), for candidates who are over 18 (341,472), and who did not receive a transplant from a living donor (287,032). We further remove individuals with unusual A1, A2, A1B, A2B blood types (283,198). In the AB vs.\ B/O blood types analysis, we also remove the A blood types (194,337). We drop individuals with a positive cPRA (180,828). We keep only individuals who are on dialysis upon entering the waitlist (168,372). All remaining reductions in sample for the analysis result from missing values in the covariate matrix. These include: $<1\%$ of missing values for education, $<0.01\%$ of missing values for BMI, $<2.8\%$ of missing values for previous malignancy. The final dataset for analysis includes 120,602 individual spells of candidates who entered the kidney transplant waitlist. We take December 1st 2014 as the 2014 reform cutoff date.

\begin{flushleft}
\textbf{CPRA analysis data selection:}
\end{flushleft}
For the cPRA sample we include observations which were exposed to the cPRA measure in the kidney allocation system. To ensure a similar sample of durations both before and after the change in reform, we include all observations of candidates who entered the waitlist after December 1st 2008. In our data, the cPRA score varies over time. In order to avoid conditioning on post-regime randomization intermediate variables, we define the cPRA score as the first registered cPRA.  For the cPRA analysis, we include all major blood types. For the cPRA missing values we have: $<0.5\%$ of missing values for education, $<0.005\%$ of missing values for BMI, , $<1.9\%$ of missing values for previous malignancy. In terms of cPRA categories, pre-December 2014 reform, we have that 89.3\% of individuals have a cPRA equal to 0, 7.3\% have a cPRA between 0 and 0.8, and 3.4\% have a cPRA above 0.8. Among these, 11\% of 0-cPRA candidates had a previous non-renal transplant, 17\% of Low-cPRA candidates had a previous non-renal transplant, and 44\% of High-cPRA candidates had a previous non-renal transplant.

In the analysis, we include all variables listed in Table \ref{ta:DescStats} as covariates. We only include binary variables as covariates. These include five categories for $Age$, and four categories for $BMI$.

\newpage

\section{Robustness Checks}
\label{sec: Robust}

\begin{table} [!h] \scriptsize
\begin{center}
\caption{Causal Effect Decomposition}
\label{ta:Effmain}
\begin{tabular}{l ccccccccc}
 \hline \hline \\
   & \multicolumn{1}{c}{(1)}     &    \multicolumn{1}{c}{(2)} &
  \multicolumn{1}{c}{(3)}  &  \multicolumn{1}{c}{(4)}  && \multicolumn{1}{c}{(5)} & \multicolumn{1}{c}{(6)}  &  \multicolumn{1}{c}{(7)} &  \multicolumn{1}{c}{(8)}  \\
  \hline
  \\[0.5ex]

$\beta_0$	&	0.754	&	0.869	&	0.815	&	0.803	&&	0.895	&	0.945	&	0.898	&	0.939		\\[-0.3ex]
	& (0.025) & (0.056) & (0.084) & (0.080) && (0.008) & (0.003) & (0.008) & (0.004)	\\[-0.3ex]
	& [0.000] & [0.000] & [0.000] & [0.000] && [0.000] & [0.000] & [0.000] & [0.000]	\\[0.3ex]
$\beta_z$	&	-0.007	&	0.039	&	0.033	&	0.057	&&	0.027	&	0.008	&	0.024	&	0.013		\\[-0.3ex]
	& (0.029) & (0.009) & (0.005) & (0.012) && (0.004) & (0.003) & (0.002) & (0.001)	\\[-0.3ex]
	& [0.801] & [0.000] & [0.000] & [0.000] && [0.000] & [0.001] & [0.000] & [0.000]	\\[0.3ex]
$\beta_{(0,s]}$	&	0.198	&	0.105	&	0.142	&	0.162	&&	0.072	&	0.039	&	0.068	&	0.045		\\[-0.3ex]
	& (0.025) & (0.044) & (0.062) & (0.064) && (0.008) & (0.004) & (0.006) & (0.004)	\\[-0.3ex]
	& [0.000] & [0.017] & [0.023] & [0.012] && [0.000] & [0.000] & [0.000] & [0.000]	\\[0.3ex]
$\beta_{z(0,s]}$	&	0.018	&	-0.030	&	-0.025	&	-0.048	&&	-0.016	&	-0.002	&	-0.014	&	-0.008		\\[-0.3ex]
	& (0.029) & (0.007) & (0.004) & (0.010) && (0.003) & (0.002) & (0.002) & (0.001)	\\[-0.3ex]
	& [0.526] & [0.000] & [0.000] & [0.000] && [0.000] & [0.336] & [0.000] & [0.000]	\\[0.3ex]
	&		&		&		&		&&		&		&		&			\\
$\alpha_{z}$	&	0.265	&	0.205	&	0.269	&	0.223	&&	0.051	&	0.147	&	0.002	&	0.007		\\[-0.3ex]
	& (0.015) & (0.115) & (0.123) & (0.068) && (0.020) & (0.014) & (0.008) & (0.008)	\\[-0.3ex]
	& [0.000] & [0.074] & [0.277] & [0.001] && [0.009] & [0.000] & [0.810] & [0.349]	\\[0.3ex]
	&		&		&		&		&&		&		&		&			\\
$N_{Z=0: no-Tr}$	&	4749	&	2052	&	4514	&	4744	&&	2286	&	1,620	&	96,973	&	73,029		\\
$N_{Z=0: Tr}$	&	5043	&	1873	&	4179	&	5690	&&	635	&	872	&	23,707	&	17,544		\\
$N_{Z=1: no-Tr}$	&	108,612	&	49,859	&	82192	&	110,721	&&	7415	&	7,588	&	7,415	&	7,588		\\
$N_{Z=1: Tr}$	&	49,968	&	12,137	&	26816	&	56,458	&&	1733	&	1,776	&	1,733	&	1776		\\[0.3ex]

\hline \hline
\end{tabular}
\end{center}
Standard errors in parenthesis. P-values in brackets. Column (1) presents pre Dec. 2014 decomposition for transplant within first two years on waitlist on survival past 7 years with $Z=0$: AB-blood types, $Z=1$: B/O-blood types, without covariates. Column (2) presents decomposition effects for survival beyond 4 years post Dec. 2014 reform, with covariates included. Column (3) presents decomposition effects pre-2014 on survival past 4 years when $Z=0$: AB-blood types, $Z=1$: O-blood types, with covariates. Column (4) presents decomposition effects pre-2014 on survival past 4 years when $Z=0$: AB-blood types, $Z=1$: B/O-blood types, with covariates and including living donor transplants. Column (5) presents pre Dec. 2014 decomposition for transplant within first two years on waitlist on survival past 3 years with $Z=0$:high-cPRA, $Z=1$:low-cPRA, without covariates. Column (6) presents same cPRA decomposition post Dec. 2014, without covariates. Column (7) presents pre Dec. 2014 decomposition for transplant within first two years on waitlist on survival past 3 years with $Z=0$:0-cPRA, $Z=1$:low-cPRA, without covariates. Column (8) presents same cPRA decomposition post Dec. 2014, without covariates.
\end{table}

\newpage

\newpage

\section{Evaluating Estimation Approach on Simulated Data}
\label{app: simResults}

To relate our dynamic treatment effect model to the literature on dynamic discrete choice models we present in this appendix a standard job search model (e.g. Mortensen, \hyperlink{Mor1986}{1986}). We then extend it to include a treatment. We loosely adapt the discussion of the search model to our kidney transplant application. For expositional purposes, the model is solved under simplistic assumptions. We then explain how the dynamic discrete choice model relates to the parameters in our proportional hazard model in Section \ref{subsec:estim}. Thereafter we describe the data generating process and present simulation results for our estimator.

\subsection{Standard dynamic discrete choice model}

Consider an agent who enters an initial state at time $t=0$ and assume that time is discrete. In each subsequent period the agent faces the choice of staying within
this state or leaving. Whenever he is in the initial state, the agent derives utility $w_0$. In our application, $t=0$ is the moment a candidate enters the kidney transplant waitlist and each period the candidate, which we consider as a unit combined of psychological and biological factors, puts in a certain amount of effort to remain alive. $w_0$ represents the combined psychological and biological utility of staying alive.

Next, with probability $\lambda$ the agent receives an offer to leave the state. An offer can be interpreted as a negative health shock on the body due to kidney failure. The agent also faces a cost $c$ corresponding to the necessary biological effort to prevent a health shock, with lower costs corresponding to higher effort. Once the agent receives an offer, he has to decide immediately whether or not to accept it. An offer is characterized by its instantaneous utility $w$ drawn from the distribution $G(w)$. Upon accepting an offer, the agent derives the same instantaneous utility $w$ for each subsequent period. So once a candidate's biological constitution receives a health shock beyond what it can fight, they die and receive the (perceived) utility of death thereafter.

We assume the agent optimization behaviour follows a dynamic discrete choice model which nests search
models and optimal stopping models.\footnote{For more details, see Rogerson, Shimer, and Wright (\hyperlink{RogEA2005}{2005}).} The agent forms expectations over future instantaneous utilities. Denoting by $\rho$ the discount rate, the present discounted combined psychological and biological value of remaining alive at the start of period $t$, $V_{0,t}$, can be described by the Bellman equation,
\[ \label{eq:searchbase}
 V_{0,t} = w_0 - c + \rho \lambda \mathbb{E}[\max\{V_1(w),V_{0,t+1}\}] + \rho (1 -\lambda) V_{0,t+1}
 \]
$V_{1}(w)$ is the discounted value that the agent
would acquire by failing to fight off a health shock, dying as a consequence, and reaping instantaneous biological utility $w$. The agent follows a stationary reservation utility strategy where $w^*$ is the minimum offer of $w$ required to induce the agent to exit in the following period. In terms of our study, $w^*$ represents the utility associated to the limit at which someone's biological constitution prefers to stop functioning, rather than exert the continued effort to keep a person alive, despite the psychological desire to remain alive. For a healthy person with a well functioning immune system, we would expect $w^*$ to be very high, since only an extreme negative health shock (large $w$) would induce someone's biological functions to give in. A lower $w^*$ implies, all else equal, that a person's health is worse, making lower utilities from death $w$ more appealing. Under a reservation utility strategy, we can reformulate the Bellman equation as
\[
 V_{0,t} = w_0 - c + \rho \lambda \int_{w^*}^{\infty}\left(V_1(w)-V_{0,t+1}\right) dG(w) + \rho V_{0,t+1}
\]

We can augment this model to include a treatment prescribed by a regime. Let us assume that the agent knows that he has been assigned to a certain regime $z$ which allocates future treatment. For simplicity we consider in this section a stochastic assignment regime where the agent faces the same probability $\pi$ to receive treatment at each period. So each regime $z$ is fully characterized by its value of $\pi$. This type of randomization can correspond as in our empirical setting to a randomization set by nature but may also be a rule imposed by a policymaker. In our empirical setting, it can best be interpreted as a situation in which agents receive different signals of how likely they are to receive treatment, and interpret this signal as a constant hazard $\pi$ to treatment each period.

In our application, candidates on the waitlist were randomized at birth to have an AB blood type or a B/O blood type. This blood type, while inconsequential during most peoples' lives, is an important determinant to the time a candidate must wait until receiving a kidney transplant, which is the treatment in our setting. From the point of view of a candidate, their blood type may be the most salient feature determining the duration until they receive a transplant, but is not the only factor influencing the timing of a kidney transplant. Given the many factors determining the waiting time until a match, the blood type randomization can be seen as a stochastic treatment assignment mechanism since it influences the chances of finding a kidney match but does not determine the exact date a candidate will receive the transplant. Receiving a kidney transplant would likely reduce the arrival of health shocks ($\lambda$), change the effort a candidate needs to put into their general health upkeep (search costs $c$), or change the distribution of the (perceived) utility from death ($G(w)$).

To simplify the exposition, assume treatment only affects the distribution of the perceived utility from death by prescribing a higher mean to $G^{tr}(w)$ relative to $G(w)$ for an agent upon receiving the treatment. Allowing agents to form expectations over treatment outcomes, the alive value functions before receiving a kidney transplant, $V_{0,t}$, and after receiving a kidney transplant, $V_{0,t}^{tr}$, are given by,
\[ \label{eq:searchtreat}
\begin{split}
V_{0,t}&=w_0-c+\rho \lambda \mathbb{E}[\max\{V_1(w),(1-\pi)V_{0,t+1}+\pi V_{0,t+1}^{tr}\}]+\rho(1-\lambda)[(1-\pi)V_{0,t+1}+\pi V_{0,t+1}^{tr}]\\
V_{0,t}^{tr}&=w_0-c+\rho \lambda \mathbb{E}[\max\{V_1^{tr}(w),V_{0,t+1}^{tr}\}+\rho(1-\lambda)V_{0,t+1}^{tr} \end{split}
\]
We can solve this model for the reservation utilities after and before treatment,\footnote{Full derivation of solutions are
presented in subsection \ref{app: reswDDC}}
\[
\begin{split}
w^*(\pi)&= \frac{(1-\rho)(1-\pi)}{1-\rho+\rho \pi} (w_0-c)+ \frac{\rho \lambda(1-\pi)}{(1-\rho)(1-\rho+\rho \pi)} \int_{w^{*}}^{\infty}\left(1-G(w)\right)dw +  \frac{\pi}{1-\rho+\rho \pi}w^{tr*}\\
w^{tr*}&=w_0-c+\frac{\rho\lambda}{1-\rho}\int_{w^{tr*}}^{+\infty}\left(1-G^{tr}(w)\right)dw
\end{split}
\]
Under the assumption that $V_{0,t}^{tr} > V_{0,t}$ one can demonstrate that the pre-treatment reservation utility $w^*(\pi)$ is increasing in $\pi$. This means that if agents foresee positive effects of receiving treatment on the $w$ offer distribution, and are in a regime with higher chances to receive treatment, then they will remain for longer in the initial state if they are not yet treated. In terms of our empirical application, this implies that if candidates on the waitlist foresee a kidney transplant to improve their biological utility of life, and they have a high probability of rapidly receiving a kidney transplant (AB blood type), then they will exert more effort each period to stay alive before receiving the transplant. Our empirical findings are not consistent with this prediction. They show that candidates with a higher probability of receiving a kidney transplant display a higher pre-transplant mortality. One way to consolidate our empirical results with the biological predictions of the model is to allow a candidate's behaviour to asymmetrically influence their biological functions in response to their probability of receiving a future kidney transplant. In particular, we must assume candidates with a high probability of treatment pay less attention to their general health, leading to lower survival.

It is worth considering the interpretation of parameters in our model of subsection \ref{subsec:estim} under different treatment assignment mechanisms prescribed by the regime. If the regime enforces a constant treatment hazard each period and there is full compliance, then $\theta_t^{S}(z)=\pi^{S}(z)$. If in addition agents know the regime, do not vary their search strategy over time, and the expected value of future variables over intermediate shocks is constant over time, then $\lambda^{T}_t(z)$ will be constant. This constant is then interpreted as the effect on the exit hazard of a constant treatment hazard regime. This is the setting considered in our dynamic discrete choice search model above.

However, in many cases the treatment hazard varies over time. In our kidney transplant setting, the hazard of receiving a kidney transplant (treatment) depends on the randomization to AB vs. B/O blood type, which may have varying effects over time. In addition, the intermediate unobserved variables may be different depending on $z$ and change in unexpected ways after treatment occurs. In these dynamic treatment assignment settings, we can interpret $\lambda^{T}_t(z)$ as the pre-treatment effect on the exit hazard of the average (or perceived average) treatment hazard under regime $z$. This effect itself depends on the average path of intermediate shocks specific to $z$. In our example, at any time prior to receiving a kidney transplant, $\lambda^{T}_t(z)$ is the average effect on death of the blood type effect path, where the path may be different for candidates with AB vs. B/O blood types. In many cases, the interpretation of estimation parameters from the proportional hazard model is secondary to the causal effects of interest.

\subsection{Data Generating Process}

The data generating process for the simulation data follows the dynamic discrete choice model presented in the above section with the following specifications,
\[ \label{eq:Sim Specification}
\begin{split}
U^{no-exit}_{it}&=w_{0it}-c_{it} \\
U^{exit}_{it}&=w_{it}\\
c_{it}&=\beta^{c}_{a}\cdot a_{i} + \beta^{c}_{e}\cdot e_{i}  \\
w_{it}&=\beta^{w}_{a}\cdot a_{i}  + \beta^{w}_{s}\cdot I(s < t) + \xi_{it} \quad \xi_{it}\sim \mathcal{N}(0,\sigma_{\xi}^2) \\
w_{0it}&=0.75 \cdot \beta^{w}_{a}\cdot a_{i} \\
\end{split}
\]
where $U^{no-exit}_{it}$ and $U^{exit}_{it}$ are the instantaneous utilities when the agent chooses to remain in the initial state or exit. $\lambda_{it}$ follows a Poisson distribution with mean $\beta^{\lambda}_{a}\cdot a_{i} + \beta^{\lambda}_{e}\cdot e_{i}$. The treatment outcome for the group under regime $Z=1$ is drawn each period from a binary distribution with probability $\pi_{Z=1}=Pr(S=s|S\geq s,Z=1)=0.03$. The regime assignment for the $Z=0$ group is  $\pi_{Z=0}=Pr(S=s|S\geq s,Z=0)=0.01$. We impose that the treatment takes place before the exit decision in period $t$.

Using this model we generate the treatment durations, exit durations, and accepted offers for a population of $5000$ agents over $5000$ periods. Within this population, $a_{i}$ and $e_{i}$ assume discrete values in the intervals $[1,6]$ and $[1,3]$ respectively. The initial randomization assigns half of the population to each regime $Z=0$ or $Z=1$. As in the discussion above, we choose to focus on a situation where the treatment affects only the offer distribution $G(w;a)$. The treatment effect is negative and is calibrated to equal one standard deviation of the offer distribution, $-\sigma_{w}$, with $\beta^{w}_{s}=0$. The full choice of parameters for each policy setting is presented below. To estimate the stationary solution for the accepted offer we simulate expectations of $w_{it}$ over 1000 draws and iterate over the value function until convergence.

\begin{table} [!h]
{ \caption{Parameter choices in simulations}
\vspace{-5mm}
\label{ta:ParamChoice}
\begin{center}
\begin{tabular}{l ccccccc } 
 \hline

 $\mu_{w}$    &     13.762,    &  $\mu_{c}$    &    0.893,        &  $\mu_{\lambda}$    &     0.092,  & $\rho$    &   0.995    \\

 $\sigma_{w}$    &      5.497,  &    $\sigma_{c}$    &    0.257,      &   $\sigma_{\lambda}$    &     0.031,   &  $\sigma_{\xi}$    &     3   \\

 $\beta^{w}_{a}$    &       4,  &   $\beta^{c}_{a}$    &      0.2,    &   $\beta^{\lambda}_{a}$    &   0.5/21,    & $\pi_{Z=0}$ &   0.01      \\

 $\beta^{w}_{s}$    &     5.497,   &  $\beta^{c}_{e}$    &    0.1,      &    $\beta^{\lambda}_{e}$   &   0.1/21,    & $\pi_{Z=1}$ &   0.03        \\[0.7ex]

   \\

\bottomrule
\end{tabular}
\end{center}
}
\end{table}

\subsection{Simulation Results and Discussion}
\label{se:SimRes}

\begin{figure} [!h]
  \begin{subfigure}{0.43\textwidth}
    \includegraphics[width=\linewidth]{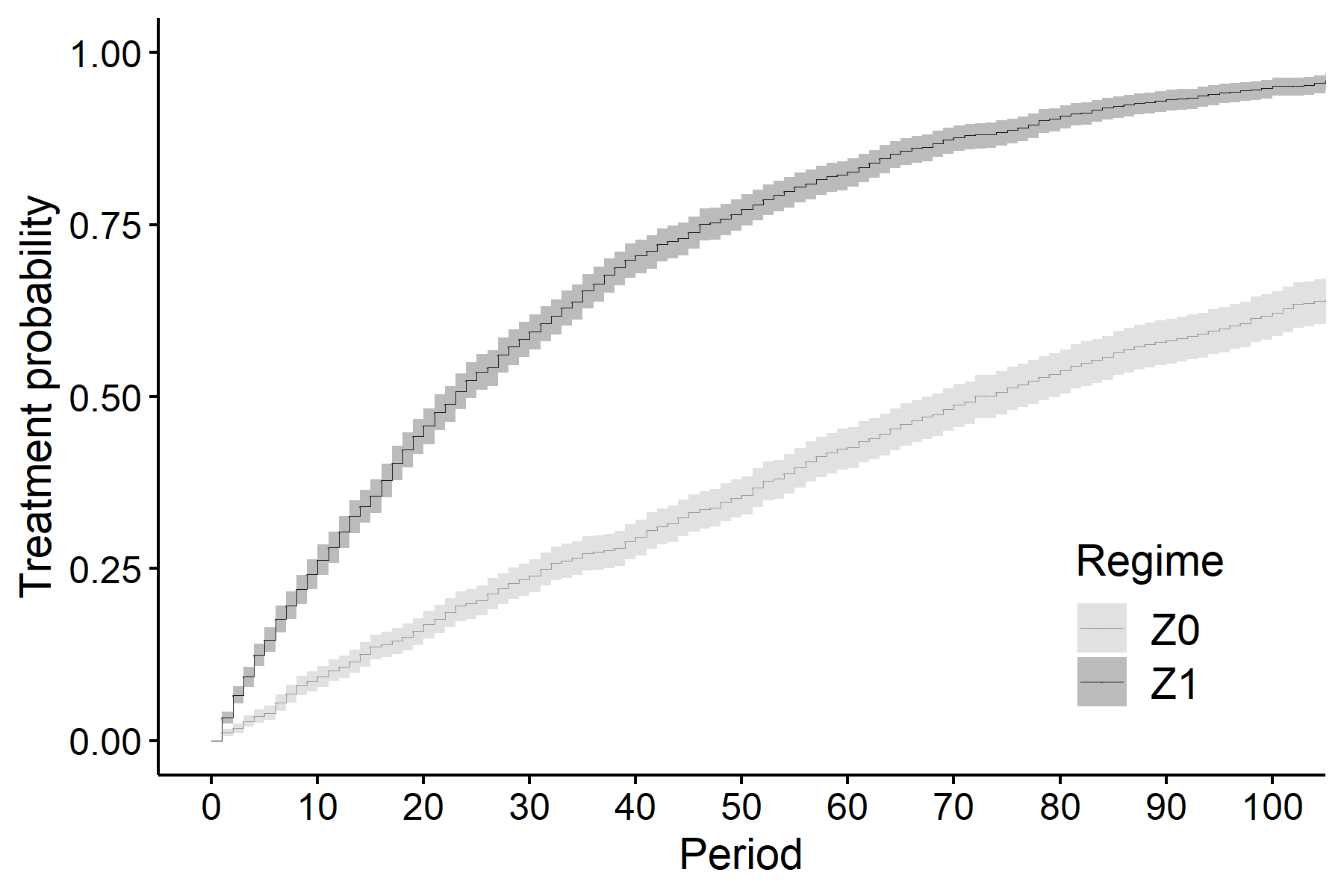}
    \caption{Transplant Probability} \label{fig:SurvT_sim}
  \end{subfigure}%
  \hspace*{\fill}   
  \begin{subfigure}{0.43\textwidth}
    \includegraphics[width=\linewidth]{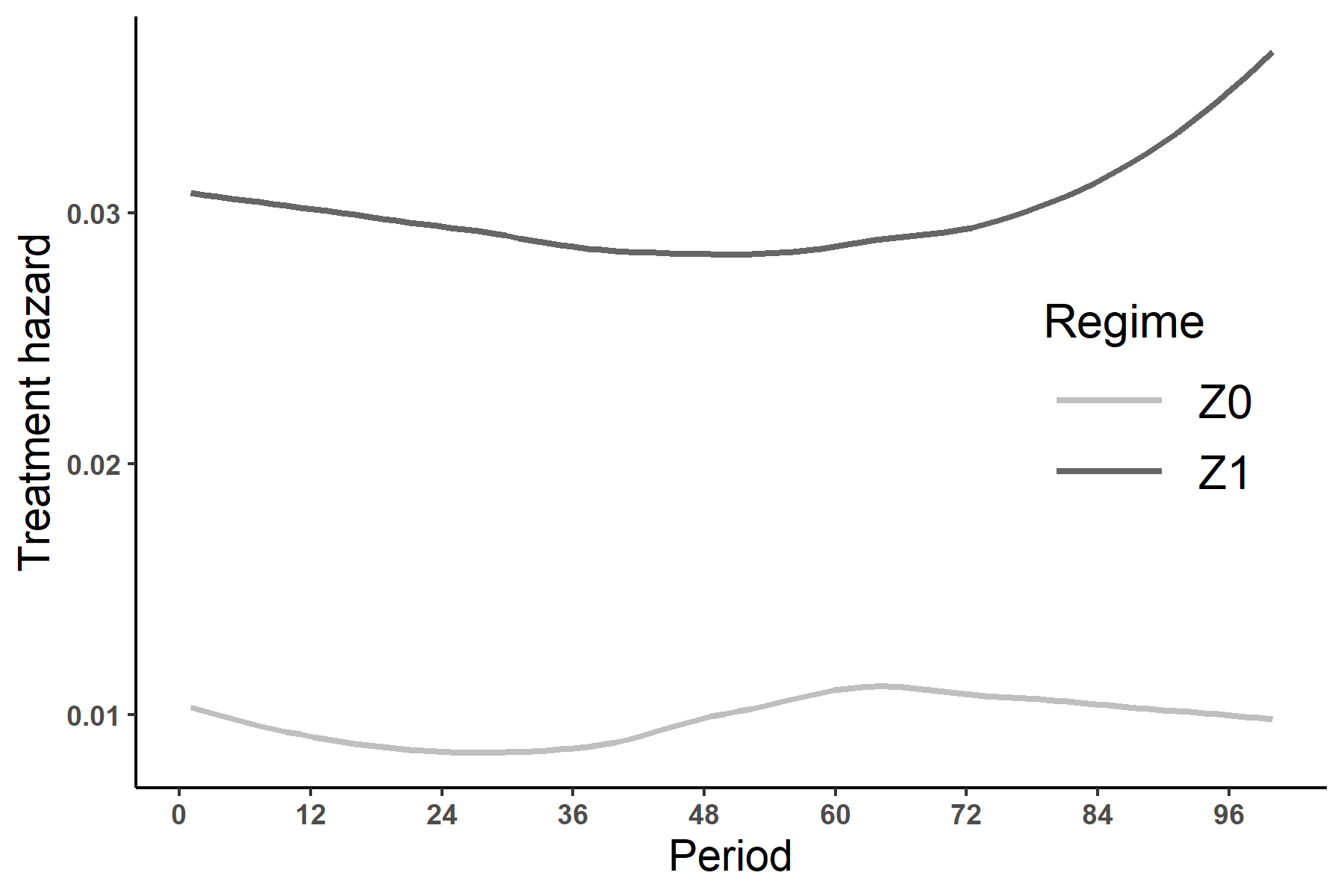}
    \caption{Transplant Hazard} \label{fig:Thaz_sim}
  \end{subfigure}%
  \newline
  \begin{subfigure}{0.43\textwidth}
    \includegraphics[width=\linewidth]{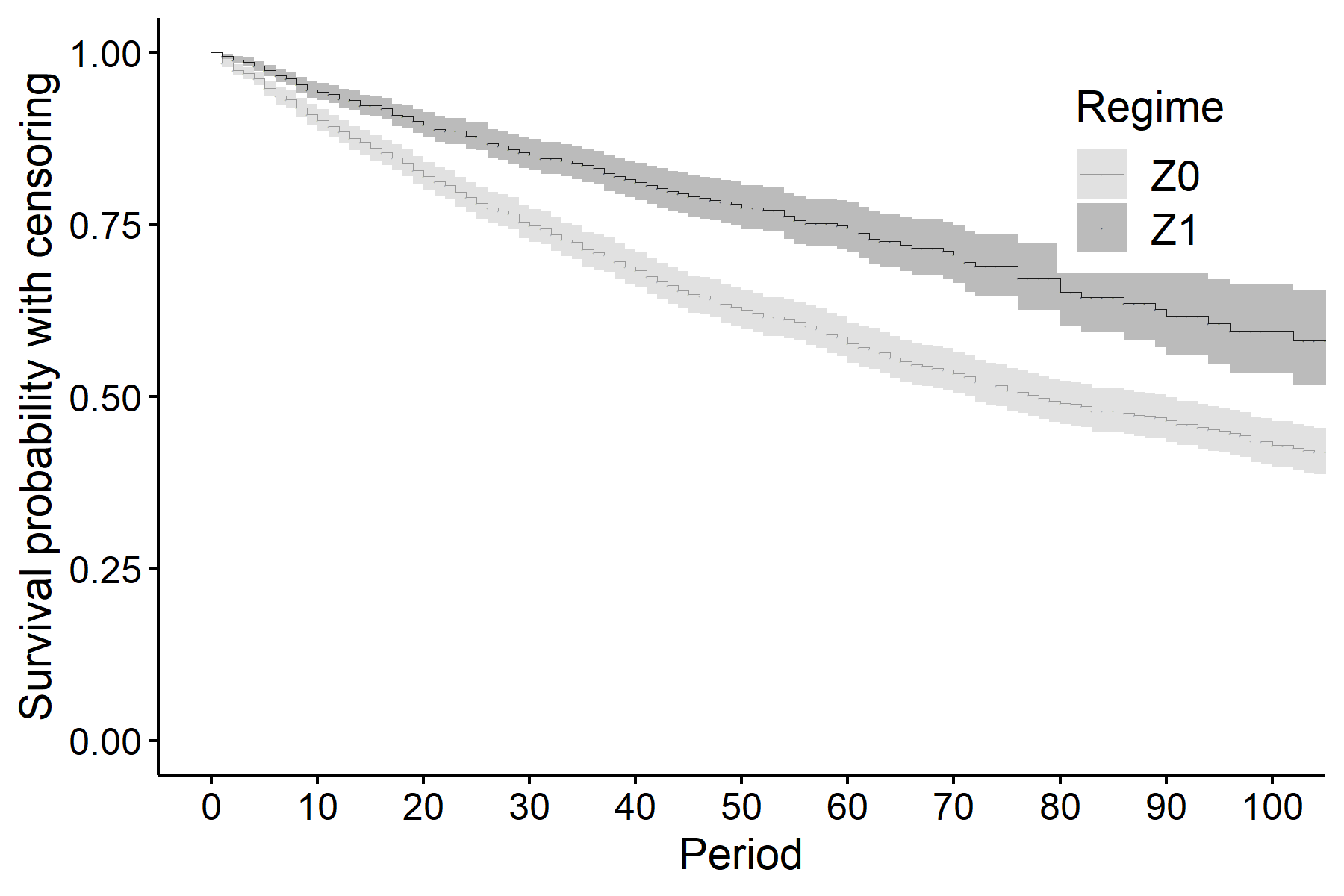}
    \caption{Pre-Transplant Survival} \label{fig:SurvC_sim}
  \end{subfigure}%
  \hspace*{\fill}   
  \begin{subfigure}{0.43\textwidth}
    \includegraphics[width=\linewidth]{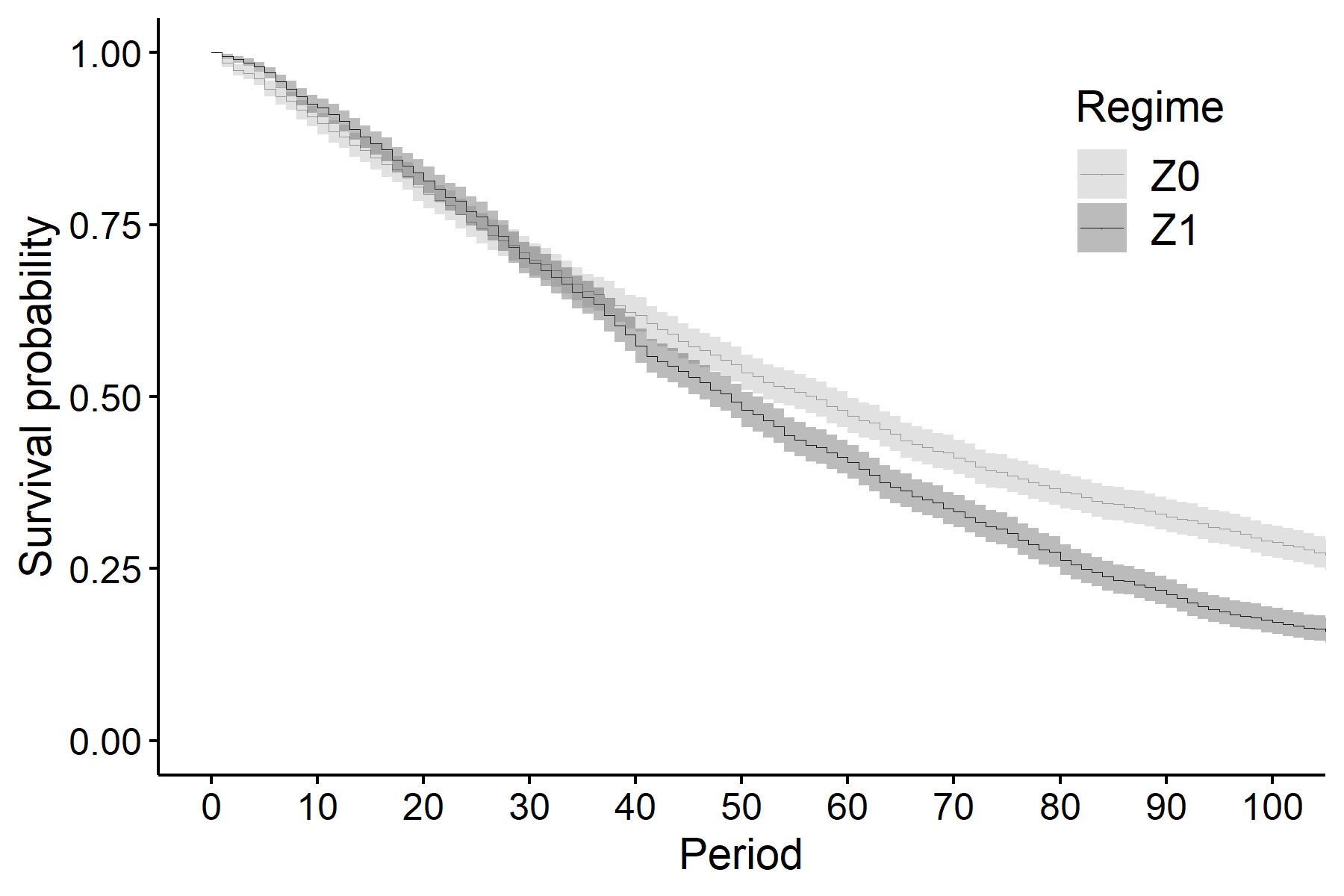}
    \caption{Survival} \label{fig:Surv_sim}
  \end{subfigure}%

\caption{Survival of candidates on the kidney transplant waitlist \\
    \scriptsize{Based on selected sample of Scientific Registry of Transplant Recipients data. Selected sample is described in section \ref{sec:data}. $N_{Z=0: no-Tr}=1013$, $N_{Z=0: Tr}=453$, $N_{Z=1: no-Tr}=441$, $N_{Z=1: Tr}=1093$.}} \label{fig:SurvAll_sim}
\end{figure}

To apply the continuous time methods under study in this paper, it is preferable to have a large dataset where the unit of time represents a relatively short period. In practice, if the unit of time is too large it may be challenging to account for dynamic selection and for the simultaneity of treatment and exit outcomes within a period. In the estimation, we treat effort $e_i$ as an unobserved characteristic for the researcher, and fully stratify $a_i$. The researcher observes individual treatment and exit duration outcomes, ability, $w_{0it}$, an indicator $Z$ for the regime assignment, and an indicator if the observation is right censored. After generating the dynamic discrete choice data we censor all observations greater than $t=60$ ($\sim 43.7\%$) and apply random right censoring to $\sim 6.3\%$ of the remaining observations. We present descriptive survival curves and hazards in Figures \ref{fig:SurvT_sim}-\ref{fig:Surv_sim}.

\begin{figure} [!h]
  \begin{subfigure}{0.48\textwidth}
    \includegraphics[width=\linewidth]{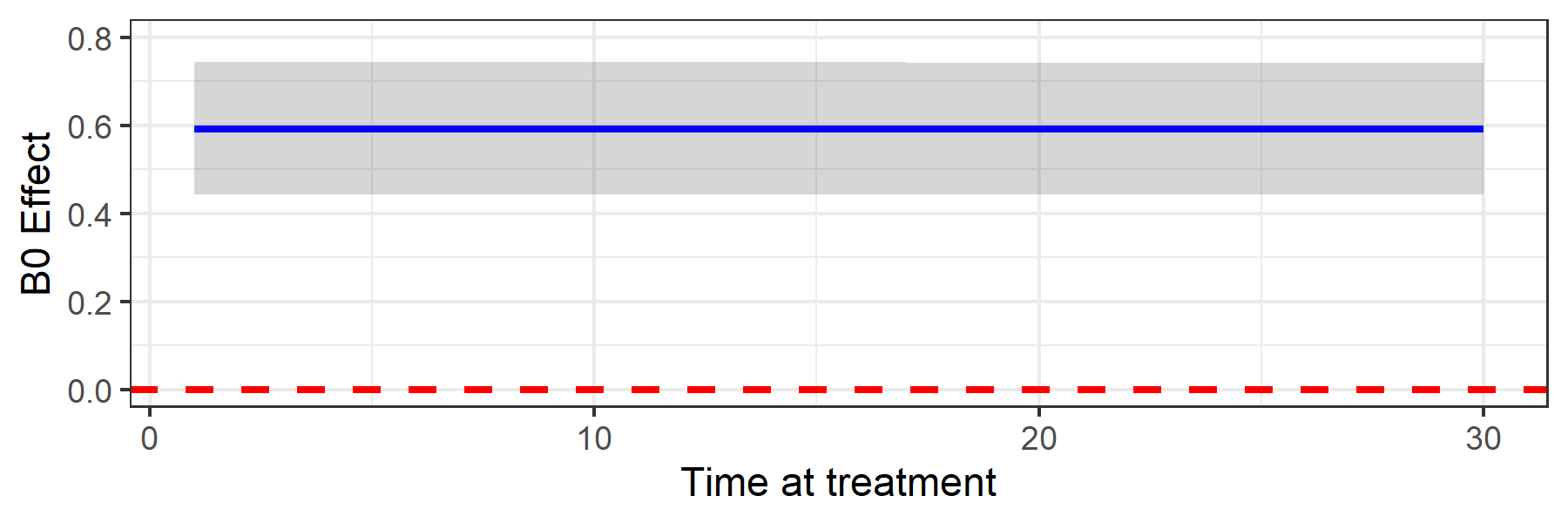}
    \caption{$\beta_0$ effect} \label{fig:B0_sim}
  \end{subfigure}%
  \hspace*{\fill}   
  \begin{subfigure}{0.48\textwidth}
    \includegraphics[width=\linewidth]{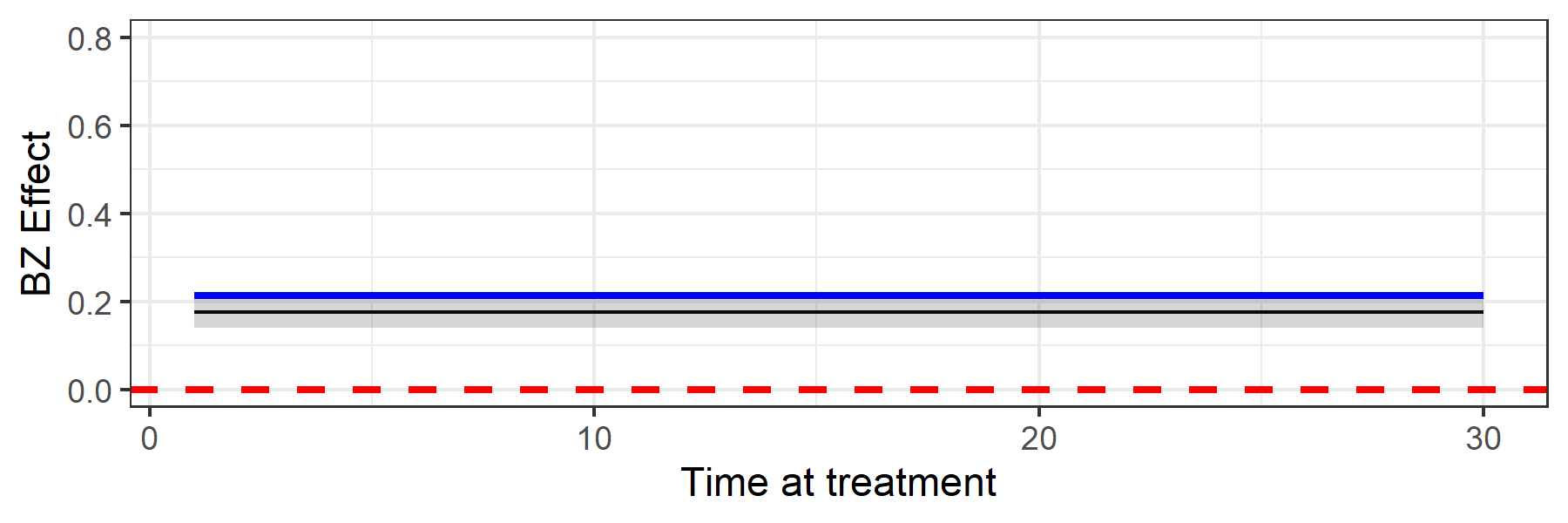}
    \caption{$\beta_z$ effect} \label{fig:BZ_sim}
  \end{subfigure}%
  \newline
  \begin{subfigure}{0.48\textwidth}
    \includegraphics[width=\linewidth]{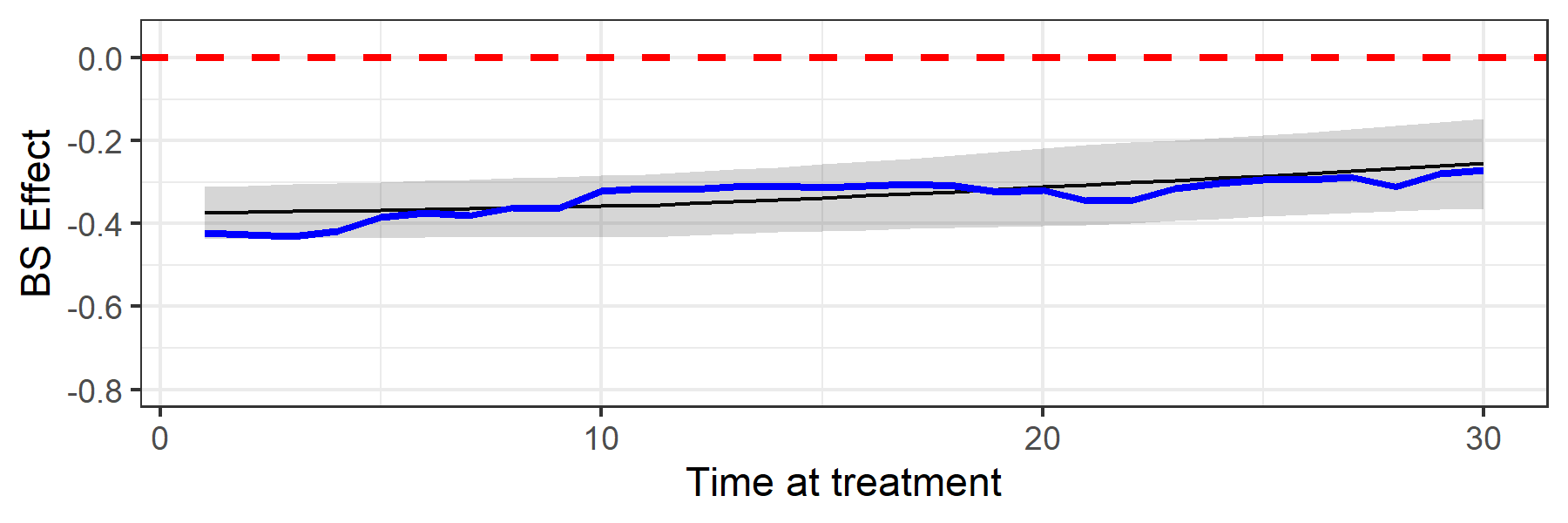}
    \caption{$\beta_s$ effect} \label{fig:BS_sim}
  \end{subfigure}%
  \hspace*{\fill}   
  \begin{subfigure}{0.48\textwidth}
    \includegraphics[width=\linewidth]{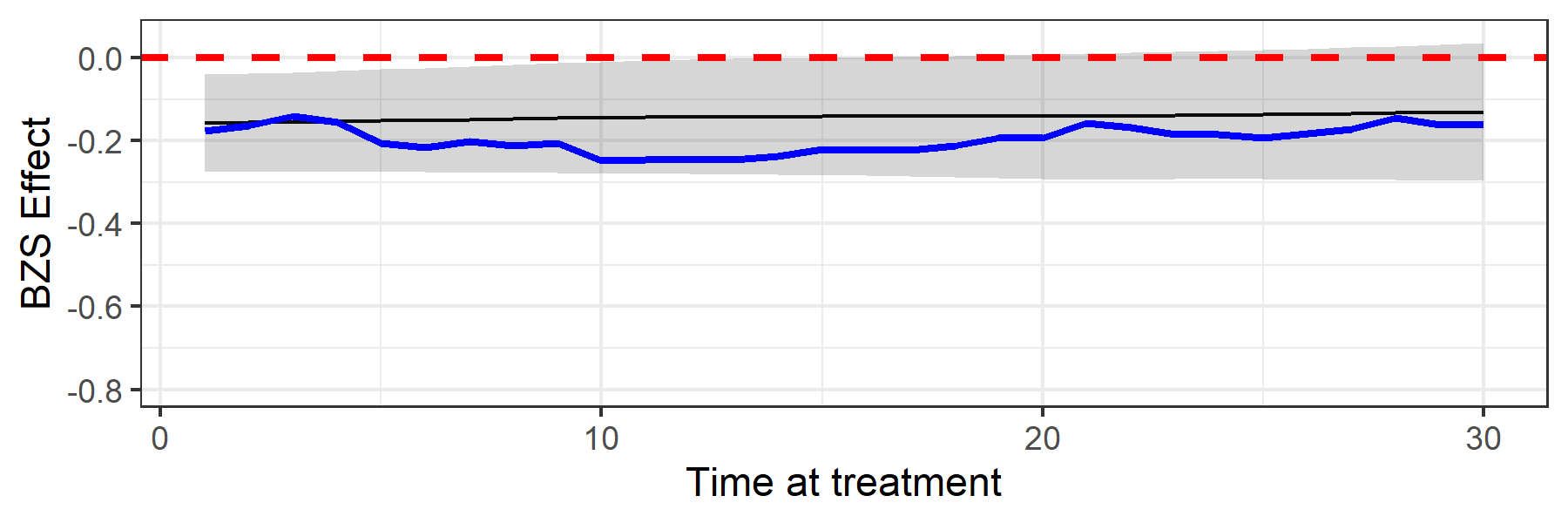}
    \caption{$\beta_{zs}$ effect} \label{fig:BZS_sim}
  \end{subfigure}%

\caption{Survival of candidates on the kidney transplant waitlist \\
    \scriptsize{Based on selected sample of Scientific Registry of Transplant Recipients data. Selected sample is described in section \ref{sec:data}. $N_{Z=0: no-Tr}=1013$, $N_{Z=0: Tr}=453$, $N_{Z=1: no-Tr}=441$, $N_{Z=1: Tr}=1093$.}} \label{fig:SurvAll_sim}
\end{figure}

Table \ref{ta:DGPsim} shows the results when applying our estimation method. We present the average effects for $\beta_0$, $\beta_z$, $\beta_{(0,s]}$, $\beta_{z(0,s]}$ over the treatment times $s=1$,\ldots,$30$ when specifying the duration dependence to six 10 period intervals. The estimator performs relatively well with all sample sizes observed. Figures \ref{fig:B0_sim}-\ref{fig:BZS_sim} provide further descriptions on the performance of our estimator. They present the estimates of $\beta_0$, $\beta_z$, $\beta_{s}$, $\beta_{zs}$ with $\tau$ fixed at $60$ over the first $30$ periods using a sample of $3000$ observations from the full population of $5000$. The estimator seems to fit the data very well for all causal estimates, and in the case of $\beta_0$, $\beta_{s}$, $\beta_{zs}$, also match closely the DGP values.

\begin{table} [!h] \scriptsize
\begin{center}
\caption{Simulation results of dynamic discrete choice model, $s$ from $1$,\ldots,$30$ and $s+\triangle$ fixed at $90$}
\label{ta:DGPsim}
\begin{tabular}{l cccc}
 \hline \hline
     &   Estimates
   & Bias &  Variance &
  MSE \\
  \hline
\\[-1ex]
\multicolumn{5}{l}{\emph{N=5000}} 										\\ [0.5ex]
$\beta_0$	&	0.590	&	0.001	&	0.006	&	0.006		\\[0.3ex]
$\beta_z$	&	0.170	&	-0.043	&	0.000	&	0.002		\\[0.3ex]
$\beta_{(0,s]}$	&	-0.297	&	0.040	&	0.002	&	0.003		\\[0.3ex]
$\beta_{z(0,s]}$	&	-0.180	&	-0.021	&	0.004	&	0.004		\\[2ex]
										
\multicolumn{5}{l}{\emph{N=3000}} 										\\ [0.5ex]
$\beta_0$	&	0.593	&	0.002	&	0.006	&	0.006		\\[0.3ex]
$\beta_z$	&	0.175	&	-0.039	&	0.000	&	0.002		\\[0.3ex]
$\beta_{(0,s]}$	&	-0.337	&	0.000	&	0.002	&	0.002		\\[0.3ex]
$\beta_{z(0,s]}$	&	-0.144	&	0.057	&	0.005	&	0.008		\\[2ex]
										
\multicolumn{5}{l}{\emph{N=1000}} 										\\ [0.5ex]
$\beta_0$	&	0.618	&	0.027	&	0.005	&	0.006		\\[0.3ex]
$\beta_z$	&	0.177	&	-0.036	&	0.000	&	0.002		\\[0.3ex]
$\beta_{(0,s]}$	&	-0.322	&	0.014	&	0.003	&	0.003		\\[0.3ex]
$\beta_{z(0,s]}$	&	-0.211	&	-0.010	&	0.004	&	0.004		\\[2ex]
										
\multicolumn{5}{l}{\emph{N=500}} 										\\ [0.5ex]
$\beta_0$	&	0.602	&	0.011	&	0.006	&	0.006		\\[0.3ex]
$\beta_z$	&	0.203	&	-0.011	&	0.001	&	0.001		\\[0.3ex]
$\beta_{(0,s]}$	&	-0.324	&	0.013	&	0.003	&	0.003		\\[0.3ex]
$\beta_{z(0,s]}$	&	-0.251	&	-0.050	&	0.007	&	0.010		\\[2ex]
										
\hline \hline
\end{tabular}
\end{center}
\end{table}

\newpage

\subsection{Reservation utilities in dynamic discrete choice model}
\label{app: reswDDC}

Consider the post-treatment Bellman equations from the setting described in section \ref{app: simResults}.
\[
\begin{split}
V_{0,t}&=w_0-c+\rho \lambda \mathbb{E}_w[\max\{V_1(w),(1-\pi)V_{0,t+1}+\pi V_{0,t+1}^{tr}\}]+\rho(1-\lambda)[(1-\pi)V_{0,t+1}+\pi V_{0,t+1}^{tr}]\\
V_{0,t}^{tr}&=w_0-c+\rho \lambda \mathbb{E}_{w^{tr}}[\max\{V_1^{tr}(w),V_{0,t+1}^{tr}\}+\rho(1-\lambda)V_{0,t+1}^{tr} \end{split}
\]
This second equation can be written as,
\[
V^{tr}_{0,t} = w_0 - c + \rho \lambda \int_{w^{tr*}}^{\infty}\left(V_1(w^{tr})-V^{tr}_{0,t+1}\right) dG^{tr}(w) + \rho V^{tr}_{0,t+1}
\]
Since all parameters and distributions in the model are time independent, we have a stationary reservation utility strategy. In a stationary strategy $V^{tr}_{0,t}=V^{tr}_{0,t+1}=V^{tr}_0$ for all $t>0$ Furthermore, the reservation utility $w^{tr*}$ is such that the agent would refuse any offer below it and accept any offer above it so $V_1(w)=\frac{w}{1-\rho}$ if $w\ge w^{tr*}$, $V^{tr}_0$ if $w^{tr}<w^{tr*}$, and $V_1(w^{tr*})=\frac{w^{tr*}}{1-\rho}=V^{tr}_0$ if $w=w^{tr*}$. It follows that,
\[
\begin{split}
 V^{tr}_{0} &= w_0 - c + \rho \lambda \int_{w^{tr*}}^{\infty}\left(V_1(w^{tr})-V^{tr}_{0}\right) dG^{tr}(w) + \rho V^{tr}_{0}\\
 &= w_0 - c + \frac{\rho \lambda}{1-\rho} \int_{w^{tr*}}^{\infty}\left(w-w^{tr*}\right) dG^{tr}(w) +  \frac{\rho w^{tr*}}{1-\rho}
 \end{split}
\]
Replacing again $V^{tr}_{0}=\frac{w^{tr*}}{1-\rho}$, rearranging this equation and using integration by parts we obtain the post-treatment reservation utility,
\[
w^{tr*}= w_0 - c + \frac{\rho \lambda}{1-\rho} \int_{w^{tr*}}^{\infty}\left(1-G^{tr}(w)\right) dw
\]
Note that this reservation utility does not depend on the regime $\pi$.

Now consider the reservation utility before treatment with a treatment assignment policy $\pi$. Since the problem is still stationary, the agent will again accept any value of $w$ higher than his reservation $w^{*}$. We can therefore rewrite\\ $\mathbb{E}_w[\max\{V_1(w)-(1-\pi)V_{0,t+1}-\pi V_{0,t+1}^{tr},0\}]=\int_{w^{*}}^{\infty}w-w^{*} dG(w)+(1-\pi)V_{0,t+1}-\pi V_{0,t+1}^{tr}$ which results in the same $V_{0}$ value function,
\[
V_{0}=w_0-c+ \frac{\rho \lambda}{1-\rho} \int_{w^{*}}^{\infty}w-w^{*} dG(w) +  \rho[(1-\pi)V_{0}+\pi V_{0}^{tr}]
\]
Since the agent will accept any value of $w$ higher than his reservation $w^{*}$ we know that  $V_1(w^*)=V_{0}=\frac{w^{*}}{1-\rho}$ which we replace in the above equation and rearrange to get,
\[
w^*=\frac{1-\rho}{1-\rho+\rho \pi}(w_0-c)+ \frac{\rho\lambda}{1-\rho+\rho \pi}(\int_{w^*(\pi)}^{+\infty}w-w^*dG(w))+\frac{\rho\pi}{1-\rho+\rho \pi}w^{tr*}
\]

We can further show prove that $w^*$ is increasing in $\pi$ if $V_0^{tr}>V_0$:

First we rewrite the previous equation to isolate $\int_{w^*}^{+\infty}w-w^*dG(w)$,
\[
\int_{w^*}^{+\infty}w-w^*dG(w)= \frac{1-\rho}{\rho\lambda}[c-w_0+\frac{1-\rho(1-\pi)}{1-\rho}w^{*}- \frac{\rho\lambda\pi+\rho(1-\lambda)\pi}{1-\rho}w^{tr*}]
\]

Let us hold $w^*$ constant in the previous equation. $w^{tr*}$ is also constant because it does not depend on $\pi$. If we increase $\pi$, the derivative of the right-hand side with respect to $\pi$ is
\[
\frac{1-\rho}{\rho\lambda}[\frac{\rho}{1-\rho}w^*- \frac{\rho}{1-\rho}w^{tr*}] \qquad \leftrightarrow \qquad \frac{1-\rho}{\lambda}[(V_0-V_0^{tr})]
\]
The last equation is negative when $V_0-V_0^{tr}<0$, so when the value of treatment is higher than that of no-treatment. Therefore, $\int_{w^*}^{+\infty}w-w^*dG(w)$ is decreasing in $\pi$. Furthermore, written as a function of $\pi$,with $w^*=w^*(\pi)$, it is also decreasing in $w^*$ which implies that $w^*$ increases in $\pi$. The agent is more willing to wait for treatment.

\vspace{10mm}

\begin{flushleft}
\textbf{Appendix References:}
\end{flushleft}

\setlength{\leftlocal}{\leftmargini}
\addtolength{\leftmargini}{-.5\leftmargini}
\begin{description}
\setlength{\labellocal}{\labelwidth} \setlength{\labelwidth}{5pt}
\setlength{\itemlocal}{\itemsep} \setlength{\itemsep}{0pt}

\item \hypertarget{Mor1986}{Mortensen, D.T. (1986). Job search and labor market analysis'', In: Ashenfelter, O., Card, D. (Eds.), \emph{Handbook
of Labor Economics. In: Handbooks in Economics}, vol. 2. Elsevier Science, New York, pp. 849–919.}

\item  \hypertarget{RogEA2005}{Rogerson, R., Shimer, R., \& Wright, R. (2005). Search-Theoretic Models of the Labor Market: A Survey. \emph{Journal of Economic Literature}, 43(4): 959-988.}

\end{description}
\setlength{\leftmargini}{\leftlocal}
\setlength{\labelwidth}{\labellocal}
\setlength{\itemsep}{\itemlocal}

\end{document}